\documentclass[preprint,12pt]{elsarticle}

\usepackage{mathtools}
\usepackage{booktabs}
\usepackage{amssymb}
\usepackage{float}
\usepackage{subcaption}
\usepackage{xcolor}
\usepackage{graphicx}
\usepackage{csquotes}
\usepackage[hidelinks, colorlinks=false, breaklinks]{hyperref}
\usepackage[nameinlink]{cleveref}
\usepackage{varwidth}
\usepackage[export]{adjustbox}
\usepackage{tikz}

\usepackage[most]{tcolorbox}

\biboptions{sort&compress}

\journal{Comput.\ Methods Appl.\ Mech.\ Engrg.}

\begin{document}

\begin{frontmatter}

\title{A survey on multi-fidelity surrogates for simulators with functional outputs: unified framework and benchmark}

\author[onera,limos]{Lucas Brunel}
\author[onera]{Mathieu Balesdent}
\author[onera]{Loïc Brevault}
\author[limos]{Rodolphe Le~Riche}
\author[eth]{Bruno Sudret}

\affiliation[onera]{organization={ONERA, Paris~Saclay University},
            city={Palaiseau},
            postcode={F-91123}, 
            country={France}}

\affiliation[limos]{organization={LIMOS (CNRS, Mines Saint-Etienne and Université Clermont Auvergne)},
            country={France}}

\affiliation[eth]{organization={ETH Zürich, Chair of Risk, Safety and Uncertainty Quantification},
            country={Switzerland}}

\begin{abstract}

    Multi-fidelity surrogate models combining dimensionality reduction and an intermediate surrogate in the reduced space allow a cost-effective emulation of simulators with functional outputs. 
    The surrogate is an input-output mapping learned from a limited number of simulator evaluations.
    This computational efficiency makes surrogates commonly used for many-query tasks.
    Diverse methods for building them have been proposed in the literature, but they have only been partially compared.

    This paper introduces a unified framework encompassing the different surrogate families, followed by a methodological comparison and the exposition of practical considerations.
    More than a dozen existing multi-fidelity surrogates have been implemented under the unified framework and evaluated on a set of benchmark problems.
    Based on the results, guidelines and recommendations are proposed regarding multi-fidelity surrogates with functional outputs.

    Our study shows that most multi-fidelity surrogates outperform their tested single-fidelity counterparts under the considered settings.
    However, no particular surrogate is performing better on every test case. Therefore, the selection of a surrogate should consider the specific properties of the emulated functions, in particular the correlation between the low- and high-fidelity simulators, the size of the training set, and the local nonlinear variations in the residual fields.

\end{abstract}


\end{frontmatter}

\section{Introduction}\label{sec:intro}

    In computational science and engineering, many-query tasks such as optimization, uncertainty quantification, and sensitivity analysis play crucial roles in decision-making processes.
    These tasks involve numerous simulations of complex phenomena.
    In classical engineering scenarios, the simulator output is functional (\textit{e.g.}, a scalar field) and discretized on a high-dimensional mesh, which can, for instance, be temporal or spatial.
    The computational cost of a single run of a complex simulator can be substantial, making a large number of evaluations unaffordable in practice.
    A common solution to this challenge is the construction of a mathematical approximation of the simulator, known as a \emph{surrogate model}, or simply \emph{surrogate}.
    Surrogate models are significantly cheaper to evaluate than the original simulators, enabling cost-effective predictions of the simulator outputs.
    In most cases, multiple fidelities of simulators are available, by varying the level of simplifications in the physics of the problem, the mesh refinement, the convergence tolerance, \textit{etc}.
    For instance, in Computational Fluid Dynamics (CFD), a high-fidelity model could be a CFD Reynolds-Averaged Navier-Stokes (RANS) simulator on a fine mesh, while a low-fidelity one could be a simpler CFD Euler simulator on a coarse mesh.
    The surrogate can be adapted to exploit the multiple fidelities available, thus becoming a \emph{multi-fidelity} surrogate.

    Among the existing approaches for constructing multi-fidelity surrogates capable of predicting functional outputs, two main families can be distinguished.
    The first combines two separate tasks: dimensionality reduction (DR) and intermediate surrogate modeling \cite{benamaraMultifidelity2017, bunnellMultifidelity2021, kerleguerMultifidelity2023, malouinInterpolation2013, perronMultifidelity2021, deckerManifold2022, rokitaMultifidelity2018, thenonMultifidelity2016, toalPotential2014, wangMultifidelity2020, mifsudVariablefidelity2016, parussiniMultifidelity2017}.
    Taking advantage of the spatial or temporal dependence of the nodes of the output field, DR describes high-dimensional data by a significantly smaller number of variables, providing a mapping between the original high-dimensional space and the new low-dimensional space.
    Then, intermediate surrogate modeling is performed between the simulator input space and this new low-dimensional space on this limited number of variables.
    The second family of approaches intertwines DR and intermediate surrogate modeling into a unified task, thanks to artificial neural networks (see, \textit{e.g.}, \cite{guoMultifidelity2022, liDeep2022, yangNeuralphysics2023}).

    Although several multi-fidelity surrogates of the first family (\textit{i.e.}, combining DR and intermediate surrogate modeling) have been proposed in recent literature \cite{benamaraMultifidelity2017, bunnellMultifidelity2021, kerleguerMultifidelity2023, malouinInterpolation2013, perronMultifidelity2021, deckerManifold2022, rokitaMultifidelity2018, thenonMultifidelity2016, toalPotential2014, wangMultifidelity2020, mifsudVariablefidelity2016, parussiniMultifidelity2017}, there has been limited theoretical and numerical comparison between them.
    Moreover, each multi-fidelity surrogate has been described in its own particular formalism.
    This paper aims at describing these existing multi-fidelity surrogates with functional outputs within a \emph{unified framework} and at benchmarking them on a set of test cases.
    The proposed framework is designed to be modular.
    It facilitates the adaptation of existing surrogates and the development of new ones.
    The benchmark involves viscous flow and airfoil design test cases of increasing complexity, considering two simulator fidelities, and is followed by thorough discussions.
    Recommendations based on the theoretical and numerical comparisons are provided to assist practitioners in choosing the appropriate surrogate for a given application.

    The outline of this paper is as follows.
    \Cref{sec:background} describes the different building blocks of the surrogates in a unified formalism.
    \Cref{sec:metamodels} presents the different families of multi-fidelity surrogates and briefly details each of them.
    \Cref{sec:test-cases} introduces the test cases making up the benchmark.
    \Cref{sec:benchmark-methodology} presents the benchmark methodology and the comparison metrics.
    Finally, \Cref{sec:results-discussions} displays and discusses the performance of the surrogates on the test cases to provide practitioners with guidelines depending on the characteristics of their problems.


\section{Background}\label{sec:background}

    In this work, several simulators with hierarchical fidelities are considered.
    The difference in fidelity may be based on different aspects such as physical simplifications (\textit{e.g.}, CFD Large Eddy Simulation (LES) for the high-fidelity and CFD Euler for the low-fidelity, which simplifies the modeling of viscosity), numerical discretization (\textit{e.g.}, fine and coarse mesh) or numerical tolerances (\textit{e.g.}, different thresholds of residuals involved in the solvers).
    The fidelities can be described by their computational cost and accuracy.
    A low-fidelity simulator presents a reduced computational cost but a low accuracy, while a high-fidelity simulator is more accurate but has a large computational cost.
    The level of fidelity is denoted by the index $l\in\{1,\dots,L\}$, sorted in decreasing fidelity order such that $l=1$ corresponds to the highest fidelity and $l=L$ to the lowest fidelity.
    It is assumed that the input variables of these simulators (\textit{e.g.}, design or environment variables) are the same for each fidelity and are gathered in the vector $\mathbf{u}\in U\subseteq\mathbb{R}^{d_{\mathbf{u}}}$.
    Each simulator $\mathcal{S}_l$ of fidelity level $l$ outputs a scalar valued field $\mathbf{y}_l \in Y_l\subseteq \mathbb{R}^{d_{\mathbf{y}_l}}$ that corresponds to a high-dimensional mesh $\mathbf{X}_l \in X_l \subseteq\mathbb{R}^{d_{\mathbf{y}_l}\times d_{\mathbf{x}}}$ (\textit{e.g.}, spatial or temporal), with $d_{\mathbf{y}_l}$ being the number of nodes, or vertices, in a $d_{\mathbf{x}}$-dimensional space, corresponding to the mesh coordinates in the associated space.
    $d_\mathbf{x}$ is assumed to be the same for all fidelities (\textit{e.g.}, all simulators output two-dimensional fields).
    The $j$-th component of the output field is the scalar $y_{j,l}\in\mathbb{R}$, with $\mathbf{x}_{j,l}$ the coordinate of the $j$-th node of $\mathbf{X}_l$.
    Note that fidelities can have a different number of mesh nodes $d_{\mathbf{y}_l}$, or the same $\forall l\in\{1,\dots,L\},d_{\mathbf{y}_l}=d_\mathbf{y}$ but with coordinates that may vary across different snapshots (see \cite{hsuSimplified2004} for illustrations).
    To summarize, the simulator $\mathcal{S}_l$ of fidelity level $l$ discretized on the mesh $\mathbf{X}_l$ is the mapping

    \begin{equation}
        \begin{split}
            \mathcal{S}_l: U & \rightarrow Y_l \\
            \mathbf{u} & \mapsto \mathbf{y}_l =\mathcal{S}_l(\mathbf{u}; \mathbf{X}_l).
        \end{split}
    \end{equation}

\begin{figure}
        \centering
        \includegraphics[width=0.65\textwidth]{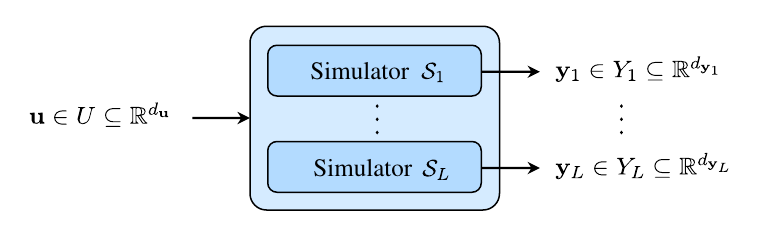}
        \caption{Sketch of a generic multi-fidelity simulation, with the input vector $\mathbf{u}\in U$ and at fidelity level $l\in\{1,\dots,L\}$ the simulator $\mathcal{S}_l$, the mesh $\mathbf{X}_l\in X_l$ and the scalar output field $\mathbf{y}_l\in Y_l$.}
        \label{fig:mf-simulators}
    \end{figure}

    Whenever the reference to the mesh $\mathbf{X}_l$ is not necessary, the above-mentioned equation is simply written as $\mathbf{y}_l = \mathcal{S}_l(\mathbf{u})$.
    The multi-fidelity simulation setup described is illustrated in \Cref{fig:mf-simulators}.

    \subsection{Overview of DR techniques}\label{sec:dr}

        \subsubsection{Single-fidelity techniques}\label{sec:dr-monofi}

            DR is first introduced for the case of a single fidelity level $l$.
            Then, an overview of the specific methods used in multi-fidelity surrogates with functional outputs is offered.
            In this work, $Y_l$ is supposed to be a vector space.
            The underlying physical model of the simulator implies that the components of the output field are usually (spatially, temporally, \textit{etc.}) dependent.
            DR assumes that the observed discretized output fields lay on or near a subspace $Z_l$ of the high-dimensional vector space $ Y_l$.
            This subspace is called the \emph{latent space} and is mathematically modeled by a $d_{\mathbf{z}_l}$-dimensional manifold.
            The coordinates on the manifold are the \emph{latent variables} and denoted by $\mathbf{z}_l$. $d_{\mathbf{z}_l}$ is the \emph{intrinsic dimension} \cite{camastraIntrinsic2016} and $d_{\mathbf{z}_l}<d_{\mathbf{y}_l}$ (often $d_{\mathbf{z}_l} \ll d_{\mathbf{y}_l}$).
            Usually, neither the dimensionality $d_{\mathbf{z}_l}$ nor the structure of the manifold (differentiable, discontinuous, \textit{etc.}) is known.
            Performing DR consists in finding the mapping $\mathcal{DR}_l^{\boldsymbol\omega _l}$ from the high-dimensional vector space $ Y_l$ to the lower-dimensional space $Z_l$, with $\boldsymbol\omega _l\in\mathbb{R}^{d_{\boldsymbol\omega _l}}$ a set of hyperparameters which the mapping may depend on.
            This mapping is summarized as:

            \begin{equation}\label{eq:dr}
                \begin{split}
                    \mathcal{DR}_l^{\boldsymbol\omega _l} :  Y_l & \rightarrow Z_l \\
                    \mathbf{y}_l & \mapsto \mathbf{z}_l
                \end{split}
            \end{equation}
            For the sake of readability, the dependence of $\mathcal{DR}_l^{\boldsymbol\omega _l}$ on $\boldsymbol\omega _l$ is omitted in the following.
            $\mathbf{z}_l$ is then a low-dimensional representation of the output field $\mathbf{y}_l$.
            In practice, assumptions on its structure and dimensionality are required in order to be able to work with $Z_l$ and compute $\mathcal{DR}_l^{\boldsymbol\omega_l}$.
            Ideally, this transformation should not yield information loss. 
            In practice though, since assumptions are made on the structure and dimensionality of the manifold, some information may be lost.
    
            DR techniques are mostly data-driven.
            In order to be accurate enough, they require a certain amount of \emph{snapshots}, \textit{i.e.}, a set of output fields of the simulator $\mathcal{S}_l$ obtained for different values of the input variables.
            The $n_l$ samples of $\mathbf{u}$ for which the simulator is evaluated are collected into the input Design of Experiments (DoE) $\mathcal{U}_l= \left\{\mathbf{u}_l^{(i)}\right\}^{n_l}_{i=1}$, with $\mathbf{I}_l= \left[\mathbf{u}_l^{(1)}, \dots, \mathbf{u}_l^{(n_l)}\right]^\top$ its associated $n_l$-by-$d_{\mathbf{u}}$ matrix form.
            The output DoE at fidelity $l$ is denoted by $\mathcal{Y}_l= \left\{\mathbf{y}_l^{(i)}\right\}^{n_l}_{i=1}$ and its $n_l$-by-$d_{\mathbf{y}_l}$ matrix form $\mathbf{O}_l= \left[\mathbf{y}_l^{(1)}, \dots, \mathbf{y}_l^{(n_l)}\right]^\top$.
            The yet to be computed collection of the latent variables corresponding to all the snapshots of the output DoE is $\mathcal{Z}_l= \left\{\mathbf{z}_l^{(i)}\right\}^{n_l}_{i=1}$ and its $n_l$-by-$d_{\mathbf{z}_l}$ matrix form $\mathbf{L}_l= \left[\mathbf{z}_l^{(1)}, \dots, \mathbf{z}_l^{(n_l)}\right]^\top$.
            The strategy for choosing these samples may have a large impact on the quality of the DR.
            A common sampling method is Latin Hypercube Sampling (LHS) \cite{santnerSpaceFilling2018}, designed to cover the input parameter space with uniform marginal distributions.
    
            There exists a wide range of DR techniques that have been developed for different fields of research and engineering.
            Multiple classifications of these techniques have been proposed, such as \cite{vandermaatenDimensionality2009, jiaFeature2022, huangReview2019}.

            Dimensionality reduction techniques can be classified according to the linearity of their mapping.
            Linear techniques are based on the hypothesis that the manifold is a vector space and the mapping between the original and the latent spaces is linear.
            The most well-known linear technique is \emph{Principal Component Analysis} (PCA), also known as proper orthogonal decomposition in the model order reduction community, and Karhunen-Loève expansion for stochastic processes \cite{liangProper2002}.
            As the scalar field is discretized, it is based on a decomposition of the covariance matrix associated with snapshots over an orthonormal basis, consisting in a set of eigenvectors and eigenvalues (Mercer's theorem \cite{mercerFunctions1909}).
            In practice, the basis is truncated by retaining the $d_{\mathbf{z}_l}$ eigenvectors corresponding to the largest eigenvalues.
            The dimensionality $d_{\mathbf{z}_l}$ of the latent space is considered to be a hyperparameter of PCA as it is not known beforehand and has to be tuned.
            It is often chosen based on the Relative Information Content (RIC) \cite{pinnauModel2008} defined as the sum of the $d_{\mathbf{z}_l}$ largest eigenvalues divided by the sum of all eigenvalues.
            Factor Analysis \cite{gorsuchBasic2014} and Independent Component Analysis \cite{comonIndependent1994} are other examples of linear DR methods.
            When the gradient of the functional output is available, methods such active subspaces \cite{constantineActive2015} can take advantage of this additional information.

            In this paragraph, the PCA of a fidelity level $l$ is briefly summarized from a practical standpoint.
            First of all, the mean of the snapshots $\bar{\mathbf{y}}_l=\frac{1}{n_l}\sum^{n_l}_{i=1}\mathbf y_l^{(i)}$ is subtracted from all snapshots to center them.
            The matrix of the centered snapshots is denoted by $\bar{\mathbf{O}}_l$.
            Note that when inverse mapping the PCA, $\bar{\mathbf{y}}_l$ must be added back to get a snapshot of the field.
            Subsequently, the $d_{\mathbf y_l}$-by-$d_{\mathbf y_l}$ sample covariance matrix $\mathbf C_l = \frac{1}{n_l}\bar{\mathbf{O}}_l^\top\bar{\mathbf{O}}_l$ is computed.
            Then, the eigendecomposition of $\mathbf{C}_l$ results in the eigenvalues $\lambda_i\in\mathbb{R}$ and eigenvectors $\psi_i\in\mathbb{R}^{d_{\mathbf y_l}}$, also called PCA modes, with $i=1,\dots,d_{\mathbf y_l}$.
            To effectively reduce the dimensionality of the output field, only a limited number $d_{\mathbf z_l}$ of eigenvectors is retained.
            As mentioned in the previous paragraph, the RIC \cite{pinnauModel2008} is a popular way of truncating the eigenbasis.
            The selected eigenvectors are collected in the projection matrix $\boldsymbol{\Psi}=[\psi_1,\dots,\psi_{d_{\mathbf z_l}}]^\top$.
            To get the latent variable values corresponding to the snapshots in $\bar{\mathbf{O}}_l$, it is projected onto the orthogonal basis formed by the eigenvectors, \textit{i.e.}, $\mathbf{L}_l=\bar{\mathbf{O}}_l\boldsymbol{\Psi}^\top$.
            The latent variables are uncorrelated, but not independent.
            Usually, instead of an eigendecomposition of the sample covariance matrix, an SVD of $\bar{\mathbf{O}}_l$ is performed \cite{liangProper2002}.

            Nonlinear techniques offer more flexibility for the structure of the manifold.
            However, some lack a closed-form inverse mapping $\mathcal{DR}_l^{-1}$ from the latent space to the original high-dimensional space.
            They involve hyperparameters in addition to the dimensionality of the latent space that have to be chosen as well.
            An example of nonlinear DR technique is Kernel PCA \cite{scholkopfNonlinear1998}, which stems from PCA, combined with the kernel trick \cite{scholkopfKernel2000}.
            The snapshots of dimensionality $d_{\mathbf{y}_l}$ are nonlinearly mapped to an even higher-dimensional feature space, where PCA is performed.
            Consequently, the mapping between the original space $Y_l$ and the latent space $Z_l$ is nonlinear.
            The kernel is chosen such that no computation is actually done in the higher-dimensional space.
            The correlation length of the kernel is another hyperparameter that needs tuning.
            The family of techniques referred to as manifold learning are also used for nonlinear DR techniques (\textit{e.g.}, isometric feature mapping \cite{tenenbaumMapping1997},or isomap for short, Local Tangent Space Alignment (LTSA) \cite{zhangPrincipal2004}).
            Autoencoders are another notable nonlinear technique \cite{bankAutoencoders2023}, which rely on artificial neural networks to compute the low-dimensional representation of the snapshots.

            This paragraph provides a brief overview of isomap and LTSA.
            Isomap \cite{tenenbaumMapping1997} seeks to map the original space $Y_l$ to the latent space $Z_l$ such that the geodesic distance in $Y_l$ between two training samples is best preserved.
            This method is based on the assumption that the geodesic distance in $Y_l$ is an accurate representation of the geodesic distance in $Z_l$.
            The process begins by building a $k$-nearest neighbors graph \cite{tenenbaumMapping1997} of the training samples.
            Each edge is assigned a weight equal to the Euclidean distance in $Y_l$ between the two samples making its vertices.
            The geodesic distance between each pair of training samples is approximated by the shortest path along this graph.
            Finally, nonmetric multidimensional scaling \cite{coxMultidimensional2001} finds a $d_{\mathbf{z}_l}$-dimensional latent space that aims to preserve these geodesic distances.
            LTSA \cite{zhangPrincipal2004}, in essence, identifies the local tangent space at each sample and then tries to align them.
            In practice, the local tangent space at each sample is approximated by the vector space spanned by the $d_{\mathbf{z}_l}$ first modes of a PCA performed on the sample and its $k$-nearest neighbors.
            This creates local latent coordinates, which are transformed into global latent coordinates through a linear transformation, which is obtained by minimizing the local reconstruction errors.
            For further details regarding the tuning of the hyperparameters for these methods, notably the number $k$ of neighbors, the reader is referred to \ref{app:numerical-settings}.

        \subsubsection{Multi-fidelity techniques}\label{sec:dr-multifi}

            The extension of DR to the multi-fidelity case can be accomplished in three different ways.
            The first consists in computing the DR mapping for one of the fidelity levels $\tilde{l}$ and to use this mapping $\mathcal{DR}_{\tilde{l}}$ to compute the low-dimensional representation of the snapshots for all fidelity levels (\textit{i.e.}, $\mathcal{DR}_l=\mathcal{DR}_{\tilde{l}}, \forall l\in\{1,\dots,L\}$) \cite{kerleguerMultifidelity2023, thenonMultifidelity2016}.
            All the output fields of the different fidelity levels are consequently mapped to the same manifold $ Z_{\tilde{l}}$.
            For this purpose, every output field must be discretized on the same mesh.
            Otherwise, a preprocessing step must be applied to project all snapshots onto a common mesh.
            It also assumes that DR computed at fidelity level $\tilde{l}$ can accurately represent snapshots of all fidelity levels, \textit{i.e.}, that the physical information captured at fidelity level $\tilde{l}$ also describes the physics of other fidelity snapshots.
            This approach enables the use of the DR techniques described in \Cref{sec:dr-monofi}.
            The choice of the fidelity level $\tilde{l}$ is important and may be driven by the amount of snapshots at certain fidelity levels \cite{kerleguerMultifidelity2023}.
            For instance, if too few higher-fidelity snapshots are available, one may use a lower-fidelity to perform DR.

            The second way is to apply a DR technique designed to be able to use the aggregated snapshots available for all fidelity levels (\textit{i.e.}, $\mathcal{DR}_l=\mathcal{DR}, \forall l\in\{1,\dots,L\}$, with $\mathcal{DR}$ computed using all snapshots at once) \cite{benamaraMultifidelity2017, rokitaMultifidelity2018, toalPotential2014, mifsudVariablefidelity2016}.
            Thus, snapshots of all fidelity levels are mapped to the same shared manifold.
            This approach requires the definition of new DR techniques, or the adaptation of the ones introduced in \Cref{sec:dr-monofi}.
            In most cases, this requires the snapshots of all fidelity levels to be discretized on the same mesh (\textit{e.g.}, in PCA to compute the covariance between each component of the discretized field, in isomap to compute the distance between the different snapshots).
            For instance, in PCA, this can be accomplished by computing the eigenvectors of the covariance matrix of the union of snapshots from all fidelity levels \cite{rokitaMultifidelity2018, mifsudVariablefidelity2016}.

            The third way is to independently perform one DR per fidelity (\textit{i.e.}, $\mathcal{DR}_{l}\neq\mathcal{DR}_{l'},\forall l\neq l'\in\{1,\dots,L\}$) \cite{bunnellMultifidelity2021, perronMultifidelity2021, deckerManifold2022, wangMultifidelity2020}.
            Consequently, there are as many DR mappings as fidelity levels.
            To be able to transfer information between the different manifolds $Z_l$, this approach requires that either manifolds have a similar structure at every fidelity level,
            or a mapping between the different manifolds can be found, such as \emph{manifold alignment} presented in \Cref{sec:manifold-alignment}.
            Again, this kind of approaches is compatible with the DR techniques described in \Cref{sec:dr-monofi}.

    \subsection{Manifold alignment}\label{sec:manifold-alignment}

        Manifold alignment is a method that provides a mapping between distinct low-dimensional representations of datasets that are associated with different manifolds.
        There are two main approaches for doing so.

        In the first approach, the DR with different fidelity snapshots is carried out such that a common manifold $ Z$ for all distinct datasets is found \cite{hamLearning2003, wangManifold2009}.
        The DR technique must be adapted so that each mapping $\mathcal{DR}_l$ defines a transformation from the original space $ Y_l$ at a fidelity $l$ to the common manifold $ Z$ ($\mathcal{DR}_l:  Y_l\rightarrow Z,\forall l\in\{1,\dots,L\}$).
        This approach can be achieved either with a set of correspondences between some input samples from the distinct datasets \cite{hamLearning2003}, or without \cite{wangManifold2009}.
        In the context of this paper, a set of correspondence means that some snapshots have been obtained for the same input variables values for different fidelity levels $l$ and $l'$, \textit{i.e.}, there are some $\mathbf{u}_l^{(i)},i=1,\dots,n_c$, with $n_c$ the number of common input variables vectors, such that $\mathbf{u}_l^{(i)}\in\mathcal{U}_l$ and $\mathbf{u}_l^{(i)}\in\mathcal{U}_{l'}$.

        In the second approach, a separate DR is performed for each fidelity, leading to $L$ mappings $\mathcal{DR}_l$ and as many manifolds $ Z_l$.
        The goal is then to find a transformation that maps one manifold onto another.
        It can be achieved through an orthogonal \emph{Procrustes analysis} \cite{wangManifold2008}.
        It is based on the minimization of the Frobenius norm between the matrices $\mathbf{L}_l$, assuming that the manifolds all have the same dimensionality $d_{\mathbf{z}_l}=d_{\mathbf{z}},\forall l\in\{1,\dots,L\}$ \cite{gowerProcrustes2010}.
        Note that this method requires some correspondence between the input variables of some samples from the different fidelity models.
        It defines a set of affine transformations (scaling, translation and generalized rotation) applied to a given manifold to match a target manifold.
        Manifold alignment by orthogonal Procrustes analysis between the fidelity levels $l$ and $l'$ is then the mapping
        \begin{equation}\label{eq:ma-orth}
            \begin{split}
                \mathcal{MA} : \mathbb{R}^{d_\mathbf{z}} & \rightarrow \mathbb{R}^{d_\mathbf{z}} \\
                \mathbf{z}_{l} & \mapsto \mathbf{z}_{l'}.
            \end{split}
        \end{equation}

        If the dimensionalities of the manifolds are different, manifold alignment by orthogonal Procrustes analysis can be extended to projection Procrustes analysis \cite{gowerProcrustes2004}.

    \subsection{Modeling the DR residuals by Gaussian processes with tensorized covariance}\label{sec:orthogonal-part}

        The DR residuals correspond to the part of information that is lost when performing DR.
        It is the difference between the original snapshots and the same snapshots projected onto the latent space and back-projected into the original space.
        These residuals computed for every snapshot,
        $\mathcal{Y}_{l,\perp} = \left\{\mathbf{y}_{l,\perp}^{(i)}\right\}^{n_l}_{i=1}$, are as follows:
        \begin{equation}
            \mathbf{y}_{l,\perp}^{(i)}=\mathbf{y}_l^{(i)}-\mathcal{DR}_l^{-1}\left(
                \mathcal{DR}_l\left(
                    \mathbf{y}_l^{(i)}
                \right)
            \right), i=1,\dots,n_l.
        \end{equation}

        Notice that $\mathcal{DR}_l^{-1}$ is not necessarily the exact inverse mapping of $\mathcal{DR}_l$ but a numerical approximation thereof.
        In Kerleguer \cite{kerleguerMultifidelity2023}, it is referred to as the \emph{orthogonal part}.
        The name ``orthogonal'' comes from the fact that in PCA, when projecting onto the reduced basis, the DR residuals correspond to the elements projected onto the space that is orthogonal to the one spanned by the PCA basis vectors.
        These residuals may be modeled by any single-fidelity surrogate with functional outputs.
        Among others, the approach introduced in \cite{kerleguerMultifidelity2023} is summarized in the following.

        Gaussian process (GP) regression with tensorized covariance \cite{rougierEfficient2008, perrinAdaptive2020} is a single-fidelity surrogate model with functional outputs.
        The DR residuals are modeled by a GP which depends on the input vector $\mathbf{u}\in U$ and the location (spatial, temporal, \textit{etc.}) in the field, denoted by $\mathbf{x}_{j,l}\in\mathbb{R}^{d_\mathbf{x}}$, $j\in\{1,\dots,d_{\mathbf{y}_l}\}$.
        Let $\mathcal{S}_{l,\perp}^{\mathbf{x}_{j,l}}(\mathbf{u})$ be the value of the DR residuals at location $\mathbf{x}_{j,l}$.
        The covariance of this GP for two different input samples $\mathbf{u}$ and $\mathbf{u}'$ and two different locations $\mathbf{x}_{j,l}$ and $\mathbf{x}_{k,l}$ is modeled by the covariance function (also called a \emph{kernel}) $\phi(\cdot)$ such that
        \begin{equation}
            \text{cov}\left(\mathcal{S}_{l,\perp}^{\mathbf{x}_{j,l}}\left(\mathbf{u}\right), \mathcal{S}_{l,\perp}^{\mathbf{x}_{k,l}}\left(\mathbf{u}'\right)\right) = \phi\left(\mathbf{u}, \mathbf{u}', \mathbf{x}_{j,l}, \mathbf{x}_{k,l}\right).
        \end{equation}

        Since the product of covariance functions is a valid covariance function, it is further assumed that this covariance structure can be separated into a covariance function for the input variables $\phi_\mathbf{u}(\cdot)$ and a covariance function for the location in the field $\phi_\mathbf{x}(\cdot)$, such that
        \begin{equation}
            \text{cov}\left(\mathcal{S}_{l,\perp}^{\mathbf{x}_{j,l}}\left(\mathbf{u}\right), \mathcal{S}_{l,\perp}^{\mathbf{x}_{k,l}}\left(\mathbf{u}'\right)\right) = \phi_\mathbf{u}\left(\mathbf{u}, \mathbf{u}'\right) \cdot \phi_\mathbf{x}\left(\mathbf{x}_{j,l}, \mathbf{x}_{k,l}\right).
        \end{equation}

        If the field is discretized on a mesh that does not vary with respect to the input vector, $\phi_\mathbf{x}(\cdot)$ can be replaced by a $d_{\mathbf{y}_l}$-by-$d_{\mathbf{y}_l}$ covariance matrix $\boldsymbol{\Phi}_\mathbf{x}$.
        Moreover, if the simulator has been evaluated on a set of $n_l$ input vectors, the $n_l$-by-$n_l$ covariance in input vector matrix $\boldsymbol{\Phi}_\mathbf{u}$ can be computed. This is obtained via a classical covariance kernel, possibly with hyperparameters that have to be tuned.
        The covariance kernel $\boldsymbol{\Phi}$ of the GP is then obtained as the tensor product
        \begin{equation}
            \boldsymbol{\Phi} = \boldsymbol{\Phi}_\mathbf{u} \otimes \boldsymbol{\Phi}_\mathbf{x}.
        \end{equation}

        This type of single-fidelity surrogate model can be used to model any simulator with functional output.
        It is used in the following as a single-fidelity surrogate for comparison purposes.

    \subsection{Intermediate surrogate modeling approaches}\label{sec:regressor}

        Surrogate models are mathematical approximations of functions whose output can only be known for a limited set of input variables values.
        They can be used to replace computationally expensive functions or, in the context of this study, to create a mapping between the input variables $\mathbf{u}\in U$ and the latent space variables $\mathbf{z}_l\in Z_l$.
        The prediction accuracy depends on different factors such as the number and distribution of samples within the input space $U$, the properties of the unknown function (\textit{e.g.}, continuity, stationarity, smoothness) or the modeling hypotheses (\textit{e.g.}, the number of coefficients for a polynomial regression, the covariance model for a Gaussian process).
        They also often involve hyperparameters, denoted by $\boldsymbol\theta\in\mathbb{R}^{d_{\boldsymbol\theta}}$, which have to be determined during the training process.
        In this paper, we define $\mathcal{IS}^{\boldsymbol\theta}$ as the mapping from the input space to the latent space $Z_l$:
        \begin{equation}\label{eq:sm}
            \begin{split}
                \mathcal{IS}^{\boldsymbol\theta} :  U & \rightarrow  Z_l \\
                \mathbf{u} & \mapsto \mathbf{z}_l,
            \end{split}
        \end{equation}
        and denote by $\hat{\mathbf{z}}_l$ a prediction of the latent variables by the intermediate surrogate, \textit{e.g.}, a Gaussian process. The dependence of $\mathcal{IS}^{\boldsymbol\theta}$ on $\boldsymbol\theta$ is omitted hereafter for the sake of conciseness.

        There exist mainly two strategies to build a surrogate of a mapping with vector outputs: either build a set of independent surrogates with scalar outputs (one for each component of the output vector), or a single surrogate capable of predicting  vector outputs, possibly using the dependence information between the output components.

        The first strategy leads to $d_{\mathbf{z}_l}$ surrogates which have to be trained independently.
        They can be single-fidelity \cite{khatouriMetamodeling2022} or multi-fidelity \cite{peherstorferSurvey2018,fernandez-godinoReview2023}.
        Gaussian process regression \cite{matheronPrinciples1963}, polynomial chaos expansion \cite{eldredRecent2009}, artificial neural networks \cite{spechtGeneral1991}, random forests \cite{cutlerRandom2012}, radial basis function (RBF) interpolation \cite{buhmannRadial2000}, or other machine learning techniques \cite{swischukProjectionbased2019} are well-known techniques to build such a single-fidelity surrogate.
        Multi-fidelity extensions include GP-based methods (\textit{e.g.}, hierarchical Kriging \cite{hanHierarchical2012}, autoregressive (AR1) co-Kriging \cite{legratietRecursive2014}, nonlinear autoregressive co-Kriging \cite{perdikarisNonlinear2017}, multi-fidelity deep Gaussian processes (DGP) \cite{damianouDeep2013,cutajarDeep2019}), a combination of Kriging and support vector regression \cite{zhaoGeneral2023}, multi-fidelity polynomial chaos expansion \cite{ngMultifidelity2012}, artificial neural networks \cite{guoMultifidelity2022}, and co-RBF \cite{durantinMultifidelity2017}.
        In \cite{beranFramework2022}, a variety of such multi-fidelity surrogates are discussed and applied to the design of air, sea and space military vehicles. Such multi-fidelity techniques mainly differ in the dependence hypothesis between the different fidelity models (\textit{e.g.}, linear, nonlinear). Another notable approach is the mixture of experts where the input space is divided in subregions in which a dedicated surrogate is built, whose predictions can be combined in different ways \cite{masoudniaMixture2014,rumpfkeilConstruction2017}.

        The second strategy involves a single multi-output surrogate that can either be single-fidelity or multi-fidelity.
        Single-fidelity examples include multi-output GP regression \cite{liuRemarks2018}, artificial neural networks \cite{spechtGeneral1991}, RBF interpolation \cite{buhmannRadial2000}.
        Similarly, an important modeling characteristic lies in the dependence hypothesis between the components of the output (\textit{e.g.}, linear, nonlinear).
        In the literature, there are few multi-fidelity surrogates capable of predicting multi-outputs. Among the existing ones, there are multi-outputs AR1 co-Kriging \cite{parussiniMultifidelity2017,toalApplications2023}, a multi-fidelity multi-output Gaussian process \cite{linMultioutput2021}, RBF interpolation with the addition of an input parameter to indicate the fidelity \cite{mifsudVariablefidelity2016} or multi-fidelity DGP \cite{damianouDeep2013,cutajarDeep2019}.


\section{Multi-fidelity surrogates for the prediction of high-dimensional field outputs}\label{sec:metamodels}

    Multiple methods for building multi-fidelity surrogates combining DR and intermediate surrogate modeling for the prediction of field outputs have been proposed in the literature.
    However, these approaches have received little comparison and have not been presented within a common framework, making it difficult to compare their pros and cons for the newcomer.
    They all share common building blocks, which were introduced in \Cref{sec:background}, and their combination can be divided into three families of approaches: corrective, mapping and fusion.
    In the original papers \cite{benamaraMultifidelity2017, bunnellMultifidelity2021, kerleguerMultifidelity2023, malouinInterpolation2013, perronMultifidelity2021, deckerManifold2022, rokitaMultifidelity2018, thenonMultifidelity2016, toalPotential2014, wangMultifidelity2020, mifsudVariablefidelity2016, parussiniMultifidelity2017}, these surrogates have all been introduced for the case when two fidelities are available (\textit{i.e.}, $l\in\{1, 2\}$).
    Therefore, the different families are subsequently introduced for the two-fidelity case, even though some could theoretically be extended to a greater number of fidelities.
    Additionally, the prediction process is only presented for the prediction of higher-fidelity simulator outputs.
    In this section, a unified framework is proposed that conveys the different paradigms in a pedagogical manner, facilitating the comparison of the different surrogates and their adaptation by practitioners.

    \subsection{Corrective approaches}\label{sec:corrective}

\begin{figure}[!ht]
            \centering
            \includegraphics[width=0.4\textwidth]{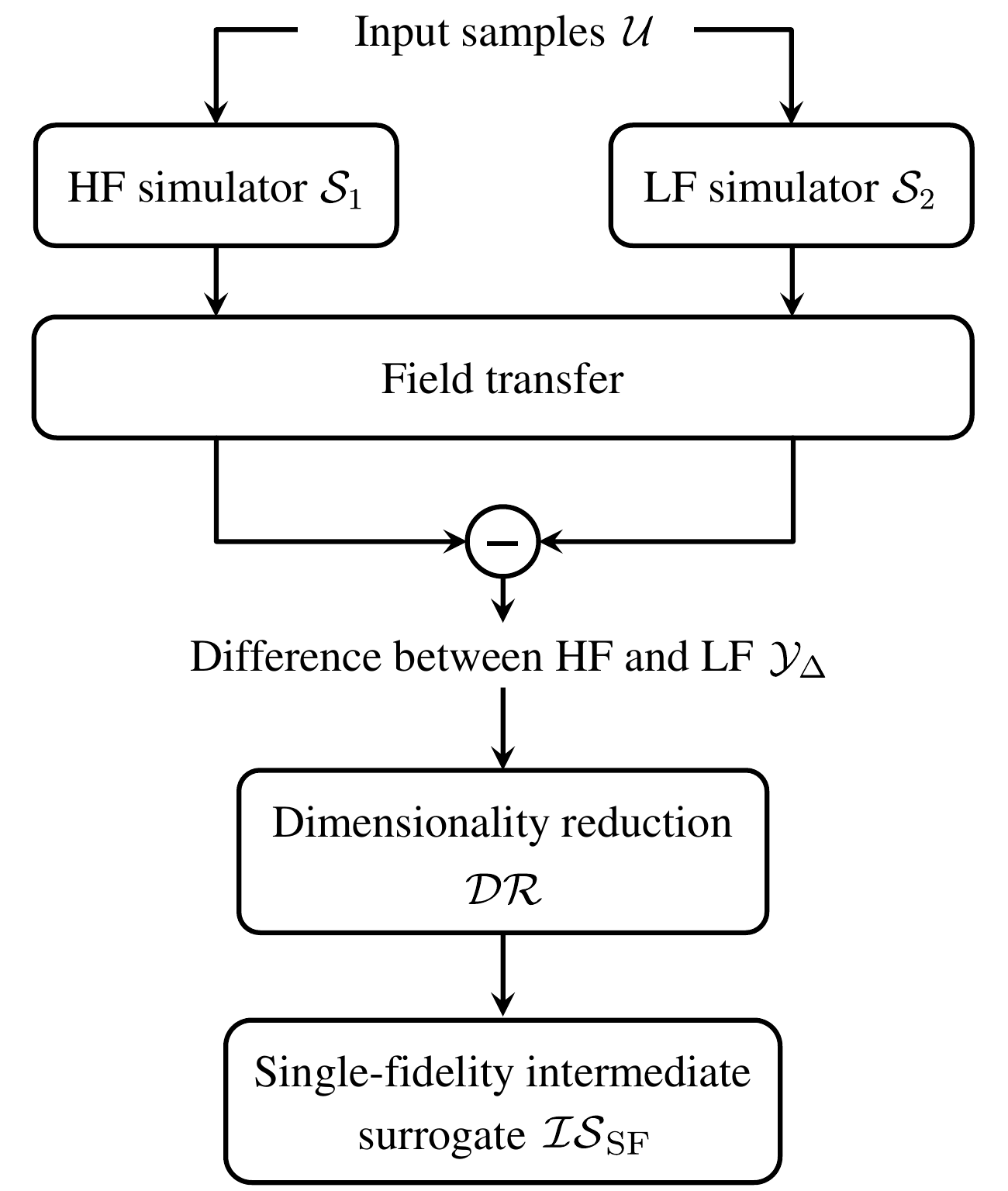}
            \caption{Diagram of the corrective approaches \cite{malouinInterpolation2013}}
            \label{fig:corrective}
        \end{figure}

        In the so-called \emph{corrective approaches}, the high-fidelity output field is not modeled directly.
        The basic idea is to build a surrogate serving as a corrective function added to the low-fidelity simulator to predict high-fidelity output fields.
        In practice, the first step of the corrective approach is to compute the difference between the high- and low-fidelity snapshots:
        
        \begin{equation}
            \mathcal{Y}_\Delta=\left\{\mathbf{y}_\Delta^{(i)}=\mathbf{y}_1^{(i)} - \mathbf{y}_2^{(i)}\right\}^n_{i=1}.
        \end{equation}
        Note that in order to be able to compute this, there must be the same number $n=n_1=n_2$ of high- and low-fidelity snapshots, corresponding to the same input variable vectors (\textit{i.e.}, $\mathcal{U}_1=\mathcal{U}_2=\mathcal{U}$).
        Consequently, if the computational cost of the high-fidelity simulator is especially high, a small number of high-fidelity snapshots limits the number of exploitable low-fidelity snapshots.
        Hence, the surrogate may not take full advantage of the multi-fidelity context.
        Furthermore, either the high- and low-fidelity output fields must be discretized on the same mesh, or a pre-processing step to transfer the fields onto a common mesh must be performed.
        Then, DR is carried out using the snapshots in $\mathcal{Y}_\Delta$ to determine the associated latent variables values

        \begin{equation}
            \mathcal{Z}_\Delta=\left\{\mathbf{z}_\Delta^{(i)}=\mathcal{DR}\left(\mathbf{y}_\Delta^{(i)}\right)\right\}^n_{i=1}.
        \end{equation}
        Finally, a single-fidelity surrogate $\mathcal{IS}_\text{SF}$ is built to map the input space $ U$ to the latent space using the samples from $\mathcal{U}$ and $\mathcal{Z}_\Delta$.
        It is then able to predict the latent variables for a given input variable vector.
        A diagram summarizing this approach is displayed in \autoref{fig:corrective}.
        Any single-fidelity DR technique and single-fidelity surrogate modeling technique may be used to construct such a corrective surrogate.

        To perform a prediction for a new input vector $\mathbf{u}^\star\notin\mathcal{U}$, the surrogate is used to predict the latent variables $\hat{\mathbf{z}}_\Delta=\mathcal{IS}_\text{SF}(\mathbf{u}^\star)$.
        Then, the DR is mapped back to get the prediction of the difference field $\hat{\mathbf{y}}_\Delta= \mathcal{DR}^{-1}\left(\hat{\mathbf{z}}_\Delta\right)$.
        Finally, $\hat{\mathbf{y}}_\Delta$ is added to the low-fidelity output field obtained by running the exact low-fidelity simulator $\mathcal{S}_2\left(\mathbf{u}^\star\right)$.
        The process of getting a high-fidelity output field prediction $\hat{\mathbf{y}}_1$ can be summarized by the following equation:

        \begin{equation}
            \hat{\mathbf{y}}_1
            = \mathcal{S}_2\left(\mathbf{u}^\star\right)
            + \mathcal{DR}^{-1}\left(
                \mathcal{IS}_\text{SF}(\mathbf{u}^\star)
            \right).
        \end{equation}

        The multi-fidelity surrogate proposed by Malouin \textit{et al}.\ \cite{malouinInterpolation2013} belongs to the corrective approaches.
        For the DR, the authors used PCA and the intermediate surrogate is a single-fidelity GP per latent variable.

    \subsection{Mapping approaches}\label{sec:mapping}

\begin{figure}[!ht]
            \centering
            \includegraphics[width=0.5\textwidth]{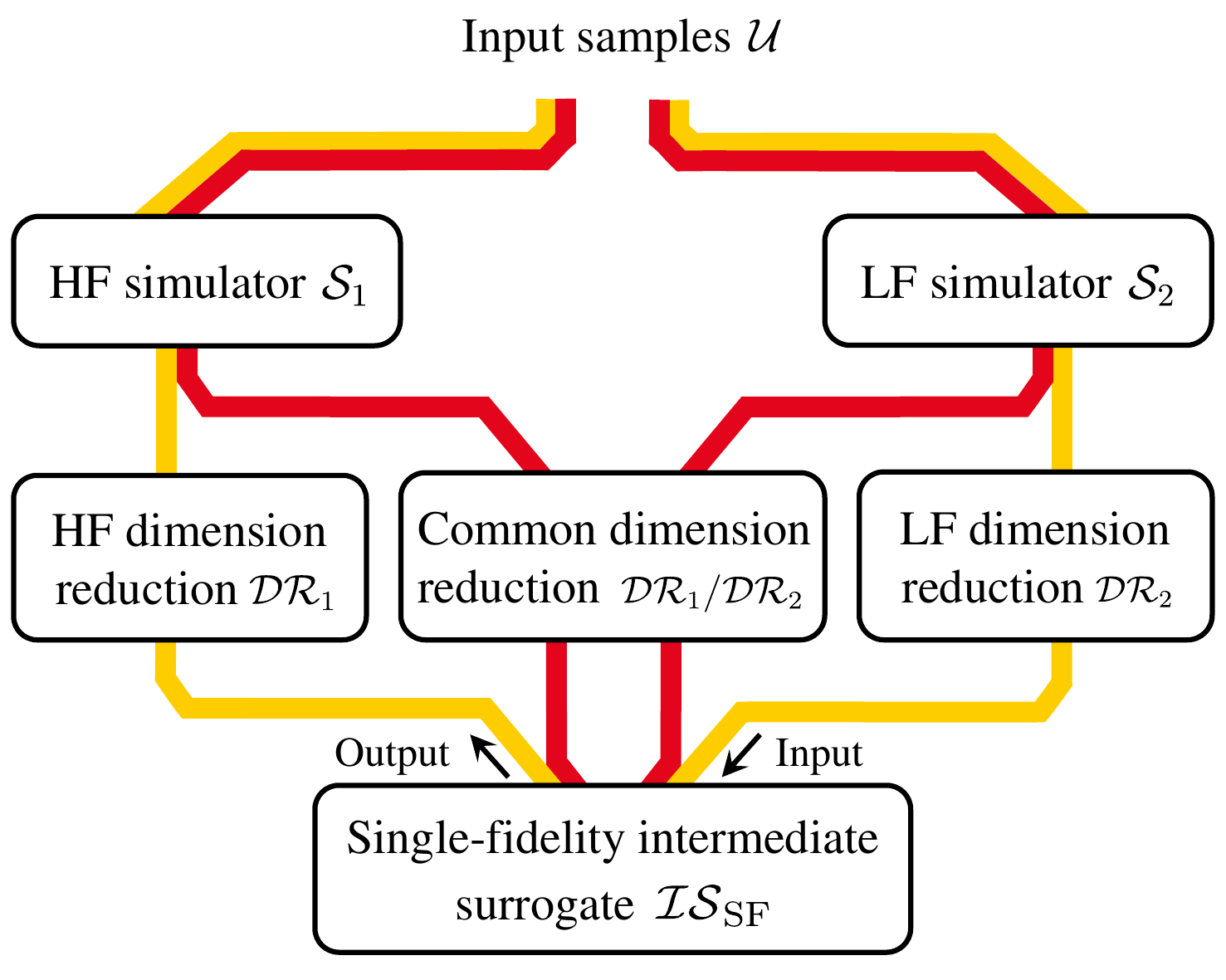}
            \caption{Diagram of the mapping approach, \cite{toalPotential2014} in red, \cite{wangMultifidelity2020} in yellow. The arrows represent the input and output of the single-fidelity intermediate surrogate $\mathcal{IS}$.}
            \label{fig:mapping}
        \end{figure}

        \emph{Mapping approaches} consist in building a multi-fidelity surrogate that maps the low-fidelity to the high-fidelity output field.
        The first step is to compute the sets of latent variable values corresponding of the high- and low-fidelity snapshots separately, respectively $\mathcal{Z}_1$ and $\mathcal{Z}_2$. They may be computed with different DR techniques, and are such that

        \begin{equation}\label{eq:dr-mapping}
            \mathcal{Z}_l = \left\{\mathbf{z}^{(i)}_l=\mathcal{DR}_l\left(\mathbf{y}^{(i)}_l\right)\right\}^n_{i=1}, l\in\{1,2\}.
        \end{equation}
        The second step is to train a single-fidelity surrogate mapping the manifold $ Z_2$, represented by the latent samples $\mathcal{Z}_2$, to the manifold $ Z_1$, represented by the latent samples $\mathcal{Z}_1$.
        This surrogate can then predict the high-fidelity latent variables if the low-fidelity latent variables are provided.
        As for the corrective approach, the same number $n$ of high- and low-fidelity latent variables snapshots (corresponding to the same input variable vectors) is required to be able to define this mapping.
        Again, if acquiring high-fidelity simulator snapshots is computationally expensive, this may limit the number of low-fidelity snapshots that can be exploited.
        Hence, the surrogate may not take full advantage of the multi-fidelity context.
        Note that this mapping resembles manifold alignment as it maps the low- to the high-fidelity manifold. The difference is that it is used in the prediction process while manifold alignment is only used in the training process.
        The different steps involved in mapping surrogates are summarized in \Cref{fig:mapping}.
        Different approaches to performing DR \cite{toalPotential2014, wangMultifidelity2020} are displayed and represented by different colors (red for common DR \cite{toalPotential2014} and yellow for separate DR \cite{wangMultifidelity2020}).

        To predict the high-fidelity output field for a new input variable vector $\mathbf{u}^\star\notin\mathcal{U}$, it is necessary to first run the low-fidelity simulator to get the corresponding low-fidelity output field $\mathbf{y}_2^\star=\mathcal{S}_2(\mathbf{u}^\star)$.
        Then, the low-fidelity latent variables corresponding to this field are computed with $\mathbf{z}_2^\star= \mathcal{DR}_2(\mathbf{y}_2^\star)$.
        Next, the single-fidelity surrogate is used to predict the high-fidelity latent variables $\hat{\mathbf{z}}_1= \mathcal{IS}_\text{SF}(\mathbf{z}_2^\star)$.
        Finally, the inverse mapping of the high-fidelity DR is performed to get a prediction $\hat{\mathbf{y}}_1$ of the high-fidelity output field.
        This process is summarized in \Cref{eq:prediction_mapping}.

        \begin{equation}\label{eq:prediction_mapping}
            \hat{\mathbf{y}}_1 =
            \mathcal{DR}_1^{-1}\left(
                \mathcal{IS}_\text{SF}(
                    \mathcal{DR}_2(
                            \mathcal{S}_2(\mathbf{u}^\star)
                    )
                )
            \right).
        \end{equation}

        The multi-fidelity surrogate proposed by Wang \textit{et al}.\ \cite{wangMultifidelity2020} belongs to the mapping approach family (yellow path in \Cref{fig:mapping}).
        The DR is performed using PCA independently for each fidelity level, hereafter referred to as single-fidelity PCA (SFPCA).
        The mapping of the latent variables between the fidelity levels is a Gaussian process per high-fidelity latent variable.
        Note that if the PCA of the low-fidelity snapshots results in a large number of latent variables, this mapping has a large input dimensionality, which might make the learning process more complex.
        Additionally, classical GP kernels use Euclidean distances in the input space to measure the distance between the input samples.
        If nonlinear DR has been selected, the input space is not an Euclidean space. Therefore, the kernel should be adapted for instance by being based on geodesic distances instead (\textit{e.g.}, see \cite{pigoliKriging2016}).
        In \cite{wangMultifidelity2020}, a GP with a kernel based on Euclidean distances is used. We make the same choice in the sequel of this paper.

        Another multi-fidelity surrogate in the mapping family has been proposed by Toal \cite{toalPotential2014} (red path in \Cref{fig:mapping}).
        It uses Gappy-PCA (GPCA, most commonly known as Gappy-POD) \cite{eversonKarhunen1995} which performs a PCA over the stacked high- and low-fidelity snapshots, \textit{i.e.}, over $\mathbf{Y}_{\text{GPCA}}=\left[\mathbf{Y}_1, \mathbf{Y}_2\right]$, which is a $n$-by-$(d_{\mathbf{y}_1}+d_{\mathbf{y}_2})$ matrix.
        As used by Toal \cite{toalPotential2014}, GPCA is not only a DR technique, but also defines a mapping from $Y_2$ to $Y_1$.
        The low-fidelity simulator has to be run in order to perform a prediction of the high-fidelity output field.

    \subsection{Fusion approaches}\label{sec:fusion}

        \emph{Fusion approaches} combine DR and an intermediate surrogate model blending together the high- and low-fidelity data in such a way that running the exact low-fidelity simulator is not needed to predict high-fidelity output fields.
        The first step, which is performing DR, is the same as mapping approaches.
        The latent samples corresponding to the high- and low-fidelity snapshots sets, respectively $\mathcal{Z}_1$ and $\mathcal{Z}_2$, are computed with

        \begin{equation}\label{eq:dr-fusion}
            \mathcal{Z}_l = \left\{\mathbf{z}^{(i)}_l=\mathcal{DR}_l\left(\mathbf{y}^{(i)}_l\right)\right\}^{n_l}_{i=1}, l\in\{1,2\}.
        \end{equation}
        To perform this DR, there are two main options. The first uses the same mapping to the latent space for both fidelity levels.
        This causes the high- and low-fidelity latent variables to belong to the same manifold.
        The second option uses two distinct DR processes for each fidelity level, resulting in two distinct manifolds.
        For fusion approaches, it is possible to have a different number of snapshots for different fidelities (\textit{i.e.}, $n_1\neq n_2$).
        An optional step is to perform manifold alignment by orthogonal Procrustes analysis that transfers the known points of the low-fidelity manifold $Z_2$ to the high-fidelity manifold $ Z_1$.
        After applying the operations described in \Cref{sec:manifold-alignment}, the low-fidelity latent variable values $\mathcal{Z}_2$ would become $\mathcal{Z}_{2'}$ ($\mathbf{z}_{2'}=\mathcal{MA}(\mathbf{z}_2)$).
        Eventually, a multi-fidelity intermediate surrogate $\mathcal{IS}_\text{MF}$ is built that maps the input space $ U$ to the high-fidelity latent space $ Z_1$.
        It mixes the high- and low-fidelity training data $\mathcal{U}_1$, $\mathcal{Z}_1$, $\mathcal{U}_2$ and $\mathcal{Z}_2$ (or $\mathcal{Z}_{2'}$ if manifold alignment has been performed).

\begin{figure}[!ht]
            \centering
            \includegraphics[width=0.5\textwidth]{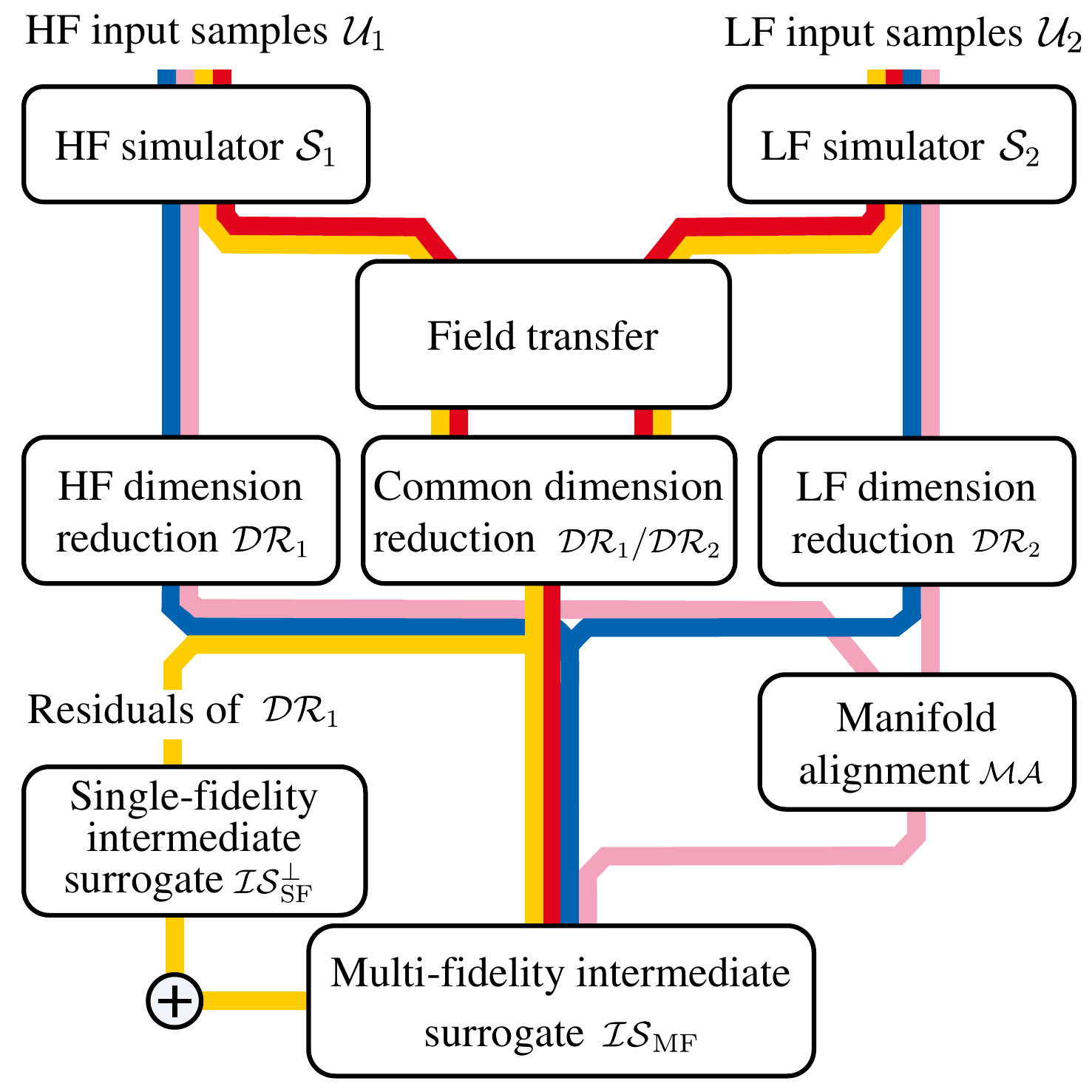}
            \caption{Diagram of the fusion approaches, \cite{rokitaMultifidelity2018,benamaraMultifidelity2017,mifsudVariablefidelity2016,thenonMultifidelity2016} in red, \cite{kerleguerMultifidelity2023} in yellow, \cite{bunnellMultifidelity2021} in blue, \cite{perronMultifidelity2021,deckerManifold2022} in pink}
            \label{fig:fusion}
        \end{figure}

        Regardless of the options chosen so far, the high- and low-fidelity latent spaces $ Z_1$ and $ Z_2$ must have the same dimensionality $d_{\mathbf{z}_1}=d_{\mathbf{z}_2}=d_\mathbf{z}$.
        This must be taken into account when performing DR.
        For instance, if DR is performed separately for each fidelity (resulting in manifolds having different dimensionalities), the common dimensionality could be the lowest of all manifold dimensionalities.
        This would lead to suboptimal DR for manifolds that would have had a higher dimensionality otherwise.
        An additional optional step is to compute the DR residuals of the high-fidelity output field and to train a single-fidelity surrogate $\mathcal{IS}^\perp_\text{SF}$ for predicting these residuals (see \Cref{sec:orthogonal-part}).
        Fusion methods are summarized in \Cref{fig:fusion}.

        To predict the high-fidelity output corresponding to a new vector of input variables $\mathbf{u}^\star\notin\mathcal{U}_1$, the multi-fidelity intermediate surrogate is used to predict the high-fidelity latent variables $\hat{\mathbf{z}}_1=\mathcal{IS}_\text{MF}(\mathbf{u}^\star)$. Then, the prediction of the high-fidelity output field $\hat{\mathbf{y}}_1$ is computed by inverse mapping the high-fidelity DR $\hat{\mathbf{y}}_1 = \mathcal{DR}_1^{-1}\left(\hat{\mathbf{z}}_1\right)$.
        Finally, if needed, the surrogate $\mathcal{IS}^\perp_\text{SF}$ can be used to predict the residuals not taken into account in the DR.
        Note that manifold alignment does not require any additional step in the prediction process, as it only modifies the training set of the multi-fidelity intermediate surrogate $\mathcal{IS}_\text{MF}$ by replacing $\mathcal{Z}_2$ with $\mathcal{Z}_{2'}$.
        The following equation summarizes the prediction process of fusion surrogate models.

        \begin{equation}
            \hat{\mathbf{y}}_1 =
            \mathcal{DR}^{-1}_1\left(
                \mathcal{IS}_\text{MF}\left(\mathbf{u}^\star\right)
            \right)
            + \mathcal{IS}^\perp_\text{SF}(\mathbf{u}^\star).
        \end{equation}

        In the following paragraphs, different multi-fidelity surrogates introduced in the literature belonging to the fusion family are briefly described.
        First, the surrogates that perform separate DR for each fidelity are presented.
        These surrogates have the advantage that the high- and low-fidelity output fields can be discretized on different meshes.
        The surrogate introduced by Bunnell \textit{et al}.\ \cite{bunnellMultifidelity2021} (blue path in \Cref{fig:fusion}), uses SFPCA and one AR1 co-Kriging per latent variable as an intermediate surrogate.
        This assumes that without any adjustment, the high- and low-fidelity manifolds are similar even though the DRs are performed separately, since AR1 co-Kriging assumes a linear dependency between the high- and low-fidelity latent variables.
        In their paper, Bunnell \textit{et al}.\ verified this similarity by plotting and visually comparing the PCA modes.
        This also assumes that there is the same number of high- and low-fidelity latent variables.
        This can either be done by \textit{a priori} setting a fixed number of modes for all fidelities, or by choosing the number of modes of one of the fidelities as the dimensionality of all latent spaces.

        The multi-fidelity surrogates proposed in Perron \textit{et al}.\ \cite{perronMultifidelity2021} and extended in Decker \textit{et al}.\ \cite{deckerManifold2022} also separate high- and low-fidelity DR (pink path in \Cref{fig:fusion}).
        The used DR techniques are PCA \cite{perronMultifidelity2021}, KPCA, isomap and LTSA \cite{deckerManifold2022} which are performed on each fidelity independently, hence called SFPCA, SFKPCA, SFIsomap and SFLTSA.
        Manifold alignment by orthogonal Procrustes analysis is performed before using one hierarchical Kriging \cite{hanHierarchical2012} per latent variable as a multi-fidelity intermediate surrogate.
        These surrogates are the only ones using manifold alignment.
        In their study, Perron \textit{et al}.\ \cite{perronMultifidelity2021} and Decker \textit{et al}.\ \cite{deckerManifold2022} considered the same number of high- and low-fidelity latent variables (this assumption could be removed, see \Cref{sec:manifold-alignment} for more details).

        Other multi-fidelity surrogates of the fusion family use a common DR for all fidelity levels.
        Consequently, they all share the same latent manifold.
        Note that the high- and low-fidelity output fields need to be discretized on the same mesh.
        The multi-fidelity surrogate introduced by Benamara \textit{et al}.\ \cite{benamaraMultifidelity2017} (red path in \Cref{fig:fusion}) uses an adaptation of Constrained PCA (CPCA).
        It consists in an untruncated PCA of the high-fidelity snapshots to which are added vectors computed from the part of the low-fidelity snapshots which is orthogonal to the basis.
        Then, intermediate surrogate modeling is performed with one AR1 co-Kriging per latent variable.

        The multi-fidelity surrogate in Mifsud \textit{et al}.\ \cite{mifsudVariablefidelity2016} performs PCA over the union of the high- and low-fidelity snapshots $\mathcal{Y}_1\cup\mathcal{Y}_2$ denoted here by MFPCA (red path in \Cref{fig:fusion}).
        Since the number of snapshots is $n_1+n_2$ and because PCA decomposes the variance of both the high- and low-fidelity output fields, the number of principal components will likely be larger than for a PCA of the high- or low-fidelity snapshots carried out seperately.
        Moreover, if the number of low-fidelity snapshots $n_2$ is much larger than the number of high-fidelity snapshots $n_1$, there is a risk of diluting the high-fidelity output field features when performing MFPCA.
        The intermediate mapping is a single-fidelity RBF with an additional discrete variable that indicates the fidelity of the corresponding snapshot (\textit{i.e.}, $\mathbf{u}_l\in U\subseteq\mathbb{R}^{d_{\mathbf{u}}}$ is replaced by $\tilde{\mathbf{u}}_l\in\tilde{ U}\subseteq\mathbb{R}^{d_{\mathbf{u}}}\times\mathbb{N}$).

        The surrogate described in Thenon \textit{et al}.\ \cite{thenonMultifidelity2016} uses a PCA over the high-fidelity snapshots (HFPCA) and one AR1 co-Kriging per latent variable as the intermediate mapping (red path in \Cref{fig:fusion}).
        This assumes that the modes of HFPCA can accurately compress the low-fidelity output field.
        This also presumes that the number of high-fidelity snapshots is large enough to build a meaningful covariance matrix to compute its eigenvectors and eigenvalues.

        The surrogate introduced by Rokita \textit{et al}.\ \cite{rokitaMultifidelity2018} performs MFPCA and uses one AR1 co-Kriging per latent variable as the intermediate mapping (red path in \Cref{fig:fusion}.
        The surrogate of Kerleguer \cite{kerleguerMultifidelity2023} uses two different DR techniques (yellow path in \Cref{fig:fusion}).
        The first one is a PCA on the low-fidelity snapshots (denoted here by LFPCA) and the other is a cross-validation based PCA over the low-fidelity snapshots (low-fidelity cross-validation based PCA, LFCVBPCA).
        An empirical distribution of the modes is obtained by considering PCAs of subsets of all snapshots as realizations of the CVBPCA.
        The intermediate surrogate is one AR1 co-Kriging per latent variable and the residuals of the high-fidelity DR is modeled by a Gaussian process with tensorized covariance.
        Note that this surrogate is the only one to model the residuals of DR.
        This surrogate either assumes that the modes computed from the low-fidelity snapshots (with LFPCA or LFCVBPCA) can accurately represent the high-fidelity output field, or that the surrogate of the residuals of DR can compensate the difference.

    \subsection{General remarks on the different approaches}\label{sec:general-remarks}

        The main drawback of mapping and corrective approaches compared to fusion approaches is that they require the evaluation of the low-fidelity simulator to make a prediction of the high-fidelity output field.
        If the low-fidelity simulator has a non-negligible cost, this can make the use of these surrogates intractable in an optimization or uncertainty quantification context. 
        To overcome this issue, the lower-fidelity simulator $\mathcal{S}_2$ could be replaced by a single-fidelity surrogate (for instance, a combination of PCA and GP regression).
        In Perron \textit{et al}.\ \cite{perronMultifidelity2021}, the surrogate using the corrective approach and the GPCA-based surrogate are compared when the low-fidelity simulator is used to make predictions of the high-fidelity output field and when it is replaced by a single-fidelity surrogate.
        In the examples considered in their study, replacing the low-fidelity simulator by an emulator does not always turn out to be detrimental.
        Hence, it is hard to predict if this procedure should improve or worsen the performance of multi-fidelity surrogate models in the general case.

        Another disadvantage of mapping and corrective methods is the fact that the same amount of snapshots for the same input samples is required for each fidelity.
        Usually, since the low-fidelity simulator has a lower computational cost than the high-fidelity simulator, more low-fidelity snapshots are available.
        Without modification, mapping and corrective methods are not able to take advantage of this situation.
        If the low-fidelity simulator is replaced by a single-fidelity surrogate model, corrective and mapping surrogates could leverage low-fidelity snapshots that exceeds the number of high-fidelity snapshots ($n_2>n_1$).

        Another point of great importance is the correspondence between the different fidelity meshes.
        The corrective and fusion approaches having a common DR for both fidelity levels (\textit{i.e.}, MFPCA, HFPCA and 
        LFCVBPCA), require that meshes are the same for each fidelity level.
        If this is not the case, a pre-processing step projecting all snapshots to a common mesh must be carried out.
        This common mesh can be the high-fidelity mesh, the low-fidelity mesh or neither of those.
        If the chosen common mesh is not the high-fidelity mesh, the prediction might have to be mapped from the common mesh to the high-fidelity mesh.

        The multi-fidelity surrogate of Benamara \textit{et al}.\ \cite{benamaraMultifidelity2017} has nested meshes for the different fidelity levels.
        The chosen common mesh is the low-fidelity mesh, hence, only the common nodes of the high- and low-fidelity meshes are used to build the multi-fidelity surrogate.
        In Malouin \textit{et al}.\ \cite{malouinInterpolation2013}, both fidelity levels are projected onto the high-fidelity mesh with a ``nearest neighbor'' strategy.
        In Thenon \textit{et al}.\ \cite{thenonMultifidelity2016}, the common mesh is also the high-fidelity mesh.
        The interpolation method used is specific to reservoir engineering \cite{cardwellAverage1945}.
        In Rokita \textit{et al}.\ \cite{rokitaMultifidelity2018}, both fidelity levels are projected onto the low-fidelity mesh.
        In Mifsud \textit{et al}.\ \cite{mifsudVariablefidelity2016}, the high-fidelity mesh is chosen as the common mesh.
        In the  work of Kerleguer \cite{kerleguerMultifidelity2023}, it is assumed that both fidelity levels are discretized on the same mesh.
        Conversely, mapping approaches do not require the mesh to be the same for all fidelity levels, except if a common DR is used.

        The total number of hyperparameters varies between the various multi-fidelity surrogates presented so far.
        First, there are hyperparameters that have to be set prior to training the surrogate.
        They can have a strong impact on the performance of both DR and intermediate surrogate modeling.
        All DR techniques reviewed in this paper have such hyperparameters:
        for instance, the RIC in the truncation of PCA, the dimensionality of the latent space in isomap and LTSA, and the kernel function of KPCA all have to be chosen beforehand, based on experience of the practitioner or any specific technique depending on the method.
        For the mapping from the input to the latent space, some hyperparameters of the selected surrogate must be set \textit{a priori}, \textit{e.g.}, the kernel function when using Gaussian processes.

        Second, some hyperparameters are tuned when the surrogate is trained.
        Note that they could be set beforehand instead.
        For instance, in DR, the number of nearest neighbors in isomap and LTSA can be chosen so as to minimize a metric (Kruskal's stress \cite{kruskalMultidimensional1964} for isomap and the variance of distance ratios \cite{shiModel2009} for LTSA).
        Similarly, the correlation length of the kernel is optimized in KPCA.
        In general, the more hyperparameters there are, the more complex the computation of the DR mapping.
        This is one of the advantages of PCA compared to nonlinear DR techniques.
        Considering the intermediate surrogate, the correlation length of the kernel function of Gaussian processes is typically determined by maximizing the likelihood \cite{rasmussenGaussian2006}.

        The above approaches could, at least theoretically, be extended to more than two fidelity levels ($l\in\{1,\dots,L\}$) in rather straightforward manners. In practice, the increase in computational cost may become a limiting factor.
        For corrective approaches, the following extension could be used: 

        \begin{equation}\label{eq:extension-correction}
            \hat{\mathbf{y}}_1=
            \mathcal{S}_L(\mathbf{u}^\star) + \sum_{l=1}^{L-1}\Delta\hat{\mathbf{y}}_{l,l+1},
        \end{equation}
        with $\Delta\hat{\mathbf{y}}_{l,l+1}$ a single-fidelity surrogate of the difference between the simulators of fidelity $l$ and $l+1$, evaluated at $\mathbf{u}^\star$.
        The lowest-fidelity simulator could be used if its computational cost is sufficiently low or it could be replaced by a single-fidelity surrogate otherwise.
        Mapping approaches could be extended by using the latent variables of all fidelity levels except the highest one as the inputs of the surrogate predicting the high-fidelity latent variables, \textit{i.e.}, $\hat{\mathbf{z}}_1=
        \mathcal{IS}_\text{SF}\left(\mathbf{z}_{[2,L]}\right)$, with $\mathbf{z}_{[2,L]}=\left[\mathbf{z}_2^\top,\dots,\mathbf{z}_L^\top\right]^\top$.
        Note that the dimensionality of the surrogate input $\mathbf{z}_{[2,L]}$ could become very large.
        If this extension is adopted, a special care should be given to the choice of the intermediate surrogate so that it supports high-dimensional inputs.

        Fusion approaches require little modification to support more than two fidelity levels.
        Indeed, only the multi-fidelity surrogate might need to be changed to be able to use training data from more than two fidelity levels.
        For instance, AR1 co-Kriging or RBF interpolation with an additional input variable indicating the fidelity level could be used.
        As discussed in \Cref{sec:manifold-alignment}, manifold alignment can be performed with more than two fidelity levels.
        Since the residuals of DR are only modeled for the highest-fidelity, this requires no modification.


\section{Description of the test cases}\label{sec:test-cases}

    Before introducing the benchmark methodology, this section presents the different test cases that are used to assess the performance of the different multi-fidelity surrogates with functional outputs.
    The physics of the problems is described and implementation details are given.

    \subsection{Viscous free fall of a ball}\label{sec:vff-case}
        
        The first test case deals with the free fall of a ball thrown in a fluid medium.
        It is derived from \cite{Viscous}.
        The goal is to build a surrogate of the trajectory of the ball, which is a time series, as a function of the initial altitude $y_0\in[0.2,0.4]$ m, the initial vertical velocity $v_0\in[10, 20]$ m.s\textsuperscript{-1}, the uniform density of the ball $\rho_{\text{ball}}\in[10,100]$ kg.m\textsuperscript{-3} and the radius of the ball $r\in[0.1,1]$ m.
        
        The ball has a mass $m=\rho_{\text{ball}}\times\frac{4\pi r^3}{3}$.
        The altitude of the ball at time $t\in[0,T]$ is denoted by $y(t)$.
        At the initial time, it is given an initial vertical velocity $\dot{y}(0)=v_0$ and an altitude $y(0)=y_0$.
        Then, it is subjected only to the acceleration of gravity $g=9.81$ m.s\textsuperscript{-2} and the viscous drag force $F_d= \frac{1}{2} C_d(\dot{y}(t))\,\rho_{\text{fluid}}\,a\,\dot{y}(t)^2$ due to the resistance of the fluid it is immersed in, with the drag coefficient $C_d(\dot{y}(t))$ which depends on the velocity $\dot{y}(t)$, the density of the fluid $\rho_{\text{fluid}}$ and the cross-sectional area $a$ ($\pi r^2$ for a sphere of radius $r$).
        The fluid is water at 20°C with a density $\rho_{\text{fluid}}=998.3$ kg.m\textsuperscript{-3} and a dynamic viscosity is $\eta=1.002\times 10^{-3}$ Pa.s.
        The governing movement equation is given by
    
        \begin{equation}\label{eq:vertical-movement-ball}
            m\ddot{y}(t) + F_d(\dot{y}(t)) + mg = 0.
        \end{equation}
    
        The input variables of the simulators are represented by the vector $\mathbf{u}=[y_0, v_0, \rho_{\text{ball}}, r]^\top$.
        The simulators of variable fidelity all solve \Cref{eq:vertical-movement-ball} for a finite number $d_\mathbf{y}$ of time steps evenly spaced on $[0,T]$, \textit{i.e.}, the mesh is of size $d_\mathbf{y}$.
        The altitude at each time step is collected in the vector $\mathbf{y}_l$, which forms the output field of the simulator of fidelity $l$.
        The difference between the fidelities stems from the modeling of the drag coefficient.
    
        In the low-fidelity simulator, the drag coefficient is supposed to be constant.
        It is chosen to be equal to $0.4$ \cite{timmermanRise1999}.
        It can therefore be shown that the drag force is $F_d(\dot{y}(t)) = 0.2\, \pi\, \rho\, r^2\, \dot{y}(t)^2$.
        \Cref{eq:vertical-movement-ball} can be solved analytically, which gives the solution
    
        \begin{equation}
            y_2(t) = y_0+v_{inf}t+\tau(v_0-v_{inf}) \left(1 - e^{-\frac{t}{\tau}}\right),
        \end{equation}
        with $\tau=\frac{m}{c}$ the characteristic time of the system, $v_{inf}=\frac{-mg}{c}$ the terminal velocity and $c=\frac{1}{2}C_d(\dot{y}(t))\,\rho_{\text{fluid}}\,a$.
        The output field of the low-fidelity simulator is $\mathbf{y}_2=[y_2(0),y_2(T/d_\mathbf{y}),\dots,y_2(T)]^\top$.
        
        In the high-fidelity simulator, the drag coefficient is supposed to be the following function of the velocity \cite{timmermanRise1999}
    
        \begin{equation}
            C_d(\dot{y}(t)) = \frac{24}{Re}+\frac{6}{1+\sqrt{Re}}+0.4,
        \end{equation}
        with $Re=\frac{\rho_{\text{fluid}}\ell \dot{y}(t)}{\eta}$ the Reynolds number, $\eta$ the dynamic viscosity of the medium and $\ell$ the characteristic length scale of the ball in the cross-sectional plane (which is here the diameter $2r$).
        With this drag model, \Cref{eq:vertical-movement-ball} cannot be solved analytically.
        Consequently, it is solved numerically with the LSODE algorithm \cite{petzoldAutomatic1983} and the solution is extracted on the mesh nodes.
        The output field of the high-fidelity simulator is $\mathbf{y}_1=[y_1(0),y_1(T/d_\mathbf{y}),\dots,y_1(T)]^\top$.
    
\begin{figure}[!ht]
            \centering
            \includegraphics[width=0.8\textwidth]{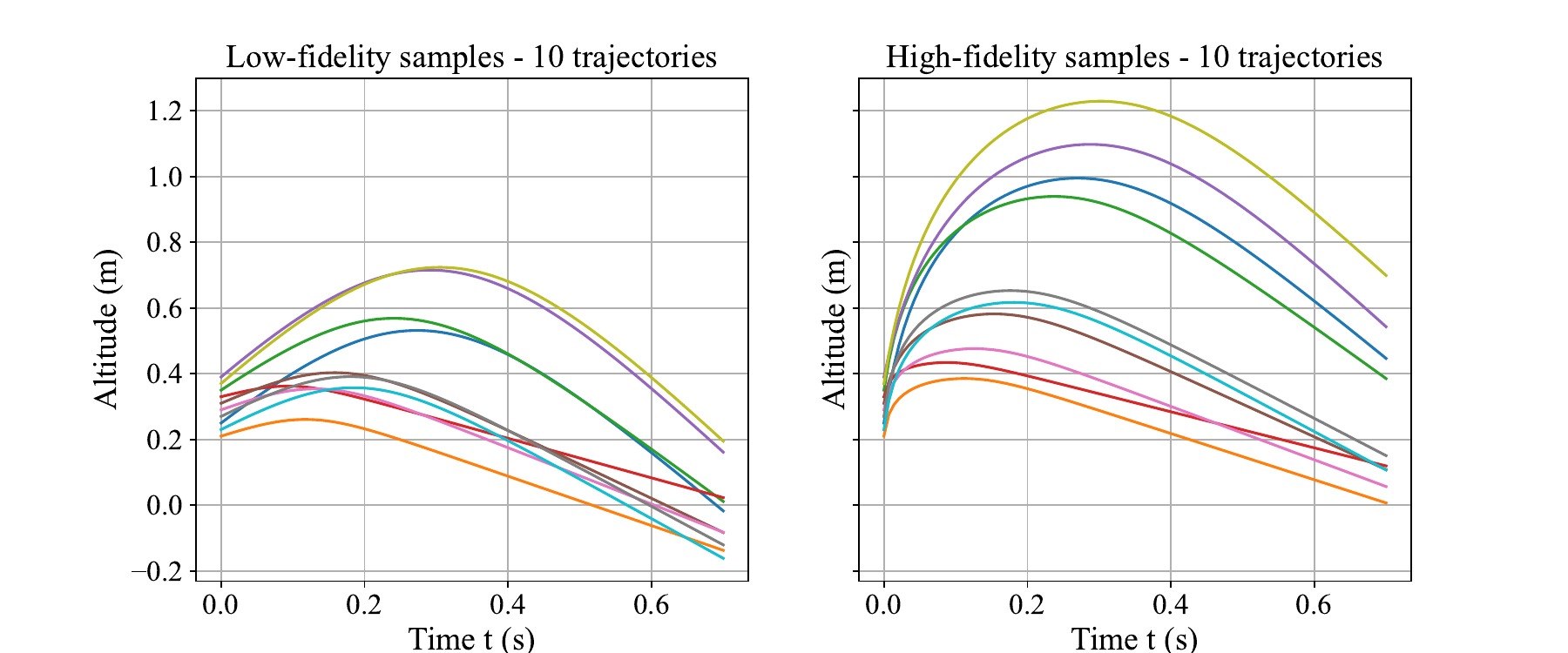}
            \caption{10 samples of the high- and low-fidelity output fields for the viscous free fall of a ball in a fluid in the absence of a ground. High- and low-fidelity fields corresponding to the same input samples are in the same color.}
            \label{fig:samples-vff-ng}
        \end{figure}

\begin{figure}[!ht]
            \centering
            \includegraphics[width=0.8\textwidth]{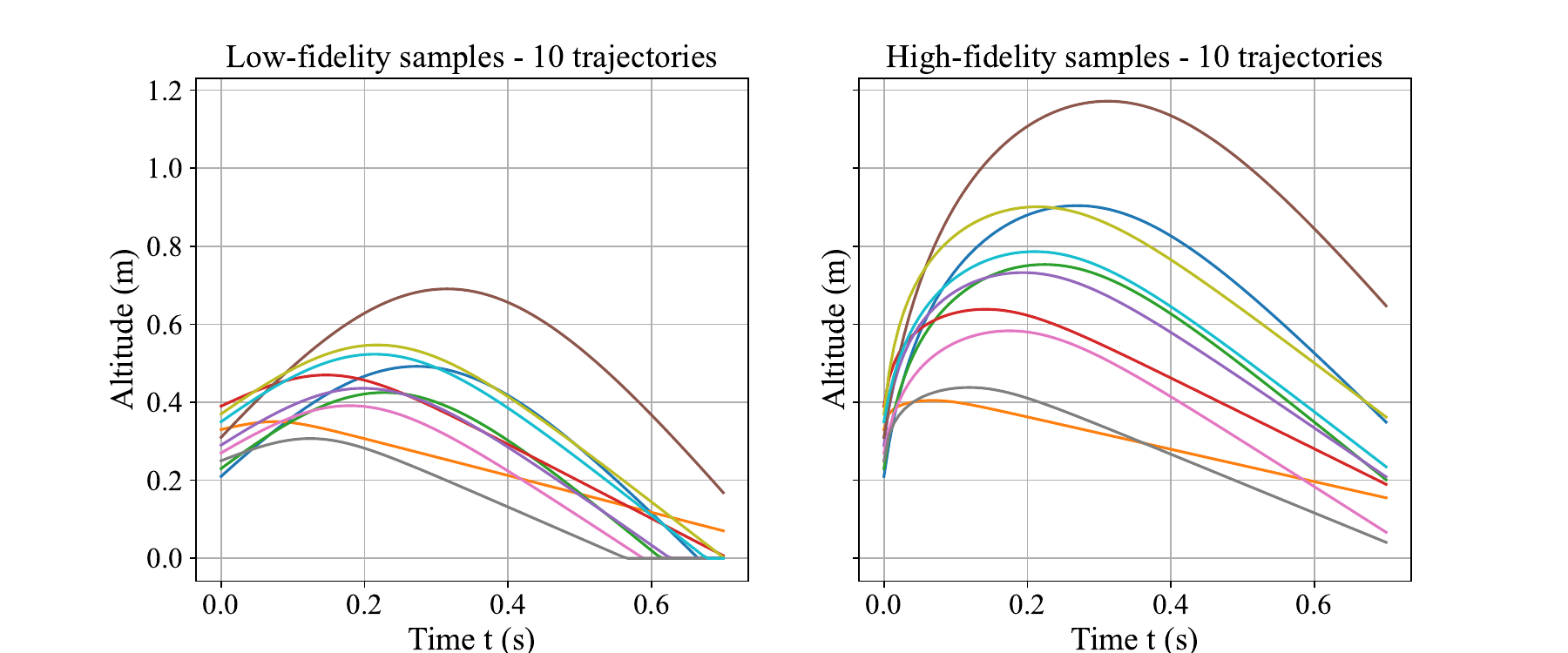}
            \caption{10 samples of the high- and low-fidelity output fields for the viscous free fall of a sphere in a fluid in the presence of a ground. High- and low-fidelity fields corresponding to the same input samples are in the same color.
            }
            \label{fig:samples-vff-g}
        \end{figure}

        From both fidelities, two related test cases are derived.
        First, the trajectory is considered in the absence of a ground.
        This configuration does not imply any modification to the problem introduced above.
        Samples of this configuration are shown in \Cref{fig:samples-vff-ng}.
        Second, a ground is added at an altitude of 0 m without rebound on impact, thus introducing a discontinuity when the sphere touches the ground.
        This is achieved by replacing every negative value of the altitude by a zero.
        Samples of this configuration are depicted in \Cref{fig:samples-vff-g}.
        
        \begin{figure}[!ht]
            \centering
            \includegraphics[width=0.8\textwidth]{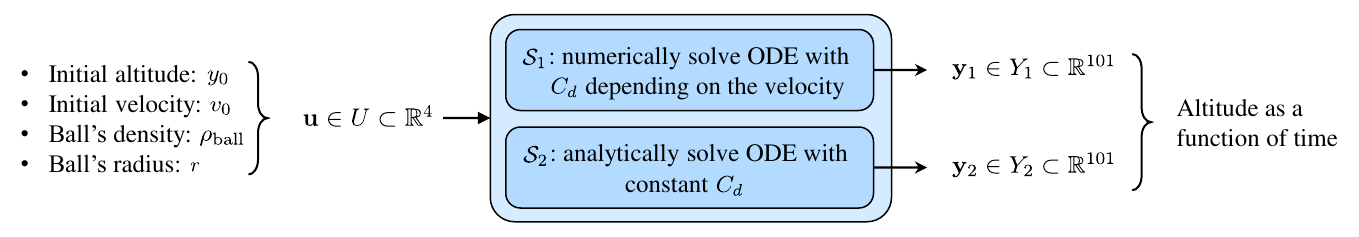}
            \caption{Simulation setup for the free fall of a ball in a fluid medium case.}
            \label{fig:simulation-setup-vff}
        \end{figure}
    
        It can be seen that the shapes of the high- and low-fidelity trajectories are similar.
        For the high-fidelity, the amplitude is higher and the terminal velocity is reached earlier.
        The impact with the ground is reached earlier in the low-fidelity simulations.
        Note that the simulation duration is chosen so that some low-fidelity simulations reach the ground while some do not.
        The time of impact is also varying among the different samples.

    \subsection{One-dimensional pressure coefficient field at the wall of a NACA~0015 airfoil}\label{sec:naca-case}

        The second test case is the study of the aerodynamic properties of an airfoil in a subsonic flying regime.
        The quantity of interest is the pressure coefficient $c_p$ field at the upper surface of the airfoil, which is defined as $(p-p_\infty)/q_\infty$, with $p$ the static pressure at the point of evaluation, $p_\infty$ the freestream static pressure and $q_\infty$ the freestream dynamic pressure.
        The goal is to build a surrogate of this spatial $c_p$ field as a function of the angle of attack and of the deformation of the airfoil geometry.

\begin{figure}[!ht]
            \centering
            \includegraphics[width=0.7\textwidth]{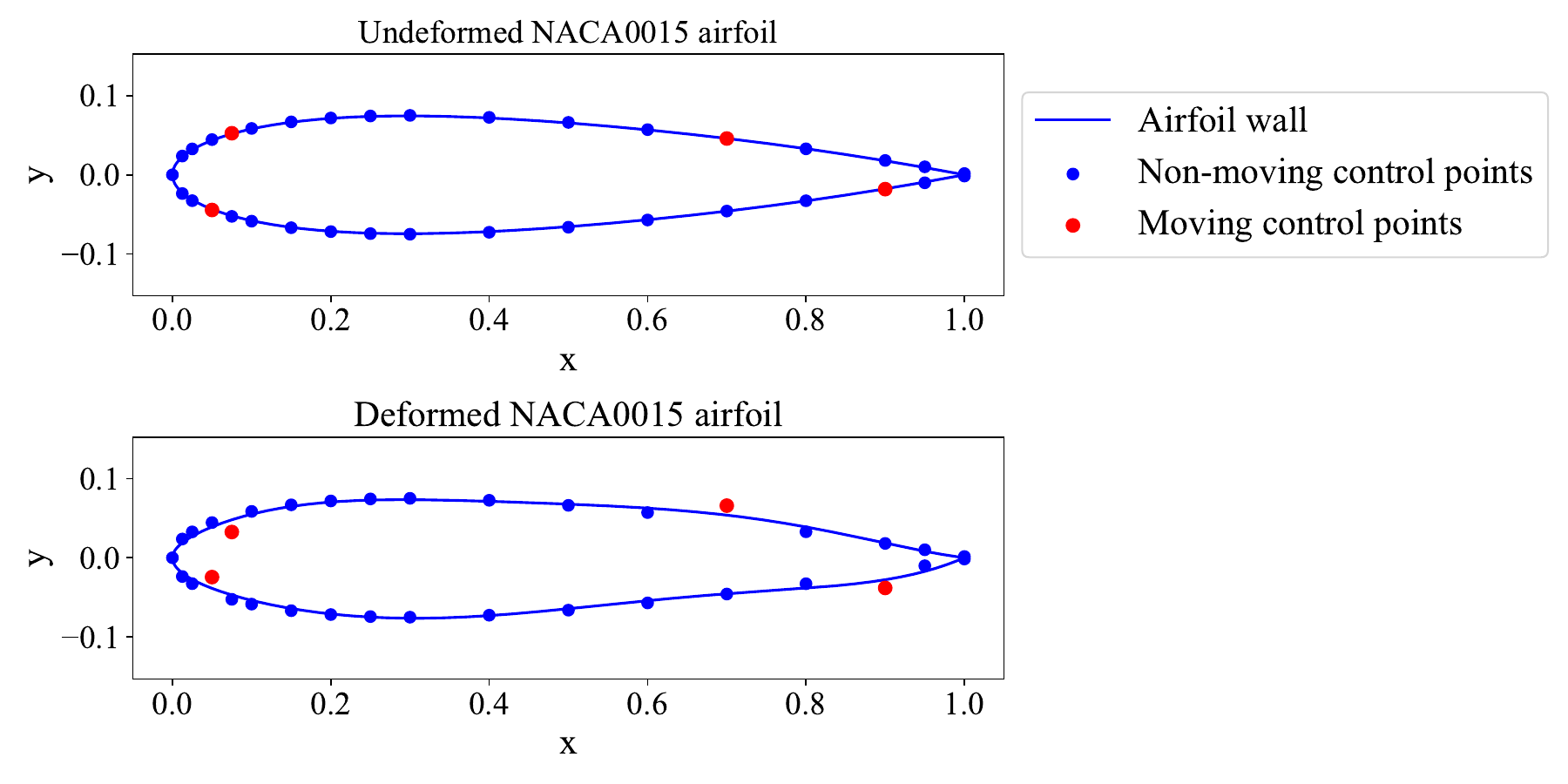}
            \caption{Undeformed airfoil (top) and deformed airfoil (bottom). The deformation of the airfoil is amplified for illustration purposes.}
            \label{fig:airfoil-onera-profil}
        \end{figure}

        The baseline airfoil is a symmetrical NACA~0015 with a chord length of 1 m.
        The geometry is parameterized by 35 control points spread out along the surface.
        They are numbered starting from the trailing edge, then going along the upper surface from right to left, then along the lower surface from left to right, finishing by the trailing edge again.
        The control points n$^\circ$5, 14, 21 and 33 are allowed to move vertically as can be seen in \Cref{fig:airfoil-onera-profil} in order to alter the shape of the airfoil. The shape of the wing is then interpolated from these control points with Bernstein polynomials.
        The angle of attack $\alpha$ varies between 4° and 6°.

\begin{figure}[!ht]
            \centering
            \includegraphics[width=0.9\textwidth]{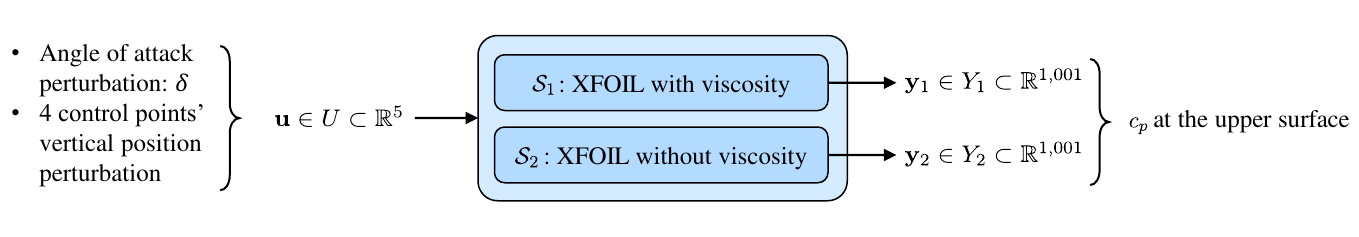}
            \caption{Simulation setup for the NACA~0015 airfoil case}
            \label{fig:simulation-setup-ao}
        \end{figure}

\begin{figure}[!ht]
            \centering
            \includegraphics[width=0.8\textwidth]{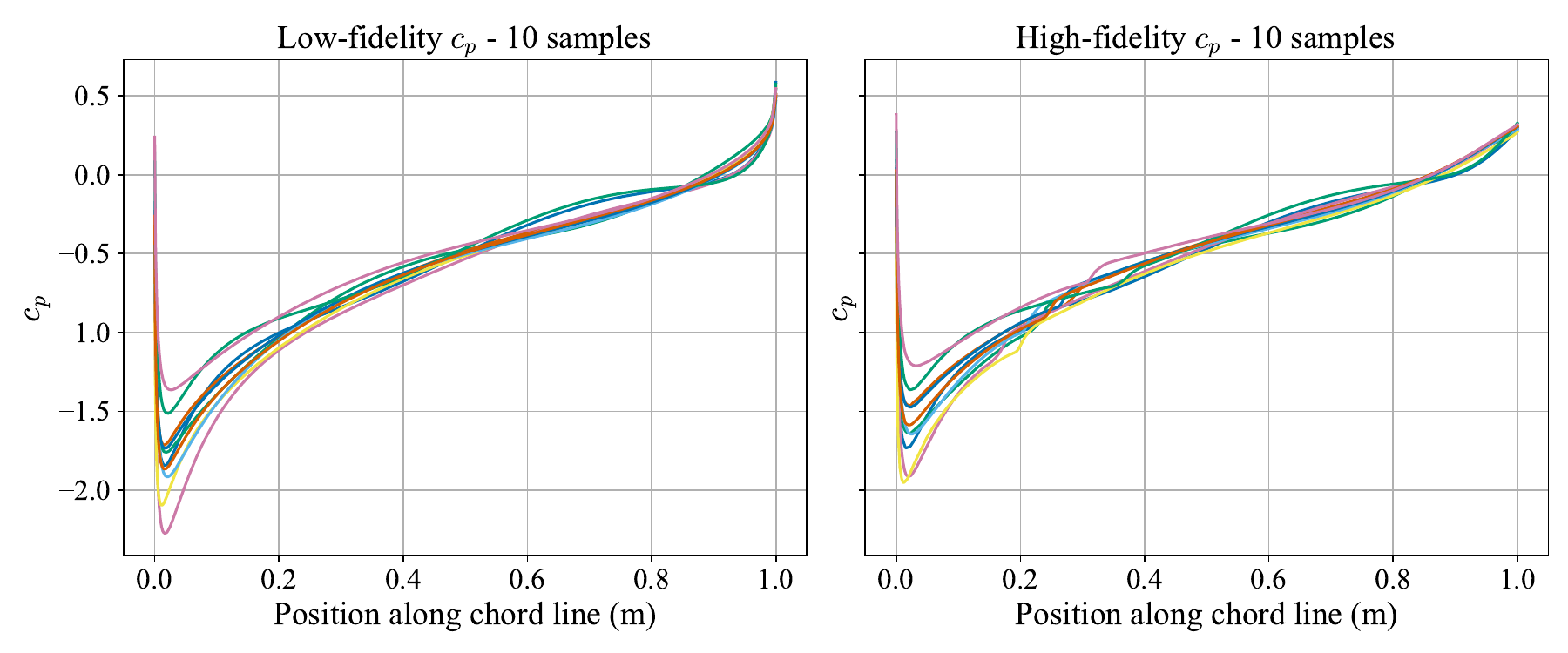}
            \caption{Ten samples of the high- and low-fidelity output fields for the NACA~0015 baseline airfoil case. High- and low-fidelity fields corresponding to the same input samples are in the same color.}
            \label{fig:samples-ao}
        \end{figure}

        The input variables of the simulators are represented by the vector $\mathbf{u}=(\alpha, \delta_5, \delta_{14}, \delta_{21}, \delta_{33})^\top$, with $\delta_i\in[-0.01,0.01]$ m the vertical displacement of the $i$-th control point from its nominal position.
        The simulations are performed using the Python package \texttt{aeropy} \cite{lealAeroPy2016} which itself uses XFOIL \cite{drelaXFOIL1989} to solve the CFD problem with a panel method.
        For both the high- and low-fidelity simulators, the upper and lower surfaces of the airfoil are discretized into 80 nodes each. 
        After running the simulators, the pressure coefficient $c_p$ is known at the 160 nodes.
        The 80 $c_p$ values on the upper surface are then interpolated on a finer mesh of 1001 nodes evenly spaced along the chord line. 
        The interpolator is the piecewise cubic Hermite interpolating polynomial of \texttt{scipy} \cite{virtanenSciPy2020}.
        For both the high- and low-fidelity, the output of the simulator is the $c_p$ field at the upper surface of the airfoil on this finer mesh.
        In the low-fidelity simulator, the fluid is supposed to be inviscid.
        In the high-fidelity simulator, the viscosity is taken into account, and the Reynolds number is $10^6$.
        The simulation setup is summarized in \Cref{fig:simulation-setup-ao} and samples of both the high- and low-fidelity simulators are depicted in \Cref{fig:samples-ao}.
    
        It can be seen that the high- and low-fidelity simulators output very similar $c_p$ fields.
        The most notable difference is the ``step'' around 0.25 m along the chord line for the high-fidelity samples.
        This step corresponds to the boundary layer separation.
        Notice that this step is not located at the same spatial position between snapshots.

    \subsection{Two-dimensional pressure coefficient field around an airfoil}\label{sec:rae-case}

        The last test case is the study of the aerodynamics of a transonic airfoil in a subsonic regime.
        This test case comes from a study of Perron \textit{et al}.\ \cite{perronMultifidelity2021} from which the data has been made available by the authors.
        The aim is to build a surrogate of the pressure coefficient $c_p$ in the 2D flow around the airfoil as a function of the angle of attack and the deformation of the airfoil.

        The baseline airfoil is the RAE~2822 which is immersed in an airflow with a freestream Mach number of 0.725.
        The angle of attack is varied between 0° and 4°.
        The airfoil is deformed by Free Form Deformation (FFD) \cite{kenwayCADFree2012}.
        It is parameterized by two FFD boxes that can only move vertically, with a displacement ranging from -0.03 to 0.03 chord length.
        The quantity of interest is the pressure coefficient field in the 2D spatial flow field.
        This simulation can typically be used to study the aerodynamic properties of the airfoil.
    
        The datasets provided by Perron \textit{et al}.\ consists of the simulation of this physical model for three different levels of fidelity.
        For each fidelity, the CFD calculations are carried out using the Stanford University Unstructured (SU2) code suite \cite{economonSU22016}.
        Additionally, the mesh at each fidelity is a structured grid, yet with varying resolution.
        When the surface of the airfoil is deformed by FFD, the deformation is propagated to the mesh using a linear elasticity approach \cite{dwightRobust2009}.
        This method ensures that for a given fidelity, the number of nodes and connectivity remain the same for every snapshot.
        Note that it also implies that the nodes have varying coordinates between the snapshots of a given fidelity.

        The simulator of higher-fidelity is a CFD RANS simulator with a Spalart-Allmaras turbulence model \cite{spalartOneequation1992}.
        The finest mesh comprises 41{,}796 nodes.
        The intermediate fidelity simulator is also a CFD RANS simulator with a Spalart-Allmaras turbulence model but with a coarser mesh of 10{,}530 nodes.
        The lower-fidelity simulator is a CFD Euler simulator with an even coarser mesh of 8{,}910 nodes with no near-wall refinement.
        For more details about the implementation, the reader is referred to the work of Perron \textit{et al}.\ \cite{perronMultifidelity2021}.

        \begin{figure}[!ht]
            \centering
            \includegraphics[width=0.7\textwidth]{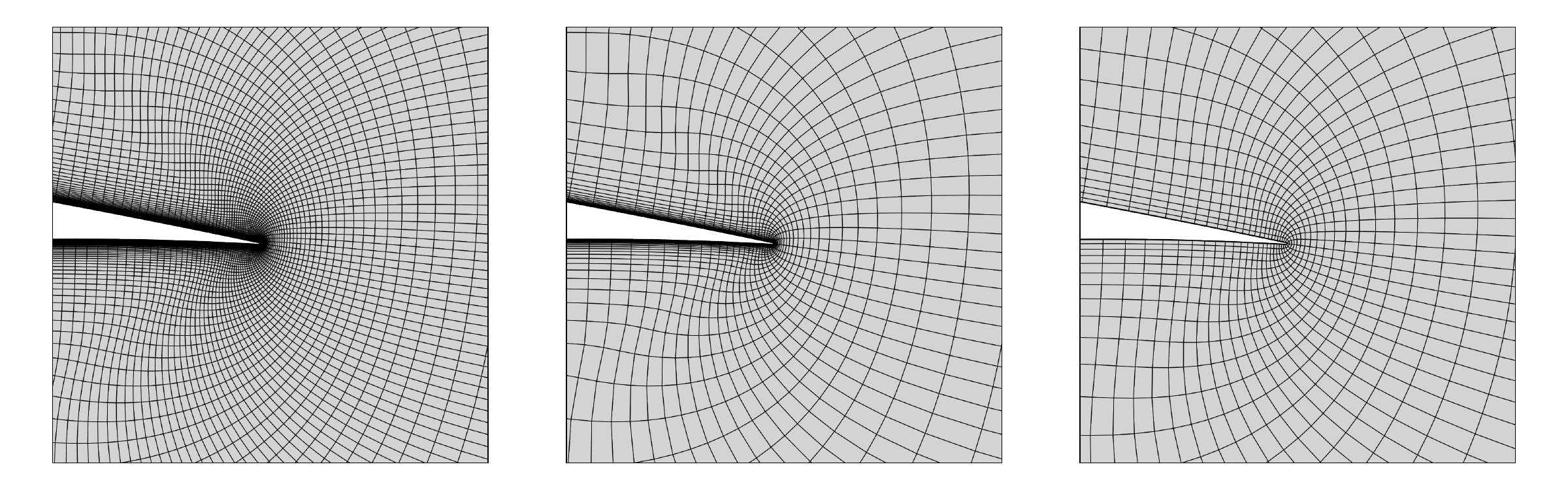}
            \caption{Zoom on the trailing edge of mesh $\mathbf{X}_1 $ (left), $\mathbf{X}_2 $ (center), $\mathbf{X}_3 $ (right)}
            \label{fig:mesh-ta}
        \end{figure}

        Since not all of the reviewed multi-fidelity surrogates can handle more than two fidelities, the benchmark is limited to the case where only two fidelities are available.
        Two possible configurations are considered in this paper.
        First, the RANS/RANS configuration is the combination of the CFD RANS with a fine mesh as the high-fidelity simulator $\mathcal{S}_1$ and the CFD RANS with a coarser mesh as the low-fidelity simulator $\mathcal{S}_2$.
        This configuration yields a high correlation between the high- and low-fidelity output fields since the physical hypotheses of CFD RANS is the same for both fidelities.
        Second, the combination of the CFD RANS with a fine mesh as the high-fidelity simulator $\mathcal{S}_1$ and the CFD Euler with a coarser mesh as the low-fidelity simulator $\mathcal{S}_2$ is referred to as the RANS/Euler configuration hereinafter.
        This second configuration has a lower correlation between the high- and low-fidelity output fields since the physical hypothesis of CFD RANS and CFD Euler are not the same (no viscosity model for CFD Euler).
    
        \begin{figure}[!ht]
            \centering
            \includegraphics[width=0.9\textwidth]{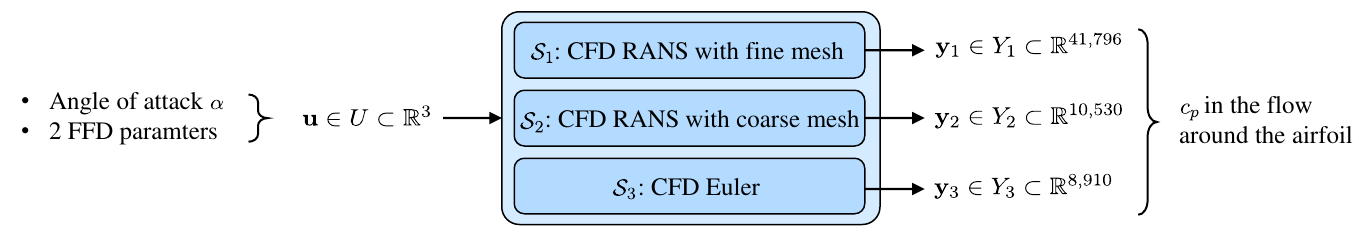}
            \caption{Simulation setup for the transonic airfoil case}
            \label{fig:simulation-setup-ta}
        \end{figure}

        \begin{figure}[!ht]
            \centering
            \includegraphics[trim={0 0 4cm 0.8cm},clip,width=\textwidth]{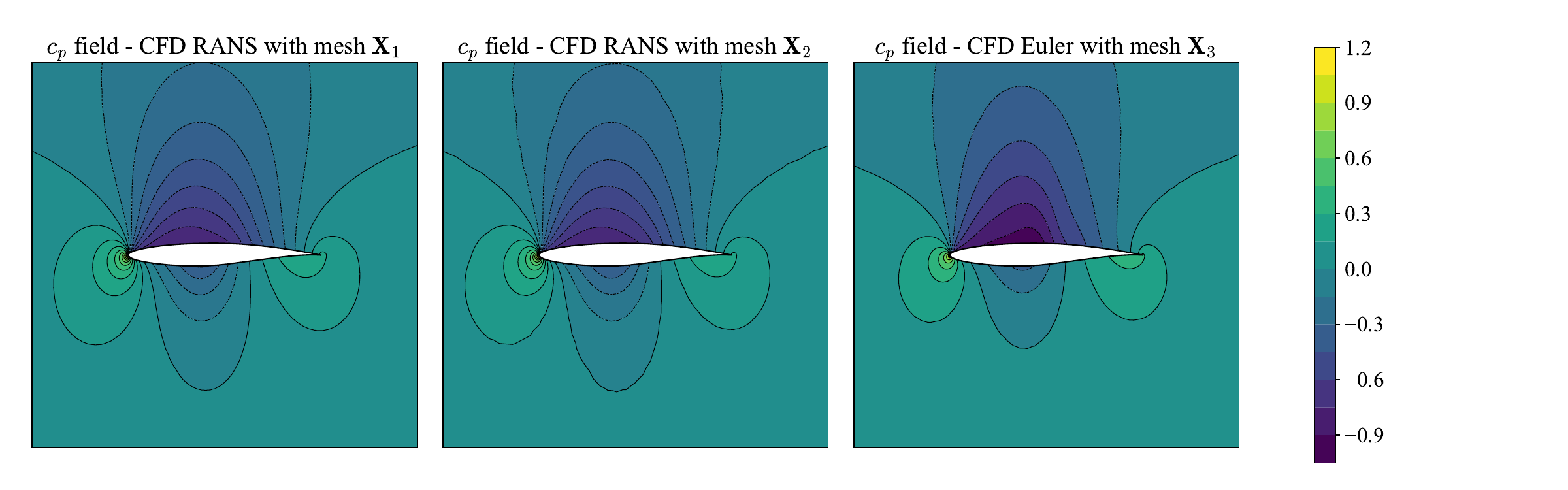}
            \caption{One sample of the CFD RANS with a fine mesh (left), CFD RANS with a coarser mesh (center) and CFD Euler (right) fields for the RAE~2822 airfoil without deformation and an angle of attack of 0°}
            \label{fig:samples-ta.pdf}
        \end{figure}


\section{Benchmark methodology}\label{sec:benchmark-methodology}

    \subsection{Training and validation data}

        The generation of the training and validation data has been designed to be consistent across all the test cases.
        The number $n_1$ of high-fidelity training samples is proportional to the dimensionality of the input space $\dim(U)$, \textit{i.e.}, there are $n_1=\{2, 5, 10\}\times\dim(U)$ high-fidelity training samples.
        Similarly, the number $n_2$ of low-fidelity training samples is chosen proportional to the number of high-fidelity samples. We choose $n_2=\{1, 5, 10\}\times n_1$.
        This gives 9~combinations of $(n_1,n_2)$.
        Usually in a multi-fidelity scenario, $n_2>n_1$.
        The configuration $n_1=n_2$ is included because the mapping and corrective surrogates are built for this specific context of the same number of high- and low-fidelity snapshots, admittedly a strong limitation of these approaches.
        This topic is further discussed in \Cref{sec:general-remarks}.

        To assess the robustness of the multi-fidelity surrogates to the training DoE, ten repetitions are considered with different DoEs for each of these combinations.
        A validation set of $n_v=1{,}000$ samples that are all different from the training samples is also created.

        For a given repetition, the value of the input training samples is chosen with a nested Latin Hypercupe Sampling (LHS) strategy, as some multi-fidelity surrogates evaluated in the benchmark require a nested DoE.
        In this work, the LHS are created with a modified version of the \texttt{smt} \cite{savesSMT2023} toolbox.
        It generates a first LHS of $n_1$ samples in $U$ that constitutes the high-fidelity input DoE $\mathcal{U}_1$.
        Then, a second LHS is built with $n_1+n_2$ samples.
        From this second LHS, the sample closest in terms of Euclidean distance to the first sample of $\mathcal{U}_1$ is removed.
        This operation is repeated for every sample in $\mathcal{U}_1$.
        $\mathcal{U}_2$ is finally obtained by the union of the remaining samples with $\mathcal{U}_1$, hence the name nested.
        Note that this nested scheme implies that $\mathcal{U}_1\subset\mathcal{U}_2$.
        The output DoE $\mathcal{Y}_1$ (respectively $\mathcal{Y}_2$) is obtained by running the high-fidelity (respectively low-fidelity) simulator over the samples in $\mathcal{U}_1$ (respectively $\mathcal{U}_2$).
        Surrogates requiring the same number of high- and low-fidelity snapshots (\textit{i.e.}, corrective and mapping approaches) only use the common part of the design of experiments, that is only the high- and low-fidelity snapshots corresponding to the input samples in $\mathcal{U}_1$.
        Consequently, their training set is made of $n_1$ high- and low-fidelity snapshots when others have $n_1$ high-fidelity and $n_2$ ($\geq n_1$) low-fidelity snapshots. The additional $n_2 - n_1$ low-fidelity snapshots are simply discarded for corrective and mapping surrogates.

        The specific case of the transonic airfoil requires an additional step since the samples are already available while generating new ones is not possible.
        For a given repetition, after the generation of the nested LHS, each input DoE sample is replaced by the closest input sample in terms of Euclidean distance from the Perron \textit{et al}.\ dataset, making sure that no sample from the Perron \textit{et al}.\ \cite{perronMultifidelity2021} dataset is used more than once.
        The output DoE samples are then replaced by the corresponding output samples from the Perron \textit{et al}.\ \cite{perronMultifidelity2021} dataset.

        To ensure a representative comparison of the surrogates, the intermediate surrogate models have been streamlined.
        More precisely, each single-fidelity intermediate surrogate has been replaced by ordinary Kriging \cite{rasmussenGaussian2006} and each multi-fidelity Kriging by AR1 co-Kriging \cite{legratietRecursive2014}.
        The practical implementation used for the benchmark and the naming of the surrogates is summarized in \Cref{tab:summary}.
        The Python code of the surrogate models as well as the training and validation data are available upon request.
        All the multi-fidelity surrogates are compared to two reference single-fidelity surrogates, trained using only the high-fidelity snapshots.
        The first one is based on PCA with one Gaussian process per latent variable.
        The second one is a Gaussian process with covariance tensorization.

        \begin{table}[!ht]
            \centering
            \caption{Summary of multi-fidelity surrogates for the prediction of field outputs (for the type of surrogate model: C=Corrective, M=Mapping, F=Fusion, TensCovGPR=Kriging with tensorized covariance, S=Single-fidelity; for dimensionality reduction: SF=Single-Fidelity, MF=Multi-fidelity, HF=High-Fidelity, LF=Low-Fidelity, PCA=Principal component Analysis, DiffPCA=PCA of the difference between the high- and low-fidelity snapshots, CPCA=Contrained PCA, GPCA=Gappy-PCA, KPCA=Kernel PCA, LTSA=Local Tangent Space Alignment, MA=Manifold Alignment; and for the intermediate surrogate: GPR=Gaussian process regression, CategGPR=GPR with categorical variable, AR1=Autoregressive co-Kriging)}
            \tiny
            \begin{tabular}{c c c c c c}
                 \toprule
                 \textbf{Name} & \textbf{Ref.} & \textbf{Type} & \begin{tabular}{@{}c@{}}\textbf{Common}\\\textbf{mesh}\end{tabular} & \textbf{DR} & \begin{tabular}{@{}c@{}}\textbf{Intermediate}\\\textbf{surrogate}\end{tabular} \\
                 \toprule
                 S-HFPCA-GPR & -- & -- & -- & PCA of $\mathcal{Y}_1$ & GP regression \\
                 \midrule
                 TensCovGPR & \cite{perrinAdaptive2020} & -- & -- & -- & GP regression \\
                 \midrule
                 C-DiffPCA-GPR & \cite{malouinInterpolation2013} & Corrective & Yes & PCA of $\mathcal{Y}_\Delta$ & GP regression \\ 
                 \midrule
                 M-GPCA & \cite{toalPotential2014} & Mapping & No & Gappy-PCA & Gappy-PCA \\
                 \midrule
                 M-SFPCA-GPR & \cite{wangMultifidelity2020,kangInvestigation2022} & Mapping & No & \begin{tabular}{@{}c@{}}Independent PCA \\ of $\mathcal{Y}_1$ and $\mathcal{Y}_2$\end{tabular} & GP regression \\
                 \midrule
                 F-CPCA-AR1 & \cite{benamaraMultifidelity2017} & Fusion & Yes & \begin{tabular}{@{}c@{}}Adapted \\ constrained PCA\end{tabular} & AR1 co-Kriging \\
                 \midrule
                 F-HFPCA-AR1 & \cite{thenonMultifidelity2016} & Fusion & Yes & PCA of $\mathcal{Y}_1$ & AR1 co-Kriging \\ 
                 \midrule
                 F-MFPCA-AR1 & \cite{rokitaMultifidelity2018} & Fusion & Yes & PCA of $\mathcal{Y}_1\cup\mathcal{Y}_2$ & AR1 co-Kriging \\ 
                 \midrule
                 F-MFPCA-CategGPR & \cite{mifsudVariablefidelity2016} & Fusion & Yes & PCA of $\mathcal{Y}_1\cup\mathcal{Y}_2$ & \begin{tabular}{@{}c@{}}GP regression with \\ a categorical \\ variable indicating \\ the fidelity \end{tabular} \\ 
                 \midrule
                 F-SFPCA-AR1 & \cite{bunnellMultifidelity2021} & Fusion & No & \begin{tabular}{@{}c@{}}Independent PCA \\ of $\mathcal{Y}_1$ and $\mathcal{Y}_2$\end{tabular} & AR1 co-Kriging \\ 
                 \midrule
                 F-SFPCA-MA-AR1 & \cite{perronMultifidelity2021} & Fusion & No & \begin{tabular}{@{}c@{}}Independent PCA \\ of $\mathcal{Y}_1$ and $\mathcal{Y}_2$ \\ with MA \end{tabular} & AR1 co-Kriging \\ 
                 \midrule
                 F-SFKPCA-MA-AR1 & \cite{deckerManifold2022} & Fusion & No & \begin{tabular}{@{}c@{}}Independent Kernel \\ PCA of $\mathcal{Y}_1$ and \\ $\mathcal{Y}_2$ with MA \end{tabular} & AR1 co-Kriging \\ 
                 \midrule
                 F-SFLTSA-MA-AR1 & \cite{deckerManifold2022} & Fusion & No & \begin{tabular}{@{}c@{}}Independent LTSA \\ of $\mathcal{Y}_1$ and $\mathcal{Y}_2$ \\ with MA \end{tabular} & AR1 co-Kriging \\ 
                 \midrule
                 F-SFIsomap-MA-AR1 & \cite{deckerManifold2022} & Fusion & No & \begin{tabular}{@{}c@{}}Independent isomap \\ of $\mathcal{Y}_1$ and $\mathcal{Y}_2$ \\ with MA \end{tabular} & AR1 co-Kriging \\ 
                 \midrule
                 F-LFPCA-AR1-Resid & \cite{kerleguerMultifidelity2023} & Fusion & Yes & PCA of $\mathcal{Y}_2$ & \begin{tabular}{@{}c@{}}AR1 co-Kriging \\ and Kriging \\ with tensorized \\ covariance \end{tabular} \\ 
                 \midrule
                 F-LFCVBPCA-AR1-Resid & \cite{kerleguerMultifidelity2023} & Fusion & Yes & \begin{tabular}{@{}c@{}}Cross-validation based \\ PCA of $\mathcal{Y}_2$ \end{tabular} & \begin{tabular}{@{}c@{}}AR1 co-Kriging \\ and Kriging \\ with tensorized \\ covariance \end{tabular} \\ 
                 \bottomrule
            \end{tabular}
            \label{tab:summary}
        \end{table}

        Some of the reviewed surrogates require that the low- and high-fidelity output fields are discretized on the same mesh.
        This is not a problem for the viscous free fall and the NACA~0015 airfoil since the low- and high-fidelity meshes are the same.
        For the RAE~2822 airfoil case, all fidelities have a different mesh.
        Since the surrogates that do not require a common mesh are able to predict high-fidelity output fields on the high-fidelity mesh, the high-fidelity mesh is chosen as the common mesh for surrogates that require the same mesh for each fidelity level.
        Whenever needed, the low-fidelity output field (either the CFD RANS with a coarse mesh or the CFD Euler) is transferred to the high-fidelity mesh with a nearest neighbor interpolation method (see \Cref{tab:summary}).
        The surrogate models of Kerleguer \textit{et al}.\ \cite{kerleguerMultifidelity2023} have been implemented but due to the very large computational cost of the CVBPCA, the surrogate using the latter could not be included in the benchmark.
        The manifold alignment technique implemented here is the Procrustes analysis as explained in \Cref{sec:manifold-alignment}.
        More details about the implementation of the methods can be found in \ref{app:numerical-settings}.

    \subsection{Comparison metrics}\label{sec:metrics}

        The performance of the different surrogates is estimated through their prediction accuracy.
        Predictions are made for the $n_v$ samples of the input validation set $\mathcal{U}_v$.
        The root-mean-square (rms) of the difference between the validation snapshots and the predictions is computed, which gives the rms error

        \begin{equation}\label{eq:rmse}
            e = \sqrt{\frac{1}{n_v}\sum^{n_v}_{i=1}\left|\left|
                \mathbf{y}_1^{(i)} - \hat{\mathbf{y}}^{(i)}_1
            \right|\right|^2},
        \end{equation}
        where $\mathbf{y}_1^{(i)}= \mathcal{S}_1\left(\mathbf{u}^{(i)}_l\right)$ is the evaluation of the high-fidelity simulator and $\hat{\mathbf{y}}^{(i)}_1$ is the prediction of the tested surrogate for the $i$-th sample of $\mathcal{U}_v$.

        To be comparable across different use cases, this error can be normalized by dividing $e$ by the rms of the difference between the validation snapshots and their mean value $\bar{\mathbf{y}}_1$, such that

        \begin{equation}\label{eq:rmse-norm}
            e^{\text{norm}}=\frac{e}{\sqrt{\frac{1}{n_v}\sum^{n_v}_{i=1}\left|\left|
                \mathbf{y}_1^{(i)} - \bar{\mathbf{y}}_1
            \right|\right|^2}}.
        \end{equation}

        For fusion-based surrogates using linear DR, the rms error can be decomposed to highlight the contributions of DR and of the intermediate surrogate, similarly to Perron \textit{et al}.\ \cite{perronMultiFidelity2020}.
        First, the DR error is computed by evaluating the rms of the difference between the validation snapshots and the validation snapshots mapped to the latent space and then mapped back.
        Hence, this error corresponds to the information lost by DR and its inverse mapping.

        \begin{equation}\label{rmse-dr}
            e_{\text{dr}} = \sqrt{\frac{1}{n_v}\sum^{n_v}_{i=1}\left|\left|
                \mathbf{y}_1^{(i)} - \mathcal{DR}_1^{-1}\left(\mathcal{DR}_1\left(\mathbf{y}_1^{(i)}\right)\right)
            \right|\right|^2}
        \end{equation}

        This DR rms error can be normalized as in \Cref{eq:rmse-norm},

        \begin{equation}\label{eq:rmse-dr-norm}
            e_{\text{dr}}^{\text{norm}}=\frac{e_{\text{dr}}}{\sqrt{\frac{1}{n_v}\sum^{n_v}_{i=1}\left|\left|
                \mathbf{y}_1^{(i)} - \bar{\mathbf{y}}_1
            \right|\right|^2}}
        \end{equation}

        Then, the intermediate surrogate modeling error is obtained by performing the following steps.
        First, the difference between the exact value of the latent variables obtained by applying DR to the validation snapshots $\mathcal{DR}_1\left(\mathbf{y}_1^{(i)}\right)$ and their prediction by the surrogate is computed for each sample $\hat{\mathbf{z}}^{(i)}_1=\mathcal{IS}_\text{MF}\left(\mathbf{u}^{(i)}_l\right)$.
        This difference cannot be computed for the case of nonlinear DR because they are not vectors but points on a manifold.
        Then, the difference is mapped back through the DR technique for each sample.
        Finally, the rms is computed,

        \begin{equation}\label{rmse-re}
            e_{\text{ism}} = \sqrt{\frac{1}{n_v}\sum^{n_v}_{i=1}\left|\left|
                \mathcal{DR}_1^{-1}\left(
                    \mathcal{DR}_1\left(\mathbf{y}_1^{(i)}\right)
                    - \hat{\mathbf{z}}^{(i)}_1
                \right)
            \right|\right|^2}.
        \end{equation}

        Once again, the intermediate surrogate modeling rms error can be normalized as in \Cref{eq:rmse-norm,eq:rmse-dr-norm}:

        \begin{equation}\label{eq:rmse-re-norm}
            e_{\text{ism}}^{\text{norm}}=\frac{e_{\text{ism}}}{\sqrt{\frac{1}{n_v}\sum^{n_v}_{i=1}\left|\left|
                \mathbf{y}_1^{(i)} - \bar{\mathbf{y}}_1
            \right|\right|^2}}.
        \end{equation}

        For the specific case of PCA, the relation between the rms error, the DR and the intermediate surrogate modeling error sum up, as shown in \cite{perronMultiFidelity2020}:

        \begin{equation}\label{eq:error-decomposition}
            e^2=e_{\text{dr}}^2+e_{\text{ism}}^2.
        \end{equation}

        These different errors allow to identify the most important contribution between DR and intermediate surrogate modeling to the global rms error.
        Additionally, the CPU-time required for training surrogates is recorded and the results are provided in \ref{app:computational-cost}.
        Information about the hardware used can be found in \ref{app:numerical-settings}.

        Although not a direct measure of the performance of the surrogate models, the dimension of the latent space may help the reader better understand the subsequent results.
        A table containing intervals of the dimension of the latent spaces for the different types of PCA over all test cases is provided in \ref{app:latent-dimension}.
        As for nonlinear DR, the dimension of the latent space for isomap and LTSA is set to $\operatorname{dim}(U)$ as is done in \cite{deckerManifold2022}, and to $n_1 - 1$ for KPCA.


\section{Results and discussions}\label{sec:results-discussions}

    The performance of the various multi-fidelity surrogates is now analyzed for the different test cases.
    The reading guide to the various plots is given with the results of the first test case in \Cref{sec:results-vffng}.
    The contribution of the different building blocks are then discussed.
    Whilst going through the subsequent results, readers should bear in mind that they depend on the choice of the numerical settings, detailed in \ref{app:numerical-settings}. The impact of changing the relative information content (RIC) is illustrated in \ref{app:ric}.

    \subsection{Viscous free fall of a ball without ground}\label{sec:results-vffng}

\begin{figure}
            \centering
            \includegraphics[trim={0 0 0 1cm},clip,width=\textwidth]{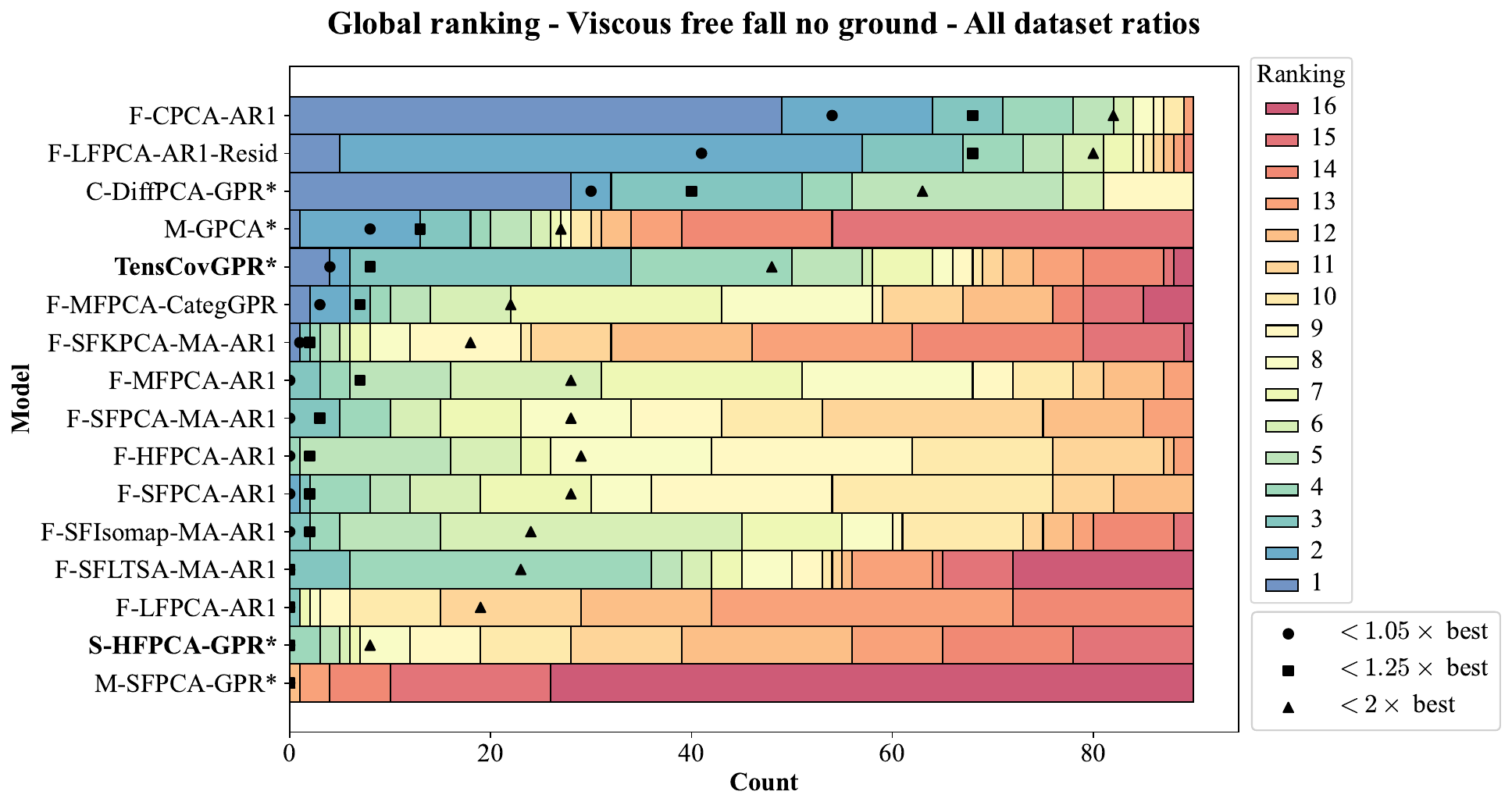}
            \caption{Global ranking of the various surrogates for the viscous free fall of a ball without ground. The colored bars show the ranking of the multi-fidelity surrogates. They are sorted vertically in decreasing number of times their normed RMSE is less than 5\%, 25\% and 100\% away from the best performing multi-fidelity surrogate.}
            \label{fig:ranking-vffng}
        \end{figure}

        The results for the viscous free fall of a sphere without ground are shown in \Cref{fig:ranking-vffng}.
        The benchmarked surrogates are listed along the vertical axis with their names given in \Cref{tab:summary}.
        Let us recall that $n_1$ (resp. $n_2$) is the number of high-fidelity (resp. low-fidelity) samples.
        For each combination of ($n_1$, $n_2/n_1$, index of repetition), each surrogate is trained, its RMSE is measured as described in \Cref{sec:metrics} and is ranked in ascending order of RMSE.
        \Cref{fig:ranking-vffng} displays a bar plot with the number of occurrences of each ranking being the length of the bars.
        For instance, the dark blue bar counts the number of times a given surrogate ranked best in terms of RMSE over all combinations of ($n_1$, $n_2/n_1$, index of repetition).
        Similarly, the dark red bar counts the number of times a given surrogate ranked worst.
        The colors between dark blue and dark red correspond to the ranking given in the legend of \Cref{fig:ranking-vffng} to the right.
        To avoid the bias of a ``strict ranking'', an additional metric is plotted in \Cref{fig:ranking-vffng}, which is depicted by the scatterplot (dots, squares and triangles).
        It allows us to measure how far apart surrogates are from the best performing one.
        For each combination of ($n_1$, $n_2/n_1$, index of repetition) the lowest RMSE of the surrogates is taken as a reference.
        Then, for each surrogate, the number of times that their RMSE is lower than 1.05 times the reference RMSE is counted.
        This number is represented by the dots and is used to sort the surrogates vertically (the first one being the best and the last one the worst).
        The procedure is the same for the square and triangle points, which accounts the number of times a RMSE is lower than 1.25 and 2 times the reference RMSE, respectively.
        This visualization is inspired by the work of L\"uthen \textit{et al}.\ \cite{luthenSparse2021}.
        Similar graphs are shown in the remainder of the paper for different test cases.

        \Cref{fig:ranking-vffng} shows that on this first case study, the fusion approach F-CPCA-AR1 performs better than any other method (it is ranking in the top 3 in 71 repetitions out of 90).
        It is closely followed by F-LFPCA-AR1-Resid, with 67 repetitions in the top 3.
        In 3\textsuperscript{rd} place, the corrective approach (C-DiffPCA-GPR) gets 51 repetitions in the top 3.
        The best mapping approach ranks 4\textsuperscript{th}, but has a large proportion of lower rankings (dark red bar, 51 repetitions over 90 among the 3~worst ranks), meaning that it is unreliable on this particular test case.
        The other mapping approach comes last, never ranking higher than 12\textsuperscript{th}.
        The single fidelity method TensCovGPR ranks higher (5\textsuperscript{th}) than a lot of multi-fidelity surrogates.
        Then, a sort of plateau of fusion surrogates performing similarly can be seen, including F-MFPCA-CategGPR, F-SFKPCA-MA-AR1, F-MFPCA-AR1, F-SFPCA-MA-AR1, F-HFPCA-AR1, F-SFPCA-AR1, F-SFIsomap-MA-AR1, and F-SFLTSA-MA-AR1, corresponding to rankings 6\textsuperscript{th} to 13\textsuperscript{th}.

        When comparing  F-SFPCA-MA-AR1 and F-SFPCA-AR1, respectively ranked 9\textsuperscript{th} and 11\textsuperscript{th}, it appears that manifold alignment brings a marginal improvement.
        The comparison between fusion surrogates that only vary by their DR technique (F-MFPCA-AR1, F-SFPCA-AR1, F-HFPCA-AR1, F-LFPCA-AR1 and F-CPCA-AR1) shows that using CPCA yields significantly better results for this test case, with 54 runs with a RMSE less than 5\% higher than the lowest RMSE, compared to 3 for the best of the others.
        As for the intermediate surrogate, the comparison of F-MFPCA-AR1 and F-MFPCA-CategGPR shows CategGPR is a better choice for this test case, as F-MFPCA-AR1 never performs less than 5\% worse than the best performing surrogate.
        Furthermore, F-LFPCA-AR1-Resid is ranked 2\textsuperscript{nd} while F-LFPCA-AR1 is ranked 14\textsuperscript{th}, illustrating a probable interest of the taking into account the dimensionality reduction residuals.

        \begin{figure}
            \begin{subfigure}{0.49\textwidth}
                \includegraphics[width=\textwidth]{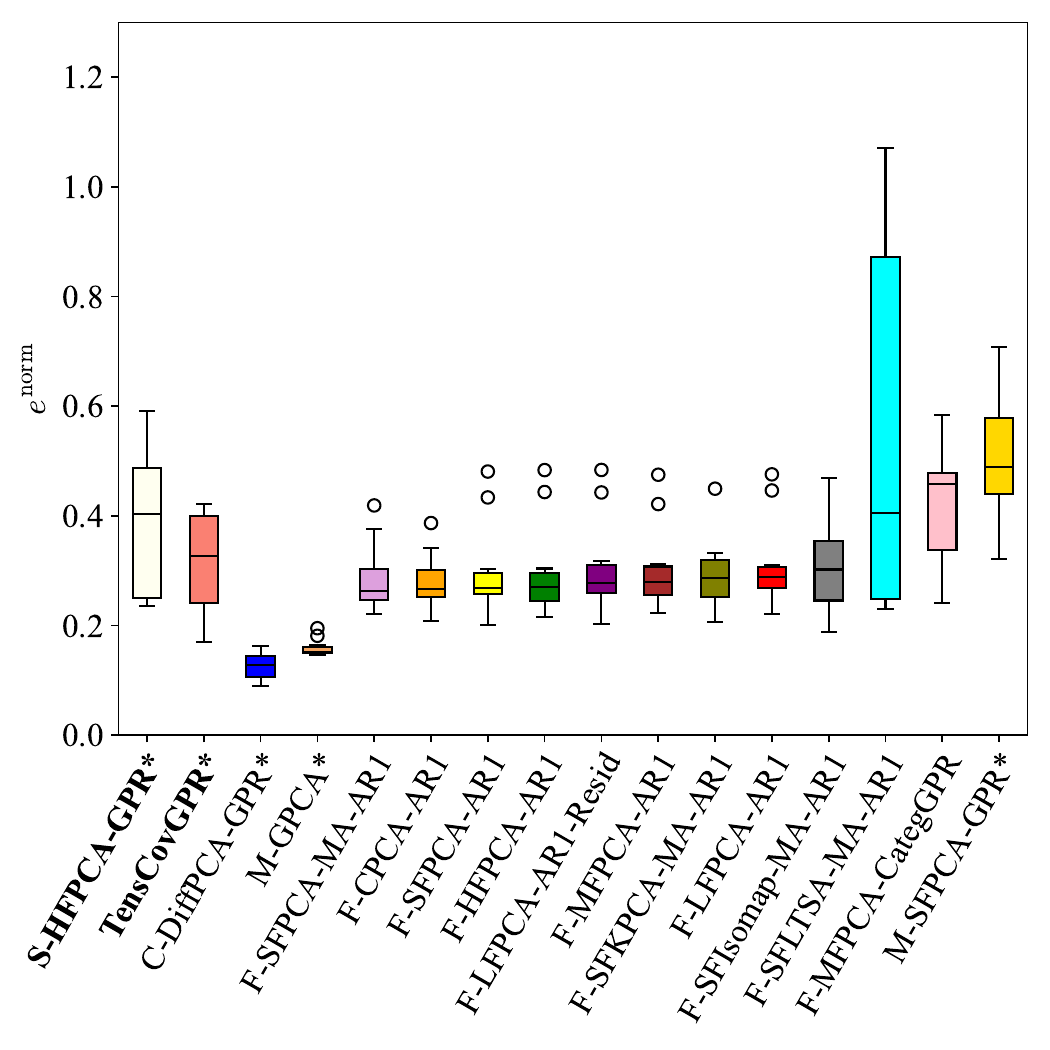}
            \end{subfigure}
            \hfill
            \begin{subfigure}{0.49\textwidth}
                \includegraphics[width=\textwidth]{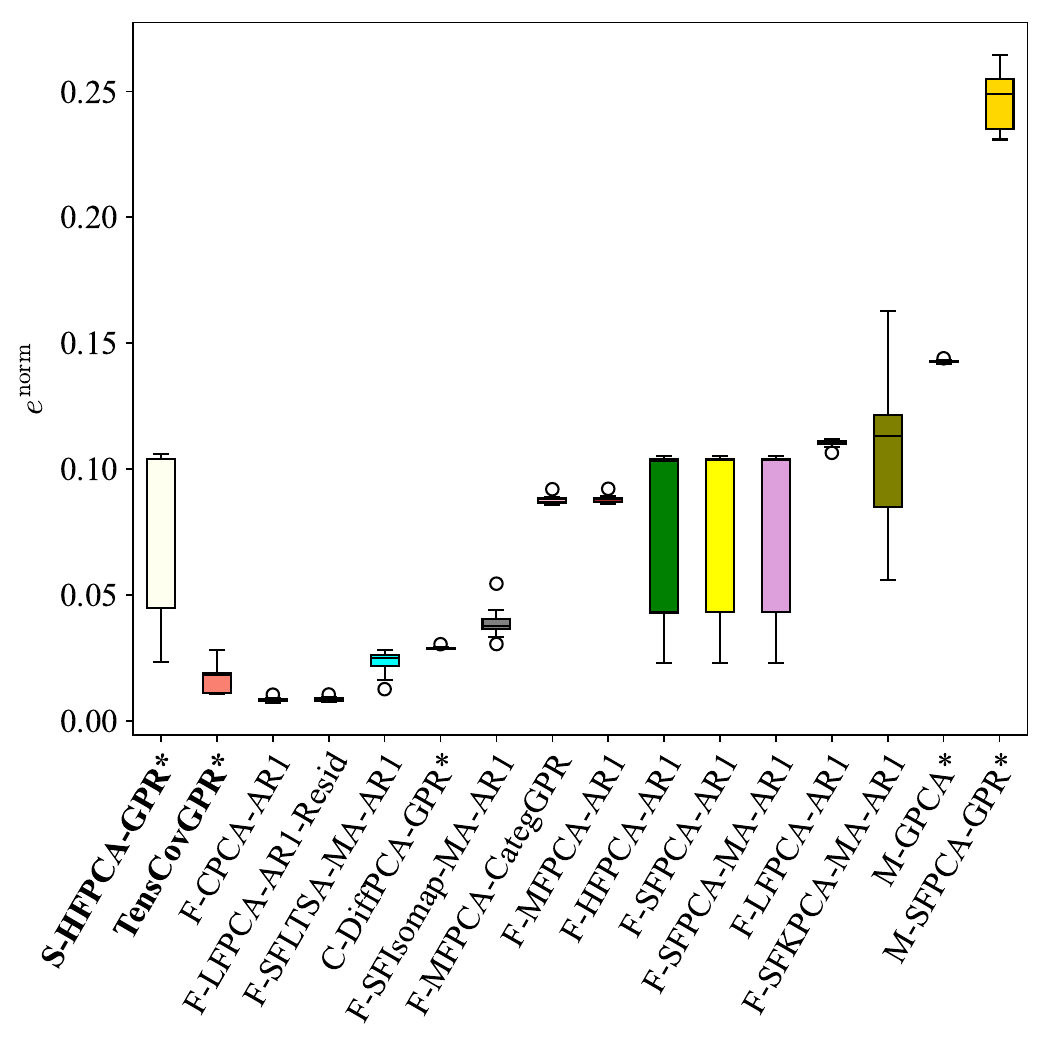}
            \end{subfigure}
            \caption{Boxplot of the normed RMSE for the viscous free fall without ground case study with $n_1=8$ and $n_2=8$ (left) and $n_1=40$ and $n_2=400$ (right). Single-fidelity surrogates (bold label) are stacked to the left and multi-fidelity surrogates are sorted by median normed RMSE.}
            \label{fig:rmse-boxplot-vffng}
        \end{figure}

        Ranking surrogates as in \Cref{fig:ranking-vffng} yields a hierarchy between them.
        However, it does not show how far apart surrogates are.
        The RMSE of the different surrogates are displayed in \Cref{fig:rmse-boxplot-vffng}.
        The two extreme cases $n_1=2\times\dim(U)$ with $n_2=n_1$ and $n_1=10\times\dim(U)$ with $n_2=10\times n_1$ are shown.
        The two single-fidelity surrogates (\textit{i.e.}, S-HFPCA-GPR and TensCovGPR) are stacked on the left for reference and multi-fidelity surrogates are sorted by increasing media RMSE.
        This graphical representation provides additional context to the rankings.
        When data is scarce, $n_1=8$ and $n_2=8$, the lowest median normed RMSE is reached by C-DiffPCA-GPR with 12.7\%.
        The largest median normed RMSE is reached by the mapping surrogate M-SFPCA-GPR with 49.0\%.
        The plateau of multi-fidelity surrogates is apparent but is different from the one seen in the rankings since results are not aggregated anymore.

        When data is abundant, $n_1=40$ and $n_2=400$, the lowest median normed RMSE is reached by F-CPCA-AR1 with 0.8\% while the largest is reached by M-SFPCA-GPR with 24.9\%.
        Some multi-fidelity surrogates that were close to the others are now falling behind, including surrogates using LFPCA, SFPCA, HFPCA and MFPCA, and mapping surrogates.
        Note that mapping surrogates do not benefit from the addition of low-fidelity snapshots that do not have corresponding high-fidelity snapshots.
        Surrogates using nonlinear DR seem to benefit more from the addition of snapshots than surrogates using linear DR.
        Anyhow, every surrogate is performing better with the addition of high- and low-fidelity snapshots.

    \subsection{Viscous free fall of a ball with ground}

        \begin{figure}
            \centering
            \includegraphics[trim={0 0 0 1cm},clip,width=\textwidth]{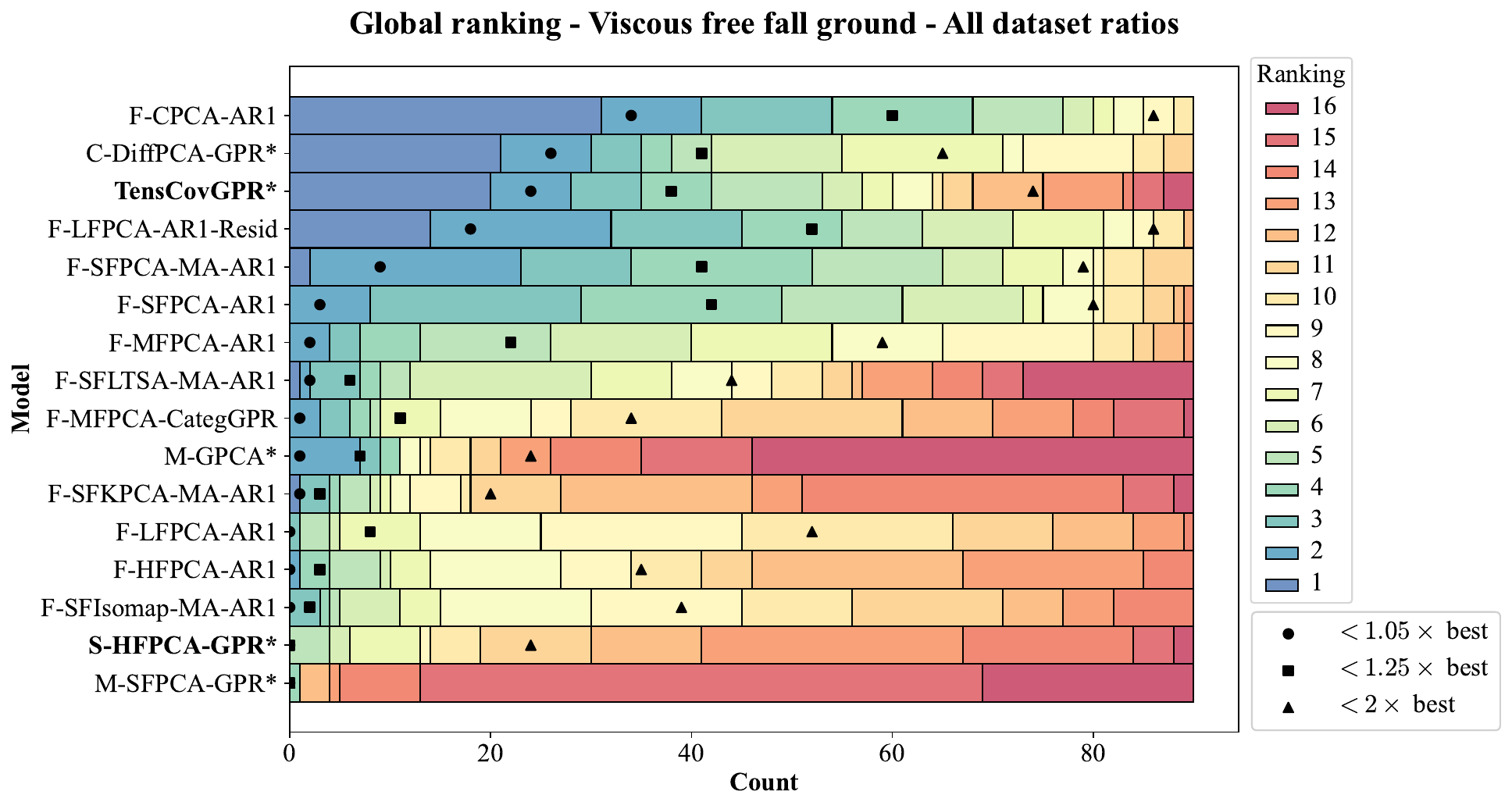}
            \caption{Global ranking of the various surrogates for the viscous free fall of a ball with ground. The colored bars show the ranking of the multi-fidelity surrogates. They are sorted vertically in decreasing number of times their normed RMSE is less than 5\%, 25\% and 100\% away from the best performing multi-fidelity surrogate.}
            \label{fig:ranking-vffg}
        \end{figure}

        \Cref{fig:ranking-vffg} shows the results for the viscous free fall with a ground test case.
        Simply adding a ground (and therefore a discontinuity in the snapshots) significantly changes the results.
        The best surrogate is still the multi-fidelity surrogate F-CPCA-AR1 with 54 repetitions over 90 among the top 3 surrogates.
        It is closely followed by the corrective surrogate C-DiffPCA-GPR coming second, with 35 repetitions in the top 3 surrogates, which was ranked 3\textsuperscript{rd} before the addition of the ground, surpassing the fusion method F-LFPCA-AR1-Resid now ranking 4\textsuperscript{th}.
        In between, in 3\textsuperscript{rd} place, the single-fidelity surrogate TensCovGPR has a moderately high proportion of bad performance as is shown by the dark red bars.
        Mapping surrogates (M-GPCA ranked 10\textsuperscript{th} with a large proportion of worst rankings, and M-SFPCA-GPR ranked last) are performing much worse than most other multi-fidelity surrogates on this test case.
        The addition of manifold alignment is still leading to a small improvement as can be seen by comparing F-SFPCA-MA-AR1 (ranked 5\textsuperscript{th}) and F-SFPCA-AR1 (ranked 6\textsuperscript{th}).
        Comparing the fusion surrogates that only vary by their DR technique (F-CPCA-AR1, F-MFPCA-AR1, F-SFPCA-AR1, F-HFPCA-AR1 and F-LFPCA-AR1) shows that CPCA still performs significantly better than the best of the others, with 34 against 3 repetitions with a good RMSE (where good means a RMSE below the best RMSE plus 5\%).
        Finally, the comparison of F-MFPCA-AR1 (ranked 7\textsuperscript{th}) and F-MFPCA-CategGPR (ranked 9\textsuperscript{th}) shows that AR1 is a better multi-fidelity approach for this test case.

        \begin{figure}
            \begin{subfigure}{0.49\textwidth}
                \includegraphics[width=\textwidth]{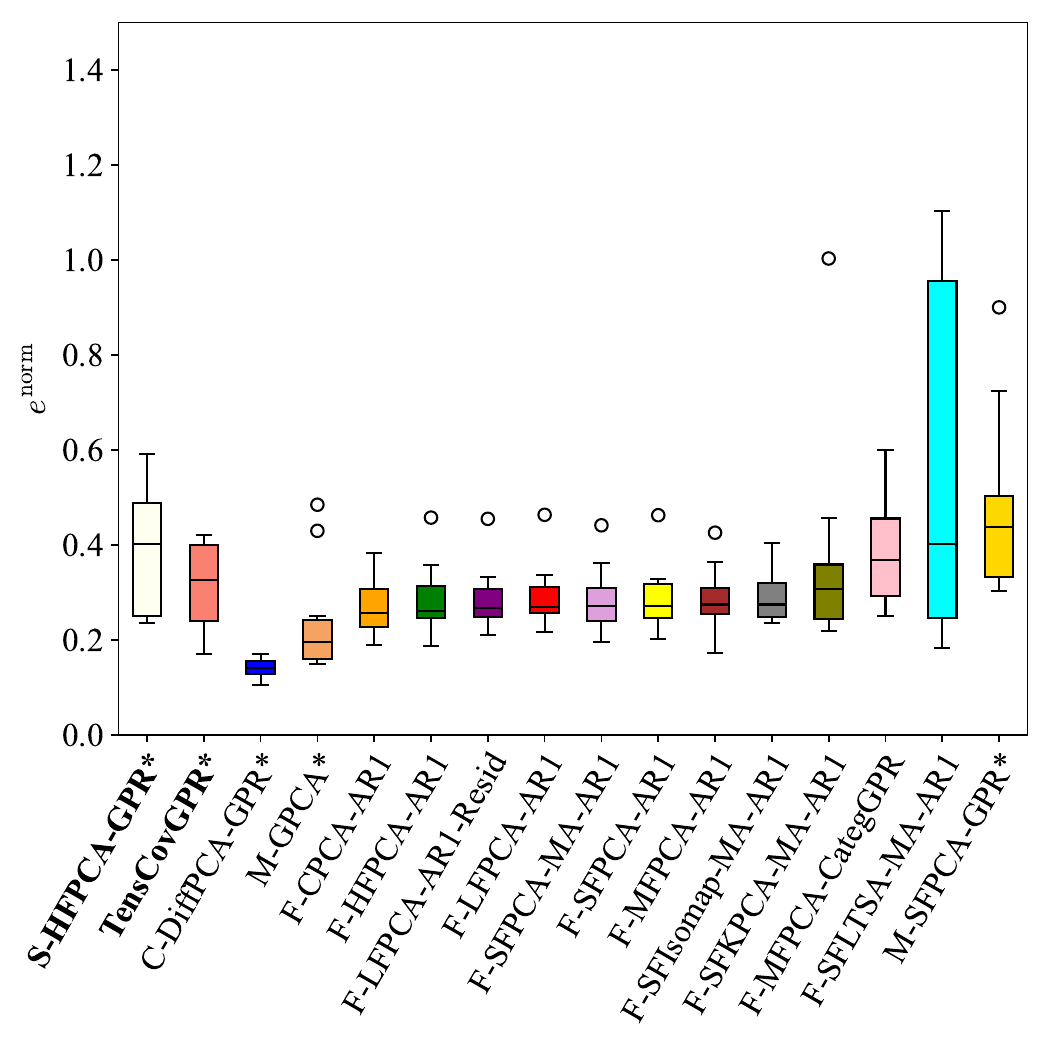}
            \end{subfigure}
            \hfill
            \begin{subfigure}{0.49\textwidth}
                \includegraphics[width=\textwidth]{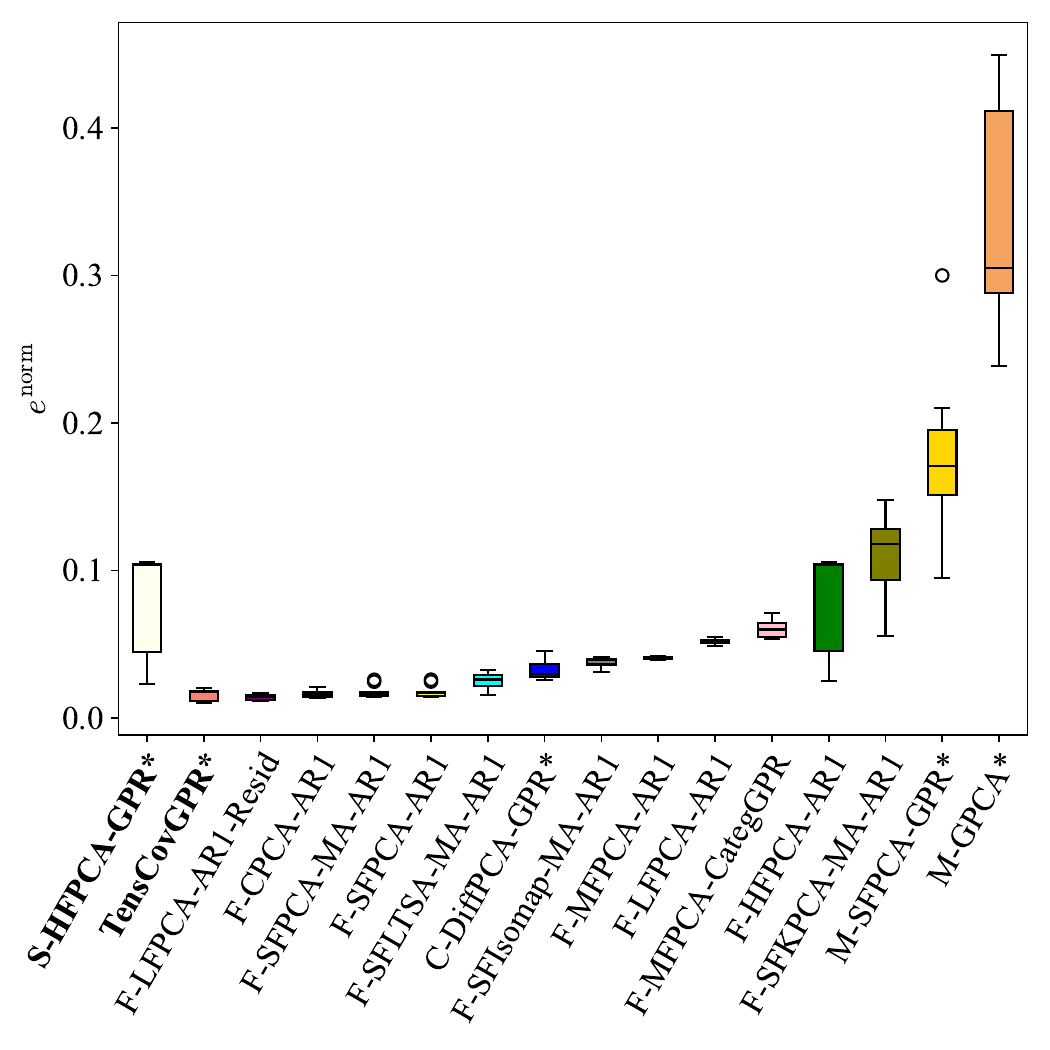}
            \end{subfigure}
            \caption{Boxplot of the normed RMSE for the viscous free fall with a ground case with $n_1=8$ and $n_2=8$ (left) and $n_1=40$ and $n_2=400$ (right). Single-fidelity surrogates (bold label) are stacked to the left and multi-fidelity surrogates are sorted by median normed RMSE.}
            \label{fig:rmse-boxplot-vffg}
        \end{figure}

        The normed RMSE are plotted in \Cref{fig:rmse-boxplot-vffg}.
        In the scarce data context, $n_1=8$ and $n_2=8$, the lowest median normed RMSE is reached by C-DiffPCA-GPR with 14.1\% while the largest is 43.8\% for M-SFPCA-GPR.
        Once again, a plateau can be seen, covering most multi-fidelity surrogates, even though the ones with a larger median normed RMSE have a larger variability.
        With more data to learn, when $n_1=40$ and $n_2=400$, the lowest median normed RMSE is 1.4\% for F-LFPCA-AR1-Resid while the largest is 30.5\% for M-GPCA.
        This time, the single-fidelity surrogate TensCovGPR is performing similarly to the best multi-fidelity surrogates with a normed RMSE of 1.7\%.
        Again, mapping surrogates benefit less than other surrogates from the addition of snapshots.

    \subsection{Pressure coefficient field around a NACA~0015 airfoil}

        \begin{figure}
            \centering
            \includegraphics[trim={0 0 0 1cm},clip,width=\textwidth]{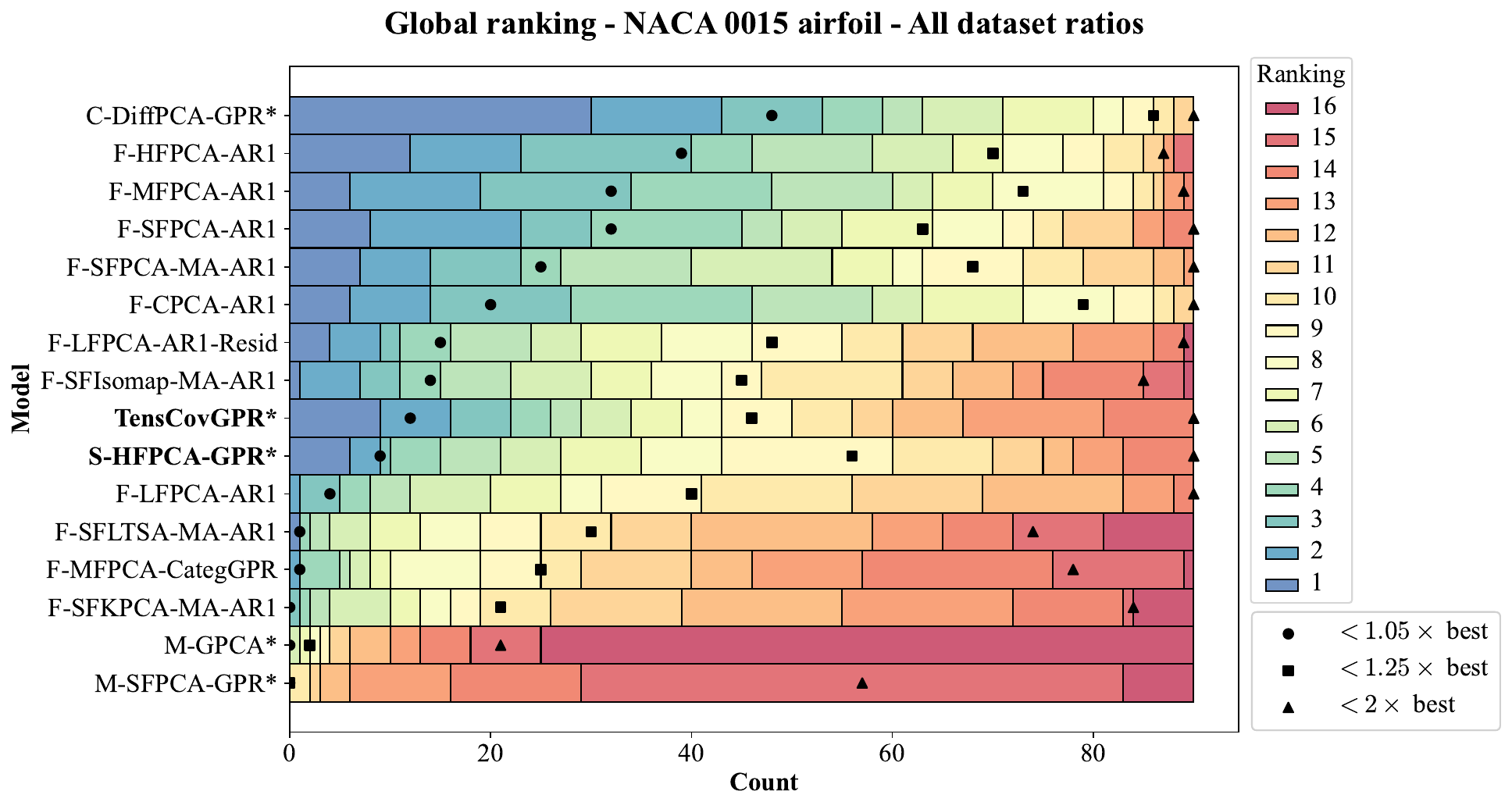}
            \caption{Global ranking of the various surrogates for the NACA~0015 airfoil. The colored bars show the ranking of the multi-fidelity surrogates. They are sorted vertically in decreasing number of times their normed RMSE is less than 5\%, 25\% and 100\% away from the best performing multi-fidelity surrogate.}
            \label{fig:ranking-ao}
        \end{figure}

        \Cref{fig:ranking-ao} depicts the results for the test case of a one-dimensional pressure coefficient field around NACA~0015 airfoils.
        C-DiffPCA-GPR is now the best performing method, with 53 repetitions out of 90 ranking in the top 3 surrogates.
        It is followed by the fusion surrogate F-HFPCA-AR1, with 40 repetitions in the top 3.
        The corrective surrogate and most of the fusion surrogates using linear DR perform better than the single-fidelity surrogates S-HFPCA-GPR and TensCovGPR.
        Most surrogates using LFPCA, nonlinear DR, CategGPR or the mapping approach rank lower than the single-fidelity surrogates (ranked 9\textsuperscript{th} and 10\textsuperscript{th}).
        The mapping surrogates are the worst performing surrogates on this test case, with 74 and 77 repetitions out of 90 in the 3 worst ranking surrogates.
        Here, comparing F-SFPCA-MA-AR1 and F-SFPCA-AR1 shows that the use of manifold alignment leads to a small decrease in the performance.
        F-MFPCA-AR1, F-SFPCA-AR1 and F-HFPCA-AR1 have very similar performances, suggesting that the choice of DR between MFPCA, SFPCA and HFPCA has no particular impact when combined with AR1 co-Kriging for this test case.
        As for the comparison of F-MFPCA-AR1 (ranked 3\textsuperscript{rd}) and F-MFPCA-CategGPR (ranked 13\textsuperscript{th}), this shows the superiority of AR1 over CategGPR on this case study.

        \begin{figure}
            \begin{subfigure}{0.49\textwidth}
                \includegraphics[width=\textwidth]{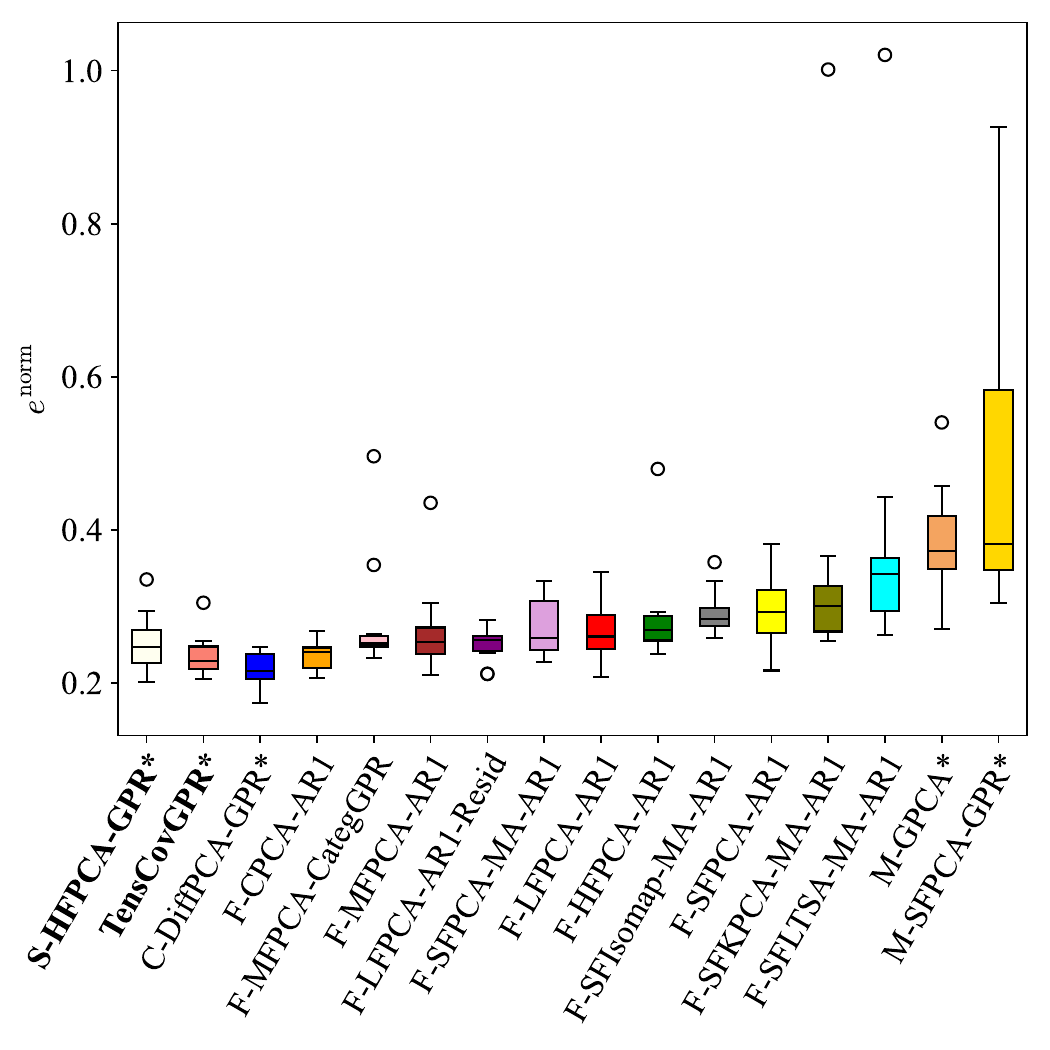}
            \end{subfigure}
            \hfill
            \begin{subfigure}{0.49\textwidth}
                \includegraphics[width=\textwidth]{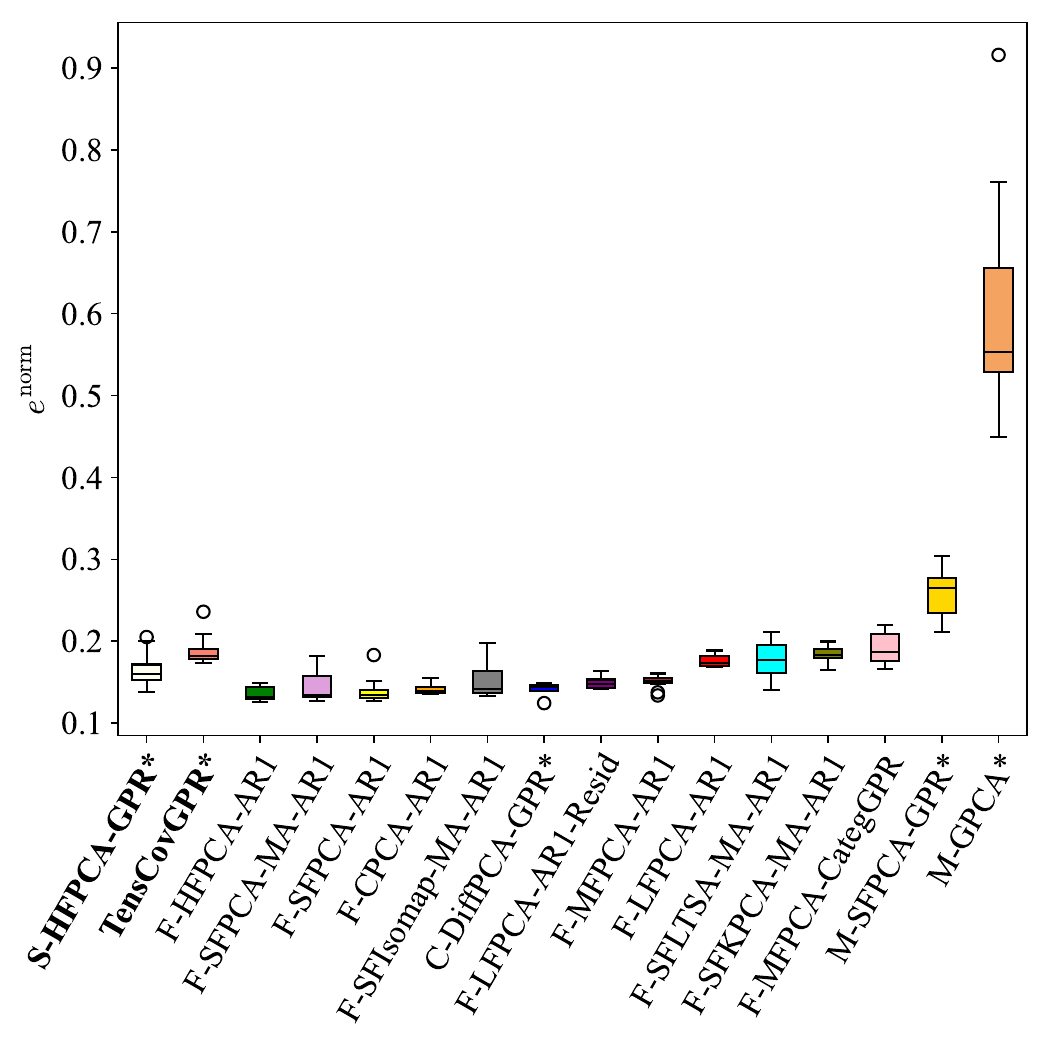}
            \end{subfigure}
            \caption{Boxplot of the normed RMSE for the NACA~0015 airfoil case with $n_1=10$ and $n_2=10$ (left) and $n_1=50$ and $n_2=500$ (right). Single-fidelity surrogates (bold label) are stacked to the left and multi-fidelity surrogates are sorted by median normed RMSE.}
            \label{fig:rmse-boxplot-ao}
        \end{figure}

        The normed RMSE are plotted in \Cref{fig:rmse-boxplot-ao}.
        When data is scarce, $n_1=10$ and $n_2=10$, the lowest median normed RMSE is 21.6\% for C-DiffPCA-GPR while the largest is reached by M-SFPCA-GPR with 38.2\%.
        Note that when both high- and low-fidelity snapshots are scarce, single fidelity-surrogates perform similarly to multi-fidelity surrogates.
        When there is much learning data, $n_1=50$ and $n_2=500$, the lowest median normed RMSE is reached by F-HFPCA-AR1 with 13.2\% while the largest is reached by M-GPCA with 55.3\%.
        With more high- and low-fidelity snapshots, multi-fidelity surrogates outperform their single-fidelity counterparts.

    \subsection{Pressure coefficient field around the RAE~2822 airfoil (RANS\slash RANS configuration)}

        \begin{figure}
            \centering
            \includegraphics[trim={0 0 0 1cm},clip,width=\textwidth]{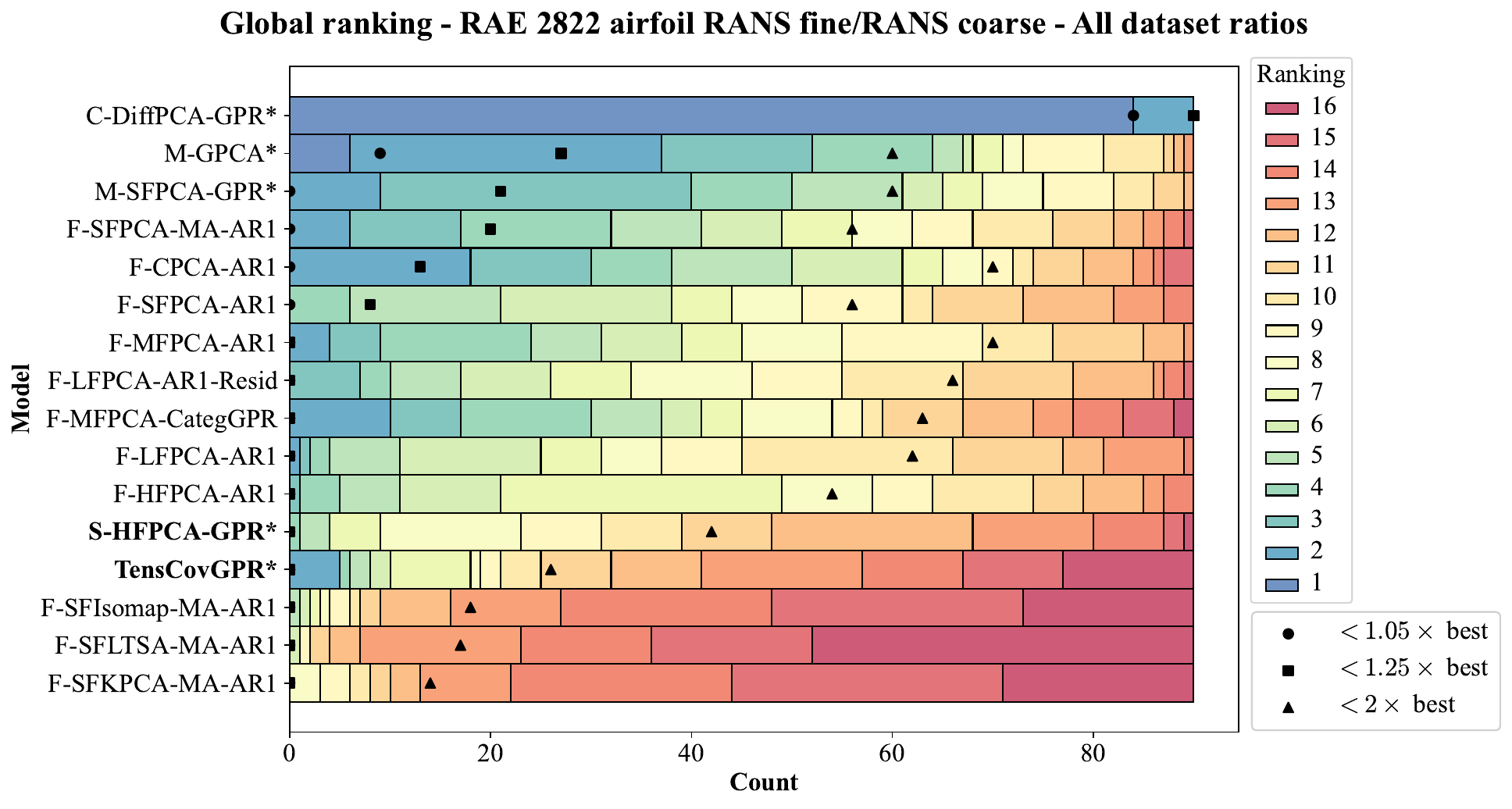}
            \caption{Global ranking of the various surrogates for the RAE~2822 airfoil, with the CFD RANS simulator with a fine mesh as the high-fidelity simulator, and the CFD RANS with a coarse mesh as the low-fidelity simulator. The colored bars show the ranking of the multi-fidelity surrogates. They are sorted vertically in decreasing number of times their normed RMSE is less than 5\%, 25\% and 100\% away from the best performing multi-fidelity surrogate.}
            \label{fig:ranking-ta2}
        \end{figure}

        \Cref{fig:ranking-ta2} shows the results for the RAE~2822 airfoil test case with the CFD RANS simulator with a fine mesh as the high-fidelity simulator and the CFD RANS with a coarse mesh as the low-fidelity simulator.
        Here, the corrective surrogate performs best.
        It is always in the top 2 and is ranking 1\textsuperscript{st} in 84 repetitions out of 90, showing a high robustness.
        The first mapping surrogate comes 2\textsuperscript{nd}, with 52 repetitions in the top 3.
        Apart from these two surrogates, none manages an RMSE within 5\% of the best RMSE.
        The histogram shows that only the corrective and M-GPCA surrogates manage to rank first for at least a combination of ($n_1$, $n_2/n_1$, index of repetition).
        The best performing fusion surrogate is F-SFPCA-MA-AR1 with 20 repetitions within 25\% of the best RMSE.
        Most fusion surrogates with linear DR exhibit relatively small performance disparities (ranked 4\textsuperscript{th} to 11\textsuperscript{th}).
        The worst surrogates for this test case are the single-fidelity surrogates and multi-fidelity surrogates using nonlinear DR.
        As for manifold alignment, it slightly improves the results (rank 4\textsuperscript{th} of F-SFPCA-MA-AR1 versus rank 6\textsuperscript{th} of F-SFPCA-AR1).
        Finally, the comparison of F-MFPCA-AR1 (ranked 7\textsuperscript{th}) and F-MFPCA-CategGPR (ranked 9\textsuperscript{th}) shows that AR1 is a marginally better intermediate surrogate when comparing the lower rankings.

        \begin{figure}
            \begin{subfigure}{0.49\textwidth}
                \includegraphics[width=\textwidth]{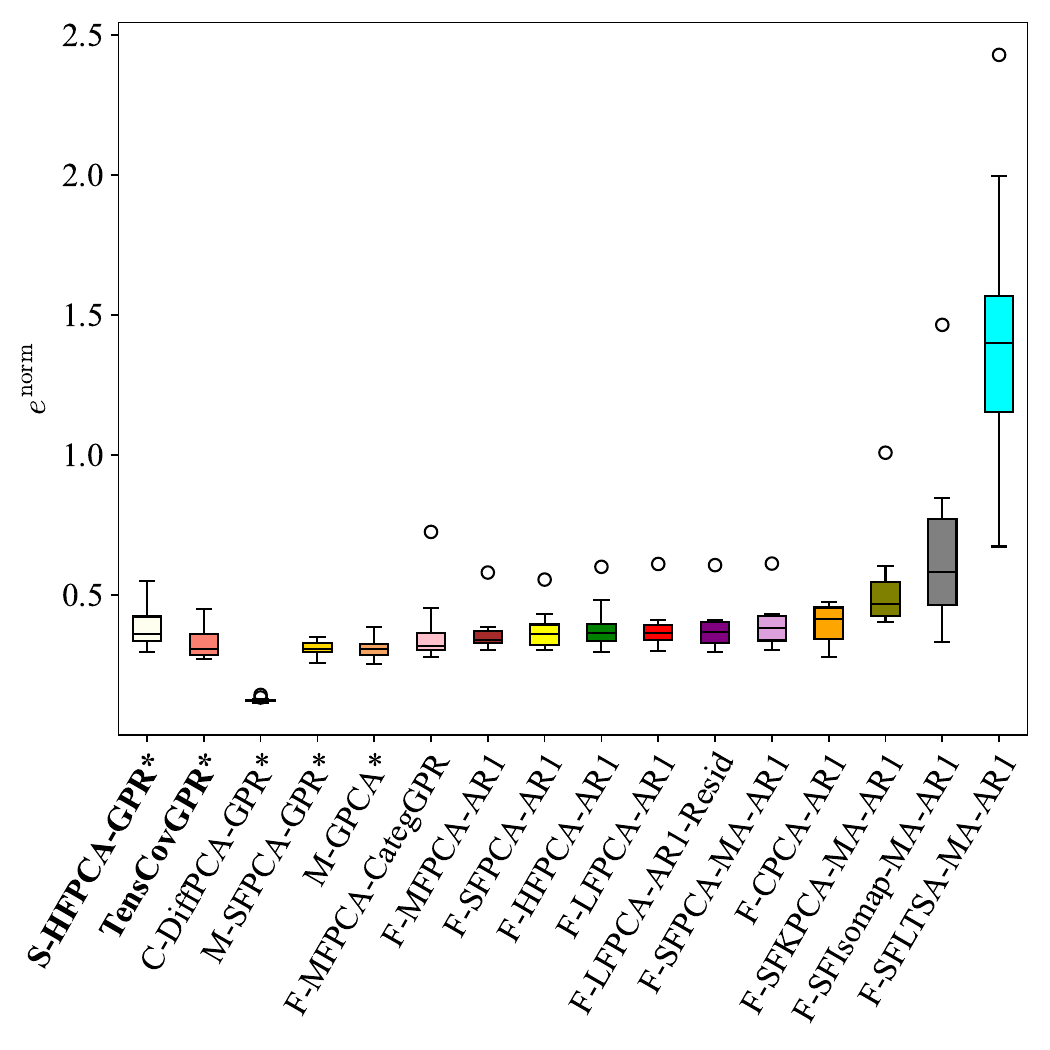}
            \end{subfigure}
            \hfill
            \begin{subfigure}{0.49\textwidth}
                \includegraphics[width=\textwidth]{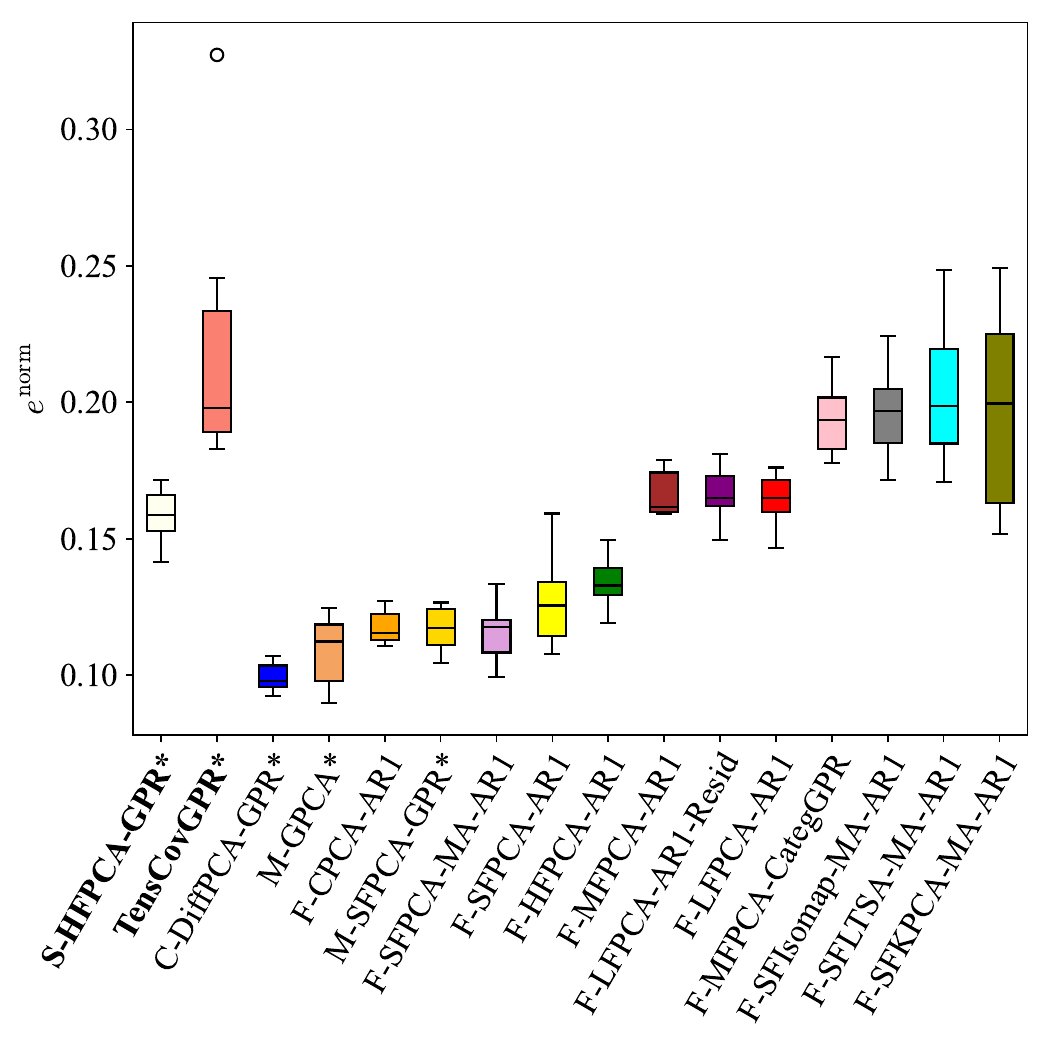}
            \end{subfigure}
            \caption{Boxplot of the normed RMSE for the RAE~2822 RANS/RANS case with $n_1=6$ and $n_2=6$ (left) and $n_1=30$ and $n_2=300$ (right). Single-fidelity surrogates (bold label) are stacked to the left and multi-fidelity surrogates are sorted by median normed RMSE.}
            \label{fig:rmse-boxplot-ta2}
        \end{figure}

        The boxplots of the normed RMSE for this test case are displayed in \Cref{fig:rmse-boxplot-ta2}.
        When $n_1=6$ and $n_2=6$, the lowest median normed RMSE is 12.6\%, achieved by C-DiffPCA-GPR, while the largest is reached by F-SFLTSA-MA-AR1 with 140.0\%.
        There are again two groups of surrogates: the multi-fidelity surrogates
        which perform better, to the exception of C-DiffPCA-GPR which outperforms the others, and surrogates using nonlinear DR which perform worse.
        Single-fidelity surrogates are performing similarly to multi-fidelity surrogates that belong to the plateau.
        With more data ($n_1=30$ and $n_2=300$), the lowest median normed RMSE is 9.8\% for M-GPCA while the largest is 20.0\% for F-SFKPCA-AR1.
        There is no longer a plateau and single-fidelity surrogates are now performing significantly worse than the best multi-fidelity surrogates.

    \subsection{Pressure coefficient field around the RAE~2822 airfoil (RANS\slash Euler configuration)}

        \begin{figure}
            \centering
            \includegraphics[trim={0 0 0 1cm},clip,width=\textwidth]{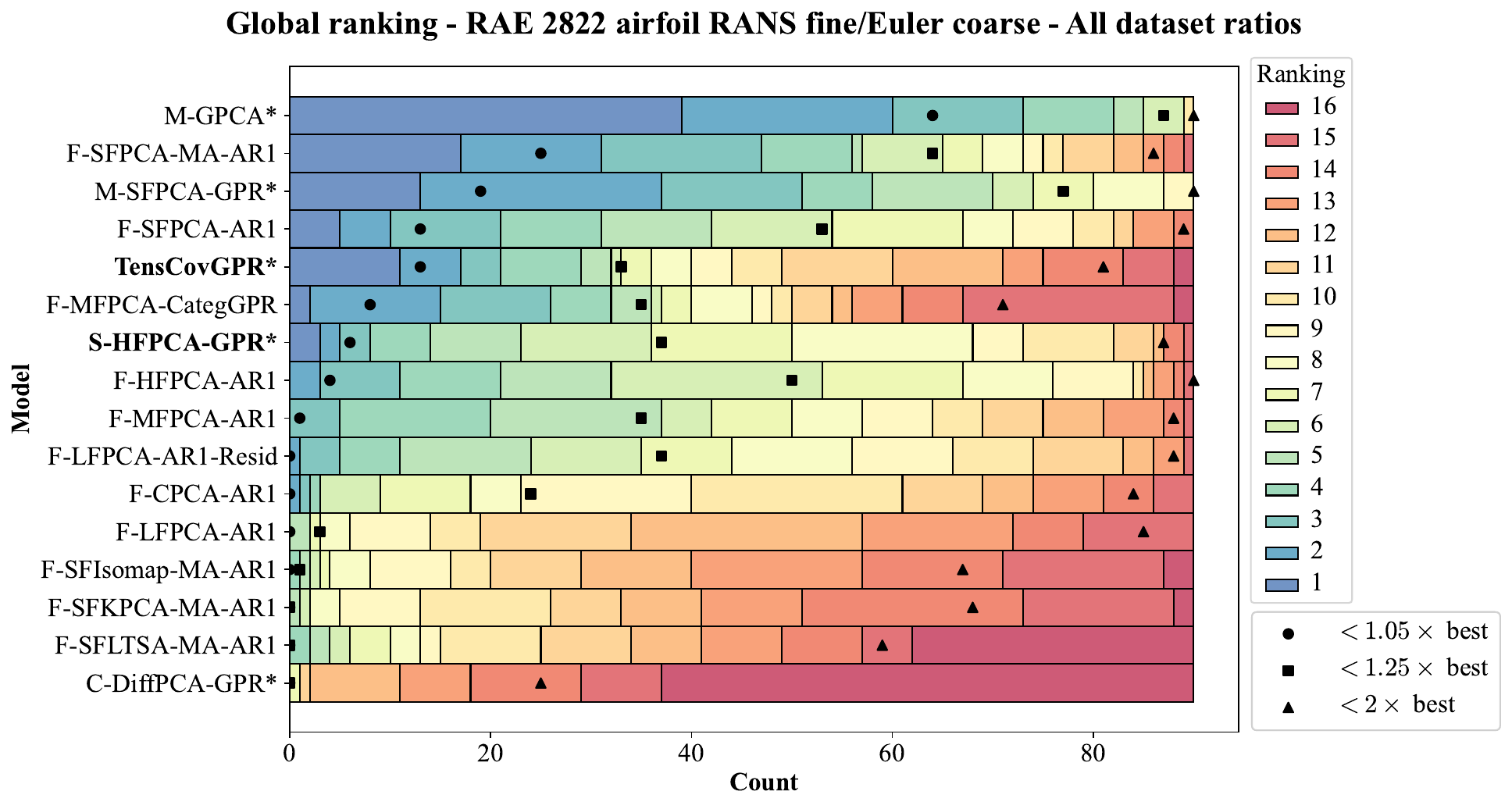}
            \caption{Global ranking of the various surrogates for the RAE~2822 airfoil, with the CFD RANS simulator with a fine mesh as the high-fidelity simulator and the CFD Euler with a coarse mesh as the low-fidelity simulator. The colored bars show the ranking of the multi-fidelity surrogates. They are sorted vertically in decreasing number of times their normed RMSE is less than 5\%, 25\% and 100\% away from the best performing multi-fidelity surrogate.}
            \label{fig:ranking-ta}
        \end{figure}

        The results for the RAE~2822 airfoil are depicted in \Cref{fig:ranking-ta} where the high-fidelity simulator is the CFD RANS simulator with a fine mesh and the CFD Euler with a coarse mesh is the low-fidelity simulator.
        The two mapping methods are both ranked in the top 3, with 73 and 51 repetitions or more among the top 3 surrogates.
        In between them, the best fusion surrogate ranks 2\textsuperscript{nd} with 47 repetition among the top 3.
        Not many fusion surrogates perform better than the reference single-fidelity surrogates.
        F-MFPCA-CategGPR is ranking among the average fusion surrogates but has a high proportion of low rankings (29 repetitions out of 90 among the 3 worst surrogates).
        Finally, the corrective approach is performing worse than any other surrogate with 72 repetitions out of 90 among the 3 worst surrogates.
        Comparing F-SFPCA-MA-AR1 (ranked 2\textsuperscript{nd}) and F-SFPCA-AR1 (ranked 4\textsuperscript{th}) shows a noticeable improvement due to manifold alignment.
        Unlike in any of the previous test cases, F-MFPCA-CategGPR (ranked 6\textsuperscript{th}) is ranked higher than F-MFPCA-AR1 (ranked 9\textsuperscript{th}).
        But as said previously, F-MFPCA-CategGPR has a high proportion of low rankings.

        \begin{figure}
            \begin{subfigure}{0.49\textwidth}
                \includegraphics[width=\textwidth]{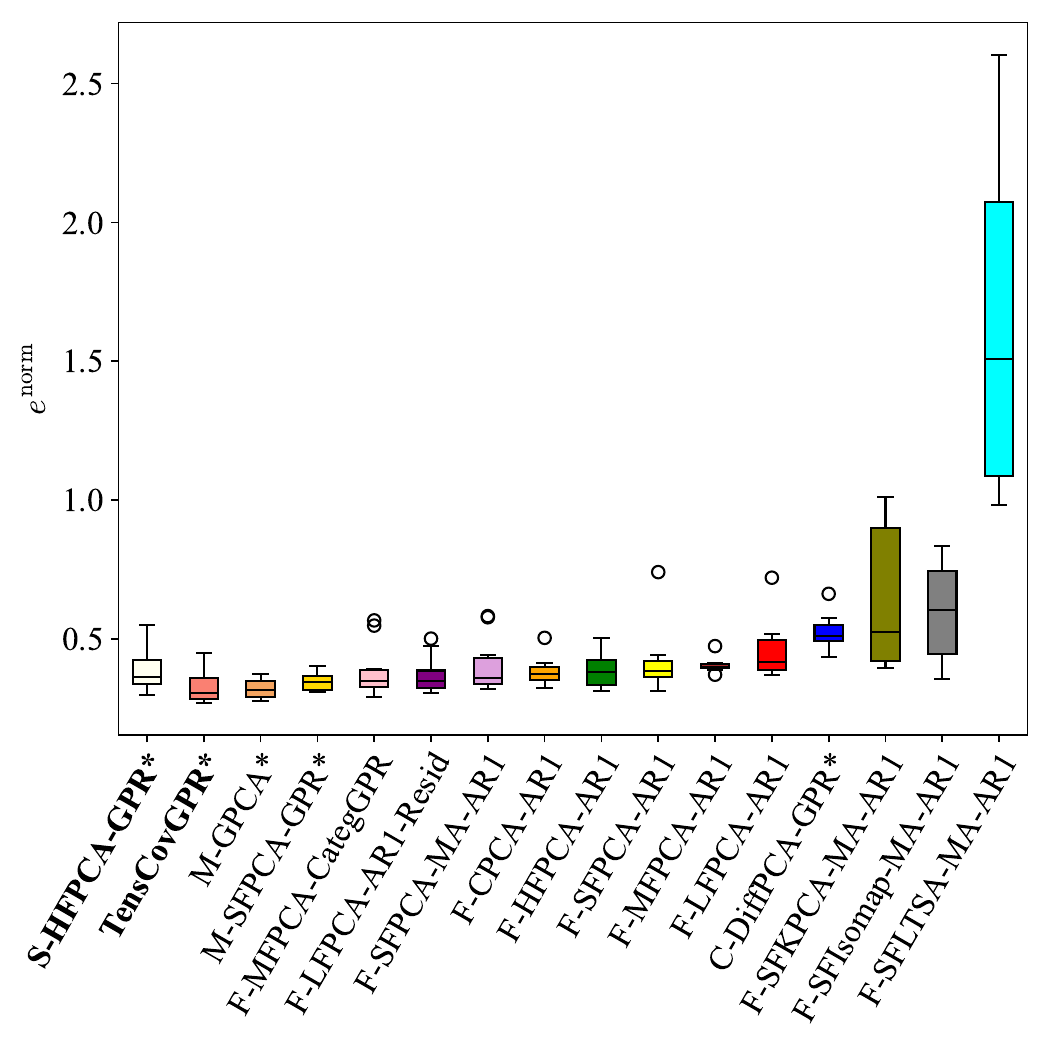}
            \end{subfigure}
            \hfill
            \begin{subfigure}{0.49\textwidth}
                \includegraphics[width=\textwidth]{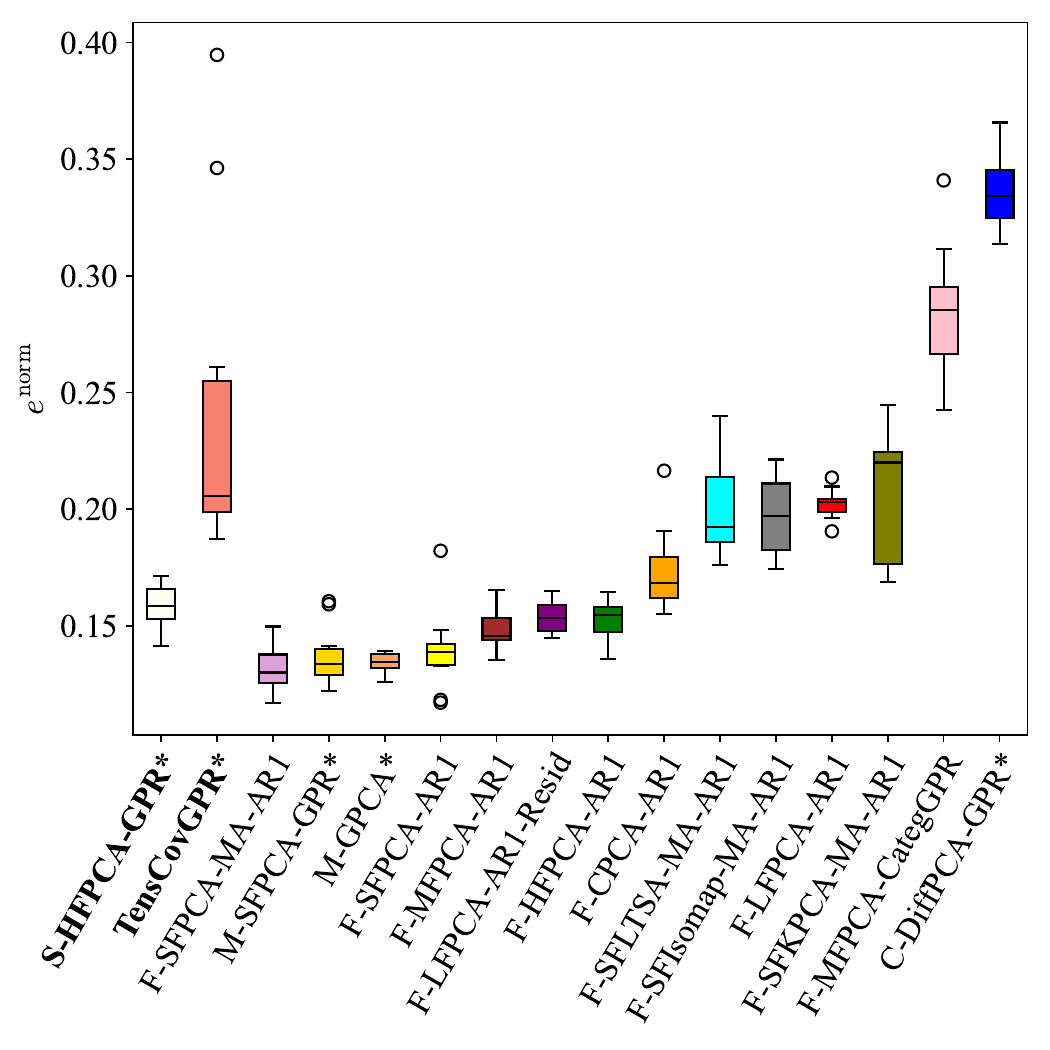}
            \end{subfigure}
            \caption{Boxplot of the normed RMSE for the RAE~2822 RANS/Euler case with $n_1=6$ and $n_2=6$ (left) and $n_1=30$ and $n_2=300$ (right). Single-fidelity surrogates (bold label) are stacked to the left and multi-fidelity surrogates are sorted by median normed RMSE.}
            \label{fig:rmse-boxplot-ta}
        \end{figure}

        To complement the ranking information, the boxplots of the normed RMSE are displayed in \Cref{fig:rmse-boxplot-ta}.
        With little data to learn, $n_1=6$ and $n_2=6$, the lowest median normed RMSE is 30.6\% for TensCovGPR while the largest is reached by F-SFLTSA-MA-AR1 with 150.7\%.
        Most multi-fidelity surrogates have a similar median normed RMSE except for surrogates using nonlinear DR and C-DiffPCA-GPR.
        Single-fidelity surrogates have a performance similar to multi-fidelity surrogates.
        When $n_1=30$ and $n_2=300$, the lowest median normed RMSE is 13.0\% for F-SFPCA-MA-AR1 while the largest is 33.4\% for C-DiffPCA-GPR.
        Single-fidelity surrogates tend to lag behind multi-fidelity surrogates in terms of median normed RMSE.
        Additionally, there is no noticeable plateau anymore.

        In this test case, the additional low-fidelity data leads to little or no improvement in the prediction capabilities of multi-fidelity surrogates, as mapping surrogates are among the best performing models.
        This is likely due to the poor correlation between the high- and low-fidelity fields.

    \subsection{Discussions}\label{sec:discussions}

        \subsubsection{Single- versus multi-fidelity surrogates}\label{sec:singlefi-vs-multifi}

            Before delving into the specificities of the different multi-fidelity surrogates, we assess first if adopting a multi-fidelity scheme is beneficial.
            In the \Cref{fig:ranking-vffng,fig:ranking-vffg,fig:ranking-ao,fig:ranking-ta2,fig:ranking-ta}, single-fidelity surrogates never rank first.
            S-HFPCA-GPR ranks 15\textsuperscript{th}, 15\textsuperscript{th}, 10\textsuperscript{th}, 13\textsuperscript{th} and 7\textsuperscript{th}, while TensCovGPR ranks 5\textsuperscript{th}, 3\textsuperscript{rd}, 9\textsuperscript{th}, 12\textsuperscript{th} and 5\textsuperscript{th}.
            Note that TensCovGPR outperform S-HFPCA-GPR on every test case.
            In \Cref{fig:rmse-boxplot-vffng,fig:rmse-boxplot-vffg,fig:rmse-boxplot-ao,fig:rmse-boxplot-ta2,fig:rmse-boxplot-ta}, when data is scarce (in the sense that both high- and low-fidelity data is limited or the high- and low-fidelity fields are poorly correlated),  single-fidelity surrogates perform similarly to multi-fidelity surrogates.
            When the low-fidelity data is scarce ($n_1=n_2$) and the correlation between the high- and low-fidelity fields is high, we observe an improved prediction capability with some multi-fidelity surrogates compared to single-fidelity surrogates.

\begin{figure}
                \begin{subfigure}{0.49\textwidth}
                    \centering
                    \includegraphics[width=0.7\textwidth]{figures/airfoil_oneraboxplot_Nhf_10_Nlf_10.pdf}
                    \subcaption{NACA~0015, $n_1=10$ and $n_2=10$}
                \end{subfigure}
                \begin{subfigure}{0.49\textwidth}
                    \centering
                    \includegraphics[width=0.7\textwidth]{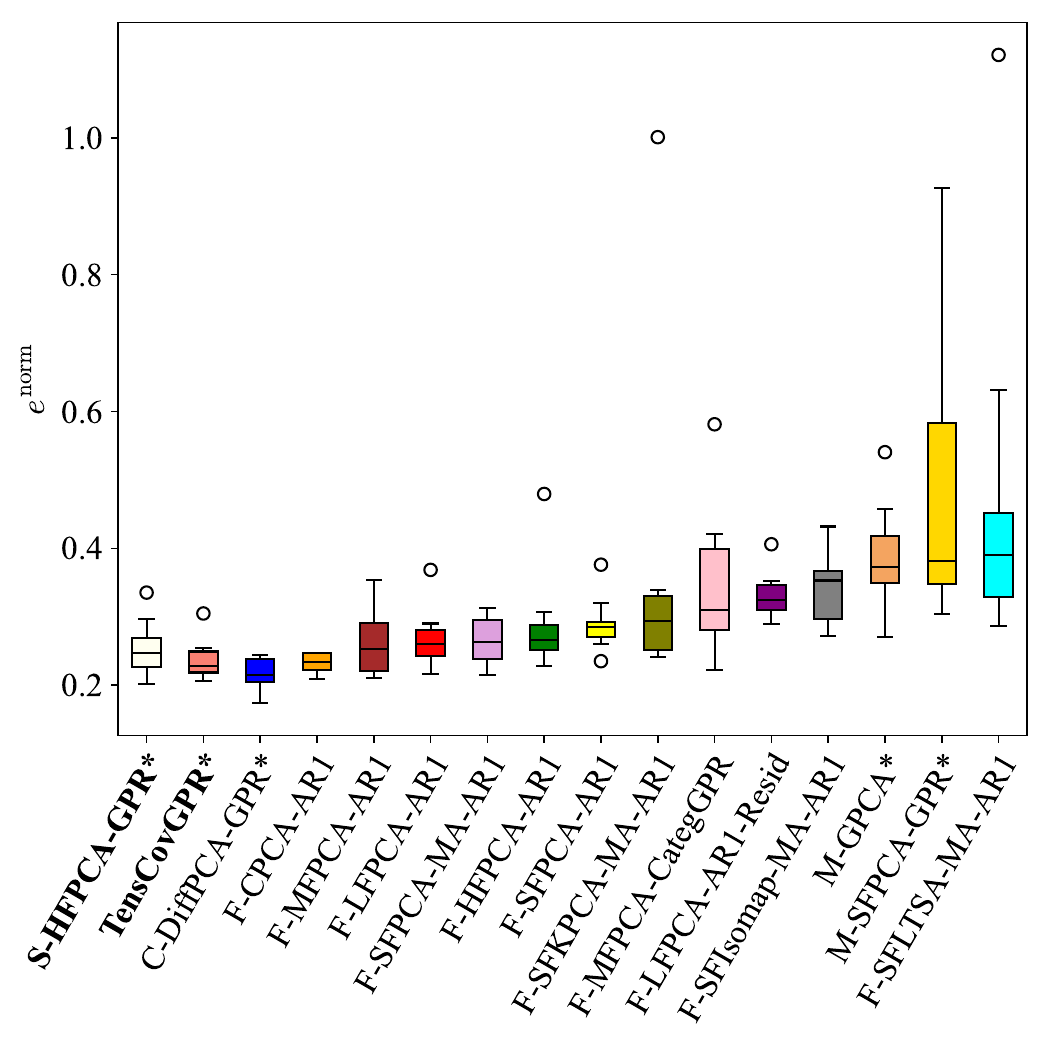}
                    \subcaption{NACA~0015, $n_1=10$ and $n_2=100$}
                \end{subfigure}
                \\
                \\
                \begin{subfigure}{0.49\textwidth}
                    \centering
                    \includegraphics[width=0.7\textwidth]{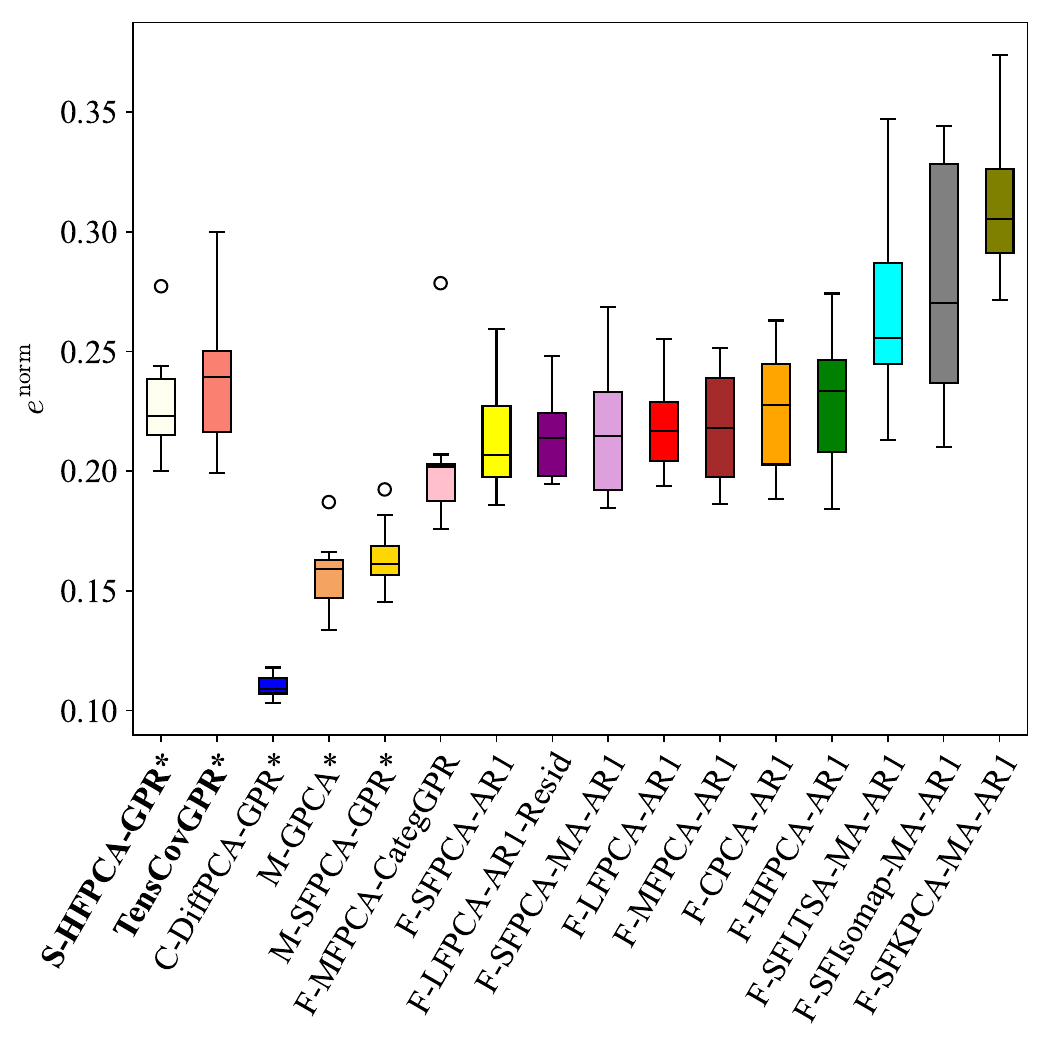}
                    \subcaption{RAE~2822 RANS/RANS, $n_1=15$ and $n_2=15$}
                \end{subfigure}
                \begin{subfigure}{0.49\textwidth}
                    \centering
                    \includegraphics[width=0.7\textwidth]{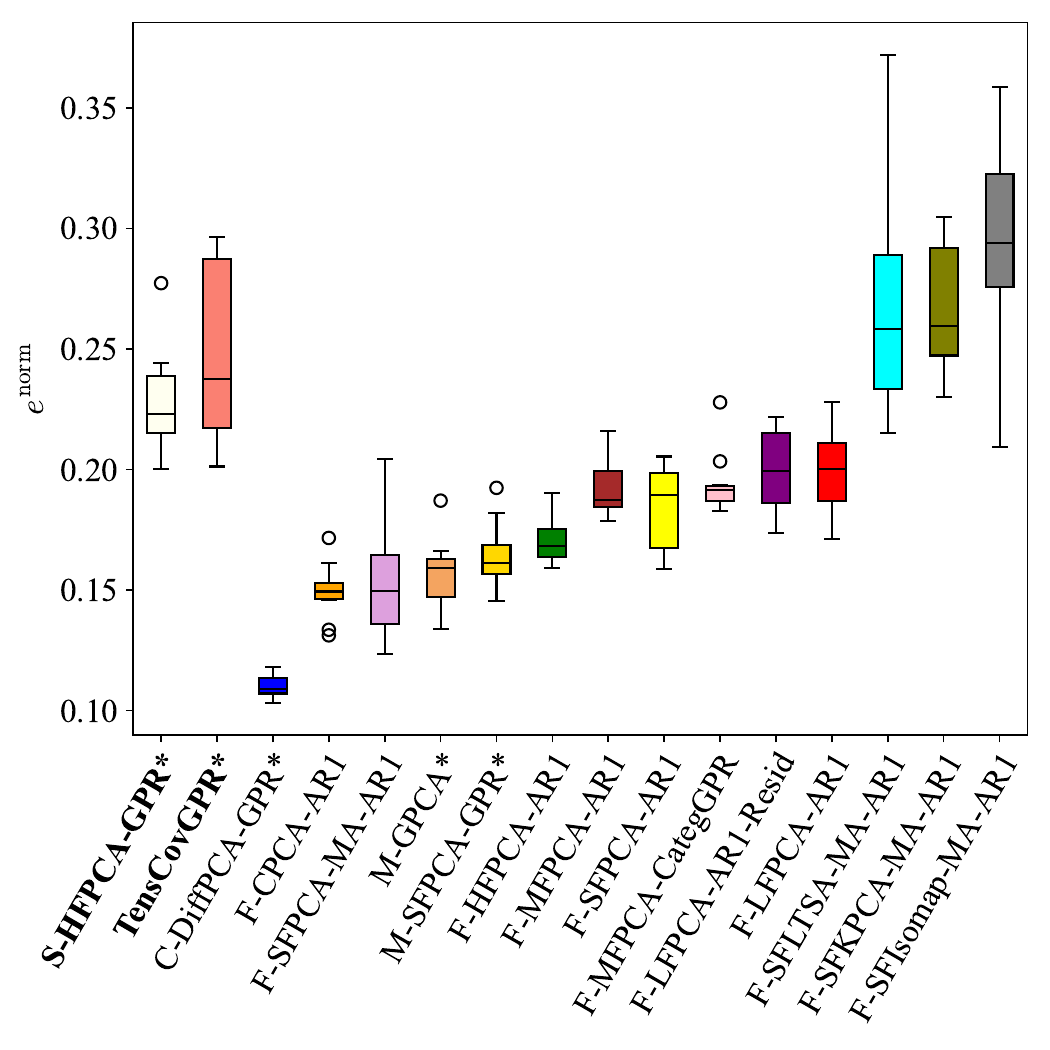}
                    \subcaption{RAE~2822 RANS/RANS, $n_1=15$ and $n_2=150$}
                \end{subfigure}
                \\
                \\
                \begin{subfigure}{0.49\textwidth}
                    \centering
                    \includegraphics[width=0.7\textwidth]{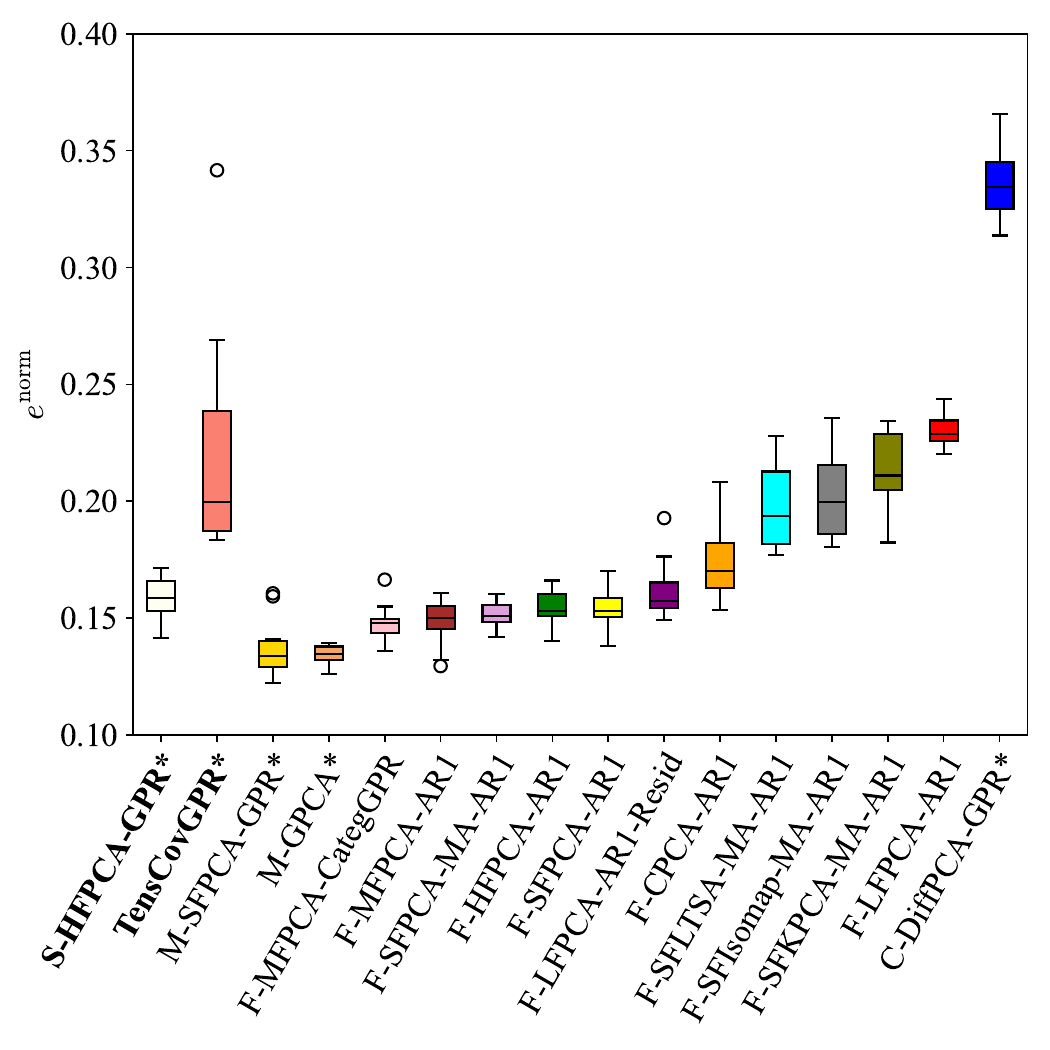}
                    \subcaption{RAE~2822 RANS/Euler, $n_1=30$ and $n_2=30$}
                \end{subfigure}
                \begin{subfigure}{0.49\textwidth}
                    \centering
                    \includegraphics[width=0.7\textwidth]{figures/transonic_airfoilboxplot_Nhf_30_Nlf_300.pdf}
                    \subcaption{RAE~2822 RANS/Euler, $n_1=30$ and $n_2=300$}
                \end{subfigure}
                \caption{Boxplots of the normed RMSE for the test cases with $n_2=n_1$ (left column) and $n_2=10\times n_1$ (right column). Single-fidelity surrogates (bold label) are stacked to the left and multi-fidelity surrogates are sorted by median normed RMSE.}
                \label{fig:singlefi-vs-multifi}
            \end{figure}

            Boxplots of the normed RMSE for different test cases are displayed in \Cref{fig:singlefi-vs-multifi} with a constant number of high-fidelity snapshots $n_1$ and a varying number of low-fidelity snapshots $n_2$.
            The first row is the NACA~0015 airfoil case with $n_1=2\times\dim(U)$, the second row is the RAE~2822 airfoil RANS/RANS case with $n_1=5\times\dim(U)$ and the third row is the RAE~2822 airfoil RANS/Euler case with $n_1=10\times\dim(U)$.
            The left column is for $n_2=n_1$ while the second column is for $n_2=10\times n_1$.

            Single-fidelity, mapping and corrective surrogates do not benefit from the addition of low-fidelity snapshots for a fixed number of high-fidelity snapshots (they can only accept $n_1=n_2$ for their training sets).
            In the NACA~0015 airfoil case with a small number of high-fidelity snapshots ($n_1=2\times\dim(U)$), the addition of low-fidelity snapshots only marginally improves the performance of multi-fidelity surrogates.
            For the RAE~2822 RANS/RANS airfoil case for the intermediate number of high-fidelity snapshots ($n_1=5\times\dim(U)$), adding low-fidelity snapshots significantly improves the performance of multi-fidelity surrogates combining linear DR with AR1 co-Kriging.
            It appears that surrogates using nonlinear DR or CategGPR are hardly affected by the data augmentation.
            For the RAE~2822 RANS/Euler airfoil case, the improvement on the performance is only marginal.
            It even worsen the results for the multi-fidelity surrogate using CategGPR as the intermediate surrogate. 
            Nevertheless, multi-fidelity surrogates perform better than single-fidelity surrogates, even though the addition of low-fidelity snapshots for a fixed number of high-fidelity snapshot only marginally improves the results.
            Note that the benefit of the addition of low-fidelity data highly depends on the correlation between the high- and low-fidelity fields.
            In particular, the RAE~2822 Euler data is, by construction, much less informative of the high-fidelity than the RANS with coarse mesh data.

        \subsubsection{Linear versus nonlinear dimension reduction}\label{sec:discussion-linear-nonlinear-dr}

\begin{figure}
                \centering
                \includegraphics[trim={0 0 0 1cm},clip,width=0.7\textwidth]{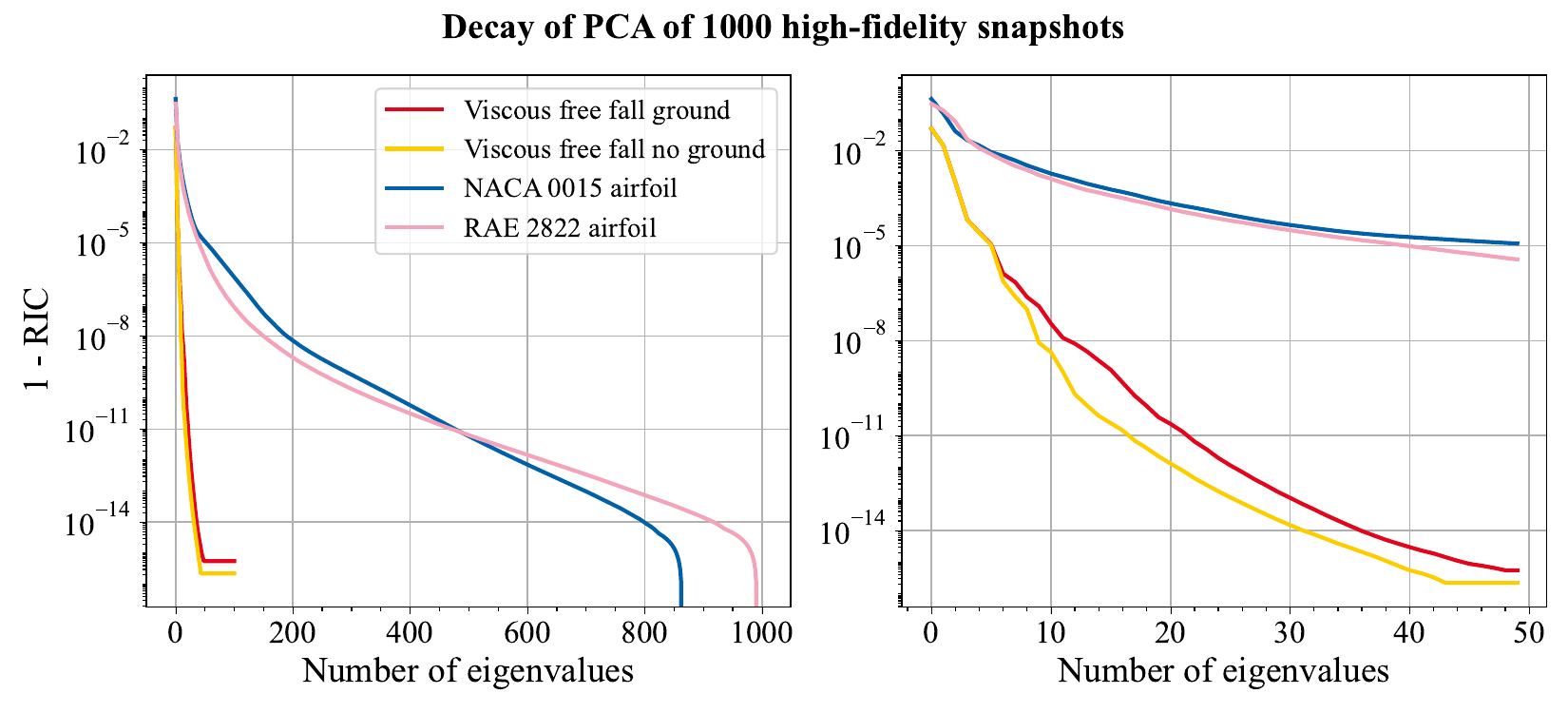}
                \caption{Rate of decay of 1~-~RIC of a PCA on the 1{,}000 high-fidelity validation snapshots for each test case. Zoom on the 50 first eigenvalues on the right. Since the high-fidelity simulator is the same for the two RAE~2822 configurations, only one rate of decay is shown for this case (pink solid line).}
                \label{fig:decay-case}
            \end{figure}

            A method to assess the effectiveness of linear DR is to measure the relative information content (RIC) of the PCA.
            \Cref{fig:decay-case} illustrates the rate of decay of 1~-~RIC of a PCA on 1{,}000 high-fidelity snapshots for each test case.
            Note that the PCA for the viscous free fall cases can have a maximum of 101 non-zero eigenvalues which is the number of nodes in the mesh, even though the number of samples exceeds the number of nodes.
            Consequently, their PCAs exhibit a faster decay in 1~-~RIC compared to the other test cases.
            In contrast, the decay of the NACA~0015 and the RAE~2822 test cases is slower, indicating that a significant number of modes is required to perform a linear DR without a substantial information loss.
            For instance, for a relative information content of 99.9\%, 3 modes are required for the viscous free fall without and with ground, 13 modes for the NACA~0015 airfoil, and 11 modes for the RAE~2822 airfoil.
            These numbers highlight that the seemingly less complex NACA~0015 airfoil poses a greater challenge for linear DR than the RAE~2822 airfoil does.

            In our experiments, the NACA~0015 and RAE~2822 airfoil cases did not benefit from nonlinear DR.
            The few multi-fidelity surrogates utilizing nonlinear DR, which are just variants of the surrogate of Perron \textit{et al}.\ \cite{perronMultifidelity2021}, do not outperform surrogates employing linear DR.
            Given the good performance of F-SFPCA-MA-AR1, this cannot be attributed to the assembly method of the surrogate F-SF\{PCA, KPCA, isomap, LTSA\}-MA-AR1.
            Hence, there are two possible explanations for the inferior performance of nonlinear DR.
            First, the chosen techniques for nonlinear DR may not be suitable for the tested use cases.
            Second, KPCA, isomap and LTSA require the selection of multiple hyperparameters, significantly complexifying the surrogate training process compared to PCA.
            Despite intensive tuning of the hyperparameters by the authors of this study, the performance of multi-fidelity surrogates using nonlinear DR did not surpass that of surrogates using linear DR.
            Details about this tuning can be found in \ref{app:numerical-settings}.
            Exploring alternative nonlinear DR techniques could be a focus for future research.

        \subsubsection{Influence of the PCA type for PCA-based fusion surrogates }\label{sec:discussion-which-pca}

            \begin{table}
                \centering
                \caption{Decomposition of the normed RMSE $e^{\text{norm}}$ into normed DR RMSE $e_{\text{dr}}^{\text{norm}}$ and normed intermediate surrogate modeling RMSE $e_{\text{ism}}^{\text{norm}}$ for F-HFPCA-AR1, F-LFPCA-AR1 and F-MFPCA-AR1.}
                \begin{tabular}{l c c c c}
                    \toprule
                    Test case & surrogate & $e^{\text{norm}}$ (\%) & $e_{\text{dr}}^{\text{norm}}$ (\%) & $e_{\text{ism}}^{\text{norm}}$ (\%) \\
                    \toprule
                    Viscous free fall & F-HFPCA-AR1 & 13.4 & 8.7 & 8.0 \\
                     without ground & F-LFPCA-AR1 & 15.8 & 11.4 & 8.2 \\
                     & F-MFPCA-AR1 & 12.3 & 7.2 & 8.1 \\
                    \midrule
                    Viscous free fall & F-HFPCA-AR1 & 13.7 & 8.7 & 8.7 \\
                     with ground & F-LFPCA-AR1 & 11.6 & 5.3 & 9.4 \\
                     & F-MFPCA-AR1 & 10.3 & 3.3 & 9.1 \\
                    \midrule
                    NACA~0015 & F-HFPCA-AR1 & 19.7 & 7.7 & 18.5 \\
                     & F-LFPCA-AR1 & 21.1 & 15.4 & 14.1 \\
                     & F-MFPCA-AR1 & 19.2 & 10.0 & 16.4 \\
                    \midrule
                    RAE~2822 & F-HFPCA-AR1 & 22.8 & 19.3 & 13.7 \\
                     RANS/RANS & F-LFPCA-AR1 & 22.3 & 13.3 & 18.5 \\
                     & F-MFPCA-AR1 & 21.4 & 10.1 & 18.7 \\
                    \midrule
                    RAE~2822 & F-HFPCA-AR1 & 24.6 & 19.3 & 17.6 \\
                     RANS/Euler & F-LFPCA-AR1 & 30.9 & 21.7 & 22.2 \\
                     & F-MFPCA-AR1 & 27.4 & 11.1 & 24.7 \\
                    \bottomrule
                \end{tabular}
                \label{tab:comparaison-erreurs-pcas}
            \end{table}

\begin{figure}
                \centering
                \begin{subfigure}{0.49\textwidth}
                    \includegraphics[trim={0 0 0 1cm},clip,width=\textwidth]{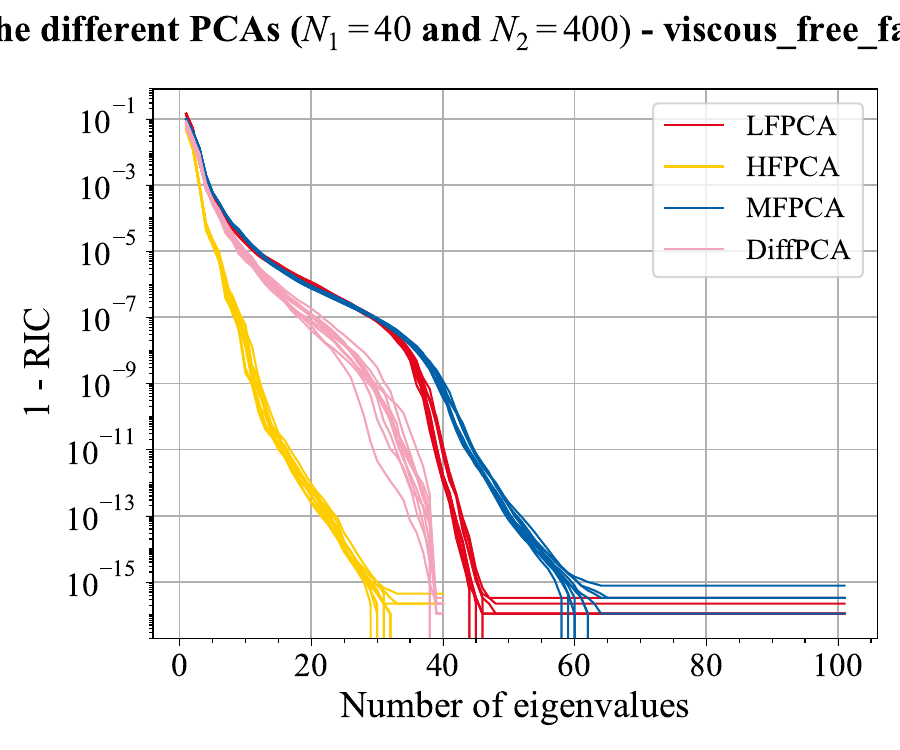}
                \end{subfigure}
                \begin{subfigure}{0.49\textwidth}
                    \includegraphics[trim={0 0 0 1cm},clip,width=\textwidth]{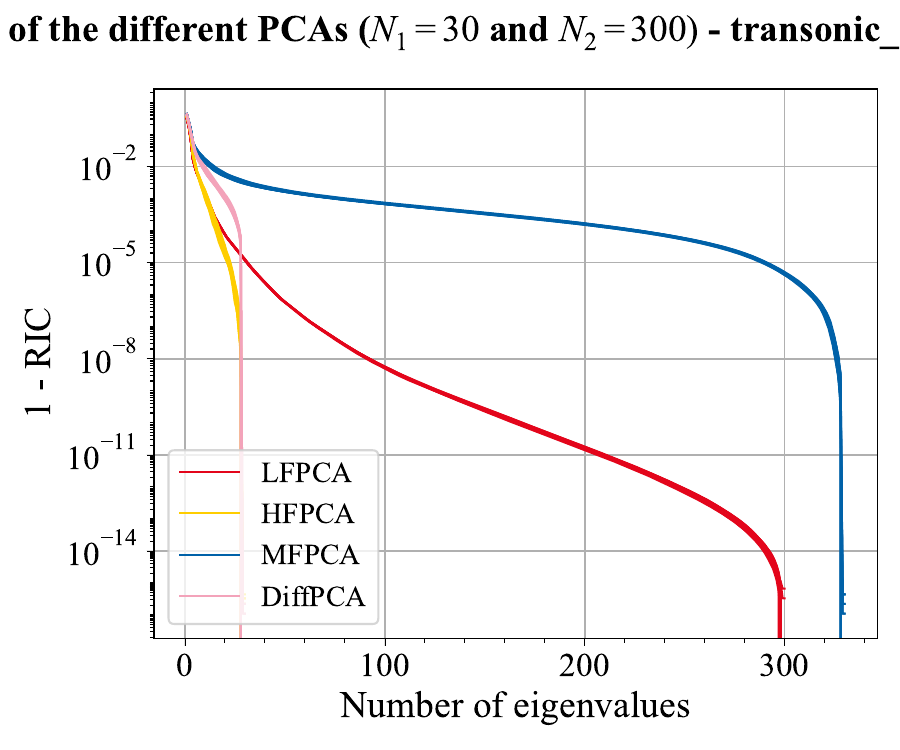}
                \end{subfigure}
                \caption{Decay of 1~-~RIC for different PCAs for the viscous free fall with ground test case with $n_1=40$ and $n_2=400$ (left), and RAE~2822 airfoil RANS/Euler with $n_1=30$ and $n_2=300$ (right), 10 repetitions each}
                \label{fig:pca-type}
            \end{figure}

            \Cref{tab:comparaison-erreurs-pcas} presents the normed RMSE, DR RMSE and intermediate surrogate modeling RMSE for the different test cases and for three different types of PCA: LFPCA, HFPCA and MFPCA.
            Recall that in LFPCA and HFPCA, the PCA is carried out on the low- or high-fidelity data, respectively, while MFPCA, the mixed-fidelity PCA, processes both data together.
            It appears that the DR error is lower with MFPCA (except for the NACA 0015 case) than with LFPCA and HFPCA.
            While this does not guarantee superior prediction capabilities for the surrogate, it is a meaningful information.
            The rankings seen in \Cref{fig:ranking-vffng,fig:rmse-boxplot-vffng,fig:ranking-vffg,fig:rmse-boxplot-vffg,fig:ranking-ao,fig:rmse-boxplot-ao,fig:ranking-ta2,fig:rmse-boxplot-ta2,fig:ranking-ta,fig:rmse-boxplot-ta} show that F-MFPCA-AR1 generally yields better results than F-HFPCA-AR1 and F-LFPCA-AR1, or is close to F-HFPCA-AR1 when the latter is better.
            \Cref{fig:pca-type} gives the decay of 1 - RIC for MFPCA, HFPCA and LFPCA in the viscous free fall with a ground and the RAE~2822 airfoil RANS/Euler test cases (see \ref{app:latent-dimension} for all test cases). Ten repetitions with different training snapshots are plotted, with $n_1=40$ and $n_2=400$ in the viscous free fall with ground case, and $n_1=30$ and $n_2=300$ in the RAE~2822 airfoil RANS/Euler case.

\begin{figure}
                \centering
                \begin{subfigure}{0.49\textwidth}
                    \includegraphics[trim={0 0 0 1cm},clip,width=\textwidth]{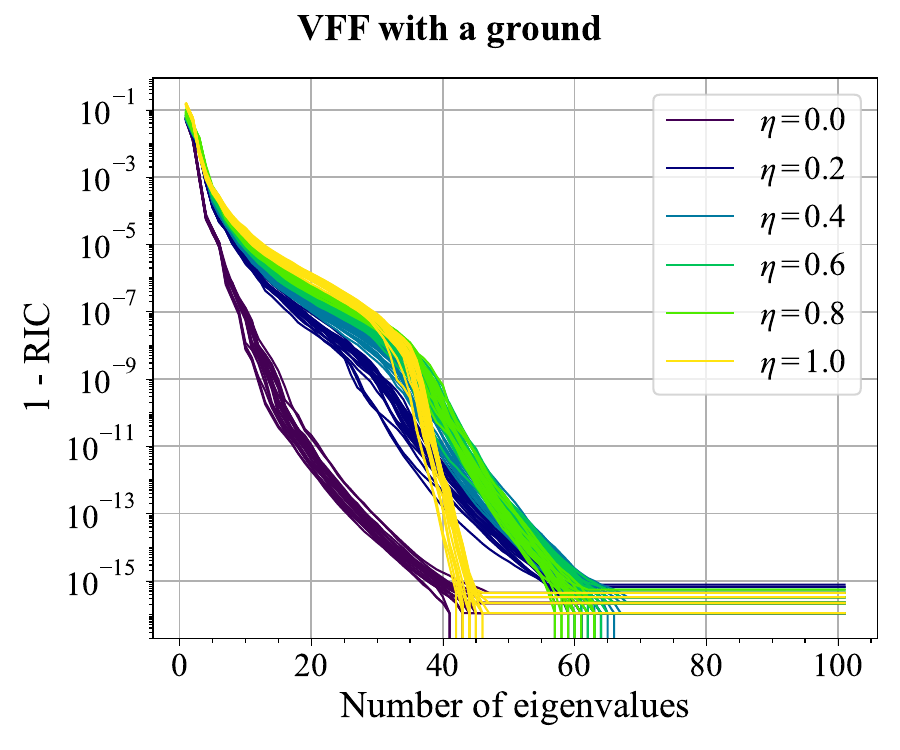}
                \end{subfigure}
                \begin{subfigure}{0.49\textwidth}
                    \includegraphics[trim={0 0 0 1cm},clip,width=\textwidth]{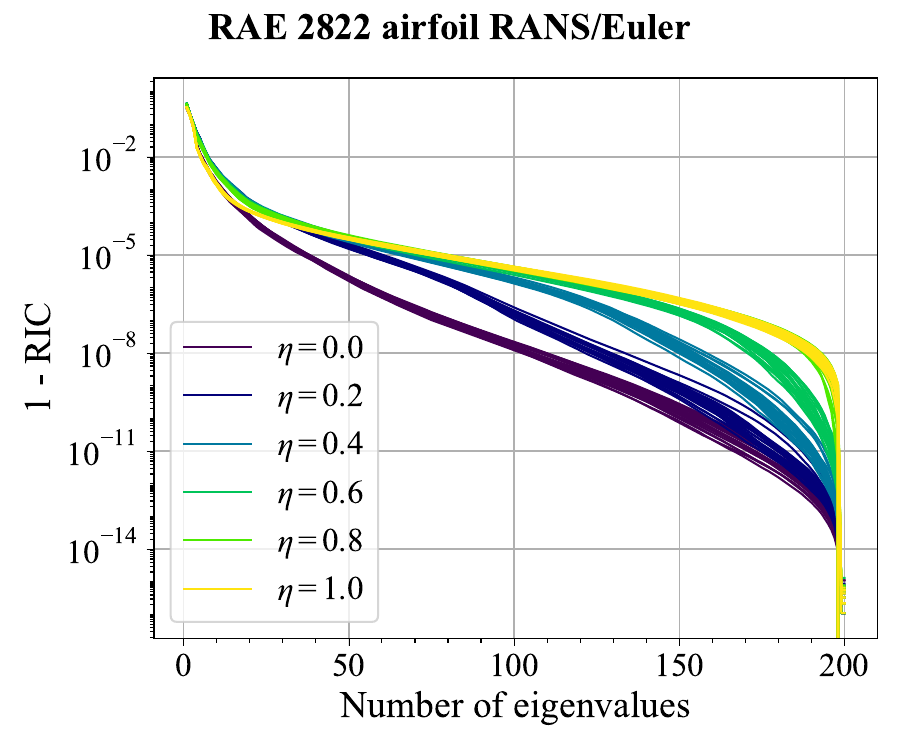}
                \end{subfigure}
                \caption{Decay of 1~-~RIC for MFPCA varying $\eta$ for the viscous free fall with ground test case (left), and RAE~2822 airfoil RANS/Euler with $n_1=30$ and $n_2=300$ (right), 10 repetitions each}
                \label{fig:mfpca-ratio}
            \end{figure}

            Let $\eta = n_2/(n_1+n_2)$ be the proportion of low-fidelity snapshots used in the PCA: $\eta=0$ in HFPCA, $\eta=1$ in LFPCA, and it varies between 0 and 1 in MFPCA.
            In \Cref{fig:mfpca-ratio} is depicted the decay of 1 - RIC with $\eta=\{0, 0.2, 0.4, 0.6, 0.8, 1\}$ for $n_1+n_2=200$.
            On the left (viscous free fall with ground), 1~-~RIC for HFPCA ($\eta=0$) and LFPCA ($\eta=1$) decays faster than for intermediate value of $\eta$.
            On the right (RAE~2822 RANS/Euler), the decay of 1~-~RIC slows as $\eta$ increases, meaning that there is more information in the first modes of HFPCA than in the first modes of LFPCA. Mixing the fidelities makes even more modes necessary in the PCA.

            A risk with MFPCA that is inherent to mixing data from multiple fidelities, is that the distinctive features of the high-fidelity field may be blended into those of the low-fidelity field.
            Note that in general, in the context of multi-fidelity, it holds that $n_1 < n_2$.
            Depending on the respective computational cost of the high- and low-fidelity simulators, it can even be $n_1\ll n_2$.
            When $n_2/n_1\rightarrow\infty$, MFPCA converges to LFPCA.

        \subsubsection{Manifold alignment}\label{sec:discussion-manifold-alignment}

            In the rankings presented above (\Cref{fig:ranking-vffng,fig:ranking-vffg,fig:ranking-ao,fig:ranking-ta2,fig:ranking-ta}), the inclusion of manifold alignment consistently proves beneficial or at worst slightly detrimental in terms of prediction accuracy.
            The computation of manifold alignment operators entails only linear algebra operations, including a singular value decomposition (SVD) in the low-dimensional latent space, which have a negligible computational overhead.
            Additionally, manifold alignment plays no direct role in the prediction process of high-fidelity output fields.
            
            Hence, based on our experiments, it is advised to perform manifold alignment whenever the high- and low-fidelity manifolds are different.
            This is of interest when the high- and low-fidelity fields are discretized on different meshes so as to avoid a pre-processing step that projects every fidelity field onto a common mesh.


\section{Concluding recommendations}\label{sec:conlusion}

    This paper provides a comprehensive review of various techniques for constructing surrogates of a set of simulators with functional outputs. The simulators have variable fidelities and they can have different meshes for distinct fidelities.
    The unified framework introduced in this paper facilitates the theoretical comparison of the various existing surrogates of simulators with functional outputs, combining dimensionality reduction and latent surrogate modeling.
    The subsequent benchmark of a dozen of these surrogates on case studies of increasing complexity allows us to provide practical recommendations. Note that they are specific to this study and depend on the numerical settings (see \ref{app:numerical-settings}) and cases considered.

    \begin{itemize}
        \item Overall, multi-fidelity surrogates tend to outperform their single-fidelity counterparts when low-fidelity data is available. The lower complexity of setup of single-fidelity surrogates makes them a good starting point but multi-fidelity surrogates usually offer better predictions (see \Cref{sec:singlefi-vs-multifi}).
        \item Starting with linear DR is advisable. It is much simpler to train and use than nonlinear techniques, and it provides satisfactory performance (see \Cref{sec:discussion-linear-nonlinear-dr}).
        Depending on the decay of 1~-~RIC, if too many modes must be retained to provide sufficiently accurate DR, nonlinear approaches could be tested.
        Special care should be taken to tuning the hyperparameters of nonlinear DR (in the reviewed methods, \textit{e.g.}, the correlation length of the kernel of KPCA, the number of nearest neighbors in isomap and LTSA backmapping).
        Note that none of the three nonlinear DR techniques reviewed in our work seem to stand out in terms of performance.
        Alternative nonlinear DR techniques could be investigated to improve the multi-fidelity surrogate models prediction accuracy.
        \item When considering fusion surrogates, CPCA or SFPCA-MA should be preferred over LFPCA, HFPCA and MFPCA if the additional training and prediction computational cost is acceptable (see \Cref{sec:discussion-which-pca}).
        \item When the high- and low-fidelity snapshots are not mapped to a common manifold, manifold alignment should be used in fusion surrogates, as it quite consistently improves the performance for a negligible additional computational cost (see \Cref{sec:discussion-manifold-alignment}).
        \item If the emulated fields have a feature moving along the mesh (\textit{e.g.}, the boundary layer separation on the NACA~0015 airfoil described in \Cref{sec:naca-case}), it might be difficult for linear DR to efficiently and accurately compress data. Nonlinear DR could offer better results. However, none of the reviewed surrogates using nonlinear DR showed better performances in this situation, which further encourages the investigation of alternative nonlinear DR techniques.
        \item Most importantly, this study shows that no surrogate is better in every case, which is another example of the No Free Lunch theorem \cite{hoSimple2002}. 
        It is therefore recommended to test several surrogates on any new task, and compare the results to opt for the most consistent solution.
    \end{itemize}

\section*{Acknowledgment}

    This work is co-funded by ONERA and the Agence de l’innovation de défense.

\section*{Reproducibility and benchmark data}

    The Python code used for the different multi-fidelity surrogate models, along with the training and validation datasets, are available upon request.

\appendix

\section{Numerical settings}\label{app:numerical-settings}

    \begin{itemize}

        \item \textbf{Hardware:} all repetitions are run on an Intel\textsuperscript{\textregistered}\ Xeon\textsuperscript{\textregistered}\ E5-2650 v4 (2.20GHz) CPU
    
        \item \textbf{Dimensionality reduction}

        \begin{itemize}
            \item  \textbf{PCA:} the Singular Value Decomposition (SVD) of \texttt{scipy} \cite{virtanenSciPy2020} is used, \{$\varnothing$, SF, HF, LF, MF, Diff, G\}PCA are truncated with a RIC of 99.9\%.
            \item \textbf{CPCA:} all modes are retained in the SVDs at all stages.
            \item \textbf{Kernel PCA:} the implementation of \texttt{scikit-learn} \cite{pedregosaScikitlearn2011} is used, the size of the latent space is set to $n_1-1$, the kernel function is RBF.
            The inverse mapping is performed with the approach described in \cite{bakirLearning2003}. Both the hyperparameters of the forward and inverse mapping are computed by minimizing the DR error with a $k$-fold cross-validation strategy with $k=\lfloor 0.5\times n_2 \rfloor$, where $\lfloor\cdot\rfloor$ is the floor function.
            \item \textbf{Isomap and LTSA:} the implementation of \texttt{scikit-learn} \cite{pedregosaScikitlearn2011} is used for the mapping from the high-dimensional space to the latent space. The training of the hyperparameters and the inverse mapping is done as explained by Decker \textit{et al}.\ \cite{deckerManifold2022}, \textit{i.e.}, the dimensionality of the latent space is set to the dimensionality of the input variable space $\dim(U)$, the number of nearest neighbors is computed by minimizing the Kruskal's stress \cite{kruskalMultidimensional1964, franzInterpolationbased2014} in isomap and the variance of distance ratios \cite{shiModel2009} in LTSA, and the inverse mapping is the one described by Roweis and Saul \cite{roweisNonlinear2000}.
        \end{itemize}
        
        \item \textbf{Intermediate surrogate modeling}

        \begin{itemize}
            \item \textbf{Single-fidelity GP regression:} single-fidelity GP regression is performed with \texttt{smt} \cite{savesSMT2023} with constant trend. The kernel function is Matérn-5/2 whose hyperparameters are computed by maximizing the likelihood with the COBYLA algorithm \cite{powellDirect1994} with 20 restarts, the default nugget value is used.
            \item \textbf{GP regression with mixed continuous and discrete variables:} the implementation of \texttt{smt} \cite{savesSMT2023} is used, the kernel function for continuous variables is Matérn-5/2 and the kernel function of discrete variables is based on Gower distance \cite{savesGeneral2022}. The hyperparameters are computed by maximizing the likelihood with the COBYLA algorithm \cite{powellDirect1994} with 20 restarts, the default nugget value is used.
            \item \textbf{Multi-fidelity AR1 co-Kriging:} a modified version of the AR1 co-Kriging of \texttt{OpenMDAO} \cite{grayOpenMDAO2019} that provides support for multistart is used. The regression function is constant and the kernel function is Matérn-5/2 whose hyperparameters are computed by maximizing the likelihood with the COBYLA algorithm \cite{powellDirect1994} with 20 restarts. The nugget value is set to $10^{-10}$.
            \item \textbf{GP regression with tensorized covariance:} it has been implemented following the description of Kerleguer \cite{kerleguerMultifidelity2023}, with a Matérn-5/2 kernel function with no nugget value. The hyperparameters are computed by minimizing the prediction error with a $k$-fold cross-validation strategy with $k=\min{(10, n_1)}$ with the COBYLA algorithm \cite{powellDirect1994}.
        \end{itemize}

        \item \textbf{Design of experiments}

        \begin{itemize}
            \item \textbf{Nested LHS:} a modified version of the implementation of \texttt{smt} \cite{savesSMT2023} is used, adding the support for variable $n_1/n_2$ ratios.
        \end{itemize}

    \end{itemize}

\section{Computational training cost}\label{app:computational-cost}

\begin{figure}[H]
    \includegraphics[width=\textwidth]{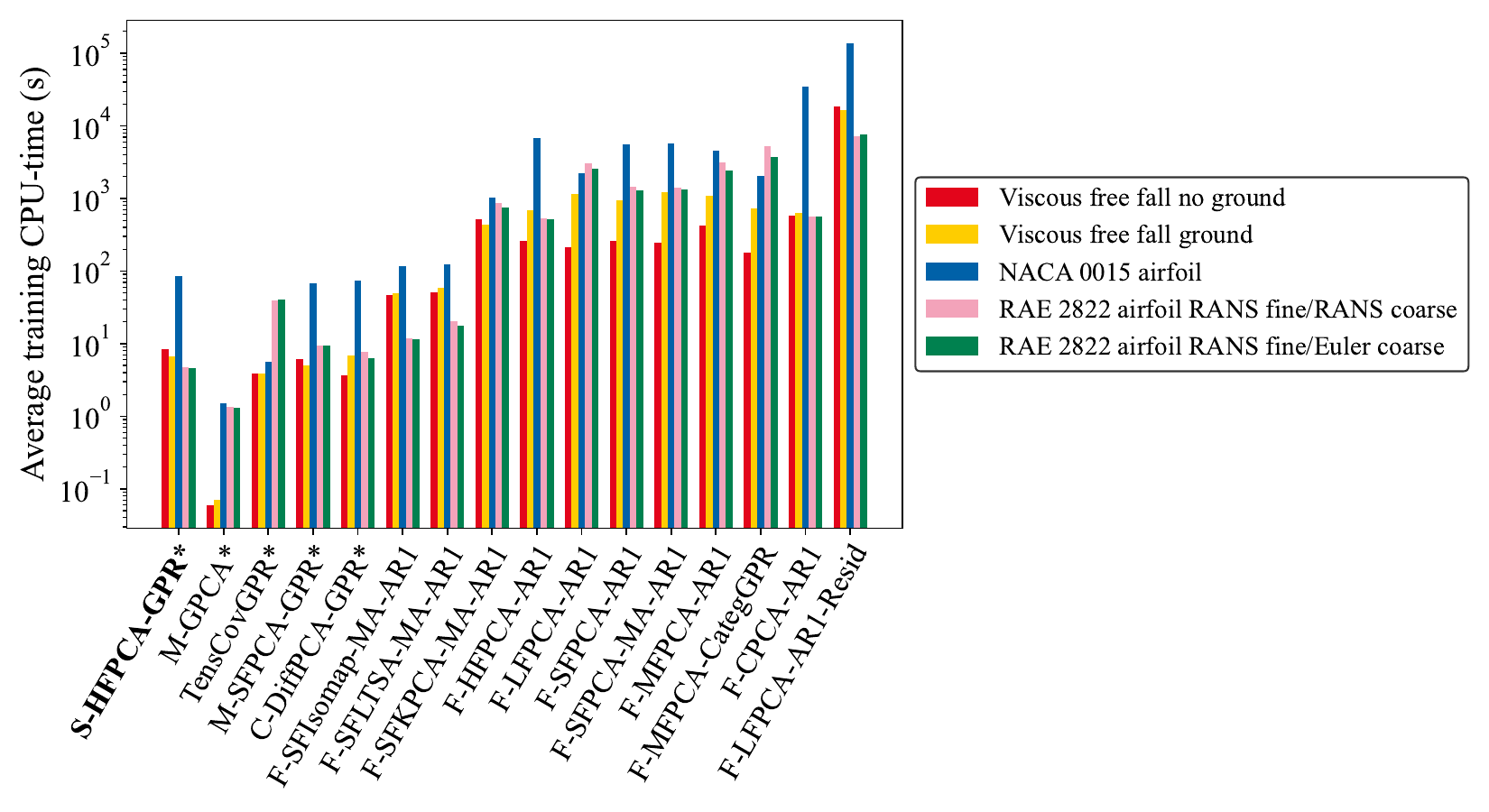}
    \caption{Histogram of the training CPU-time in seconds. The colors of the bars correspond to the different tests cases. Details about the implementation and hardware used can be found in \ref{app:numerical-settings}.}
    \label{fig:cpu-time-training}
\end{figure}

The computational training cost in terms of CPU-time of the different surrogates across the different test cases are displayed in \Cref{fig:cpu-time-training}.
This CPU-time does not include the simulation time needed to get the training samples.
Added to the fact that the DoEs have the same size, the given computational costs are direct translations of the computational complexities of the surrogate models.
To ensure a fair comparison, the different surrogates are using similar implementations (see \ref{app:numerical-settings} for more details on the chosen libraries and the tuning of hyperparameters).
The training CPU-time is averaged across all combinations of ($n_1$, $n_2/n_1$, index of repetition).

It can be seen that the NACA~0015 airfoil case incurs a significantly larger training cost than all other test cases.
This discrepancy suggests that the training computational cost is not only influenced by the dimensionality of the fields. Indeed, the order of magnitude of the number of nodes in the NACA~0015 is $10^3$, while it is $10^4$ in the RAE~2822 airfoil.
In this benchmark, one of the main contributors to the overall training computational cost is the dimensionality of the latent space.
In fact, one surrogate per latent variable must be built and trained.
The dimension of the latent space of the NACA~0015 is larger than that of the RAE~2822 airfoil.
Regarding the comparison of the training computational cost of the different surrogates, the M-GPCA surrogate has the lowest training computational cost (lower than a second on average for all test cases).
This efficiency can be attributed to the fact that building GPCA involves only a few linear algebra operations with one hyperparameters that needs tuning, unlike the surrogates using GP-based regression (\textit{i.e.}, GPR, AR1 and CategGPR).
The single-fidelity surrogate TensCovGPR is the second fastest surrogate.
Then, there are multi-fidelity surrogates using Isomap and LTSA.
This can be explained by the fact that the dimensionality of the latent space is small, leading to the training of few intermediate surrogates (see \ref{app:numerical-settings} for more details about the numerical settings).
C-DiffPCA-GPR has a training computational cost similar to the reference single-fidelity surrogate.
Subsequently, the next multi-fidelity surrogate incurs a computational cost approximately 4.5 times larger than the reference single-fidelity surrogate.
The longest multi-fidelity surrogate has an average training CPU-time about 64 times larger than the single-fidelity surrogate.
F-SFPCA-MA-AR1 is slightly faster to train than F-SFPCA-AR1. Manifold alignment involves only a few linear algebra operations in the low-dimensional latent space, resulting in a negligible additional computational cost.
This suggests that manifold alignment, by increasing the correlation between the high- and low-fidelity data, accelerates the training of the multi-fidelity intermediate surrogate model.
The small increase in training time of the F-LFPCA-AR1-Resid over the F-LFPCA-AR1 except for the NACA~0015 airfoil indicates that the modeling of the DR residuals Kriging with tensorized covariance should usually not be too costly.

\section{Comparison of the decay of 1 - RIC for the different types of PCA}\label{app:latent-dimension}

\begin{figure}[H]
    \centering
    \begin{subfigure}{0.49\textwidth}
        \includegraphics[trim={0 0 0 1cm},clip,width=\textwidth]{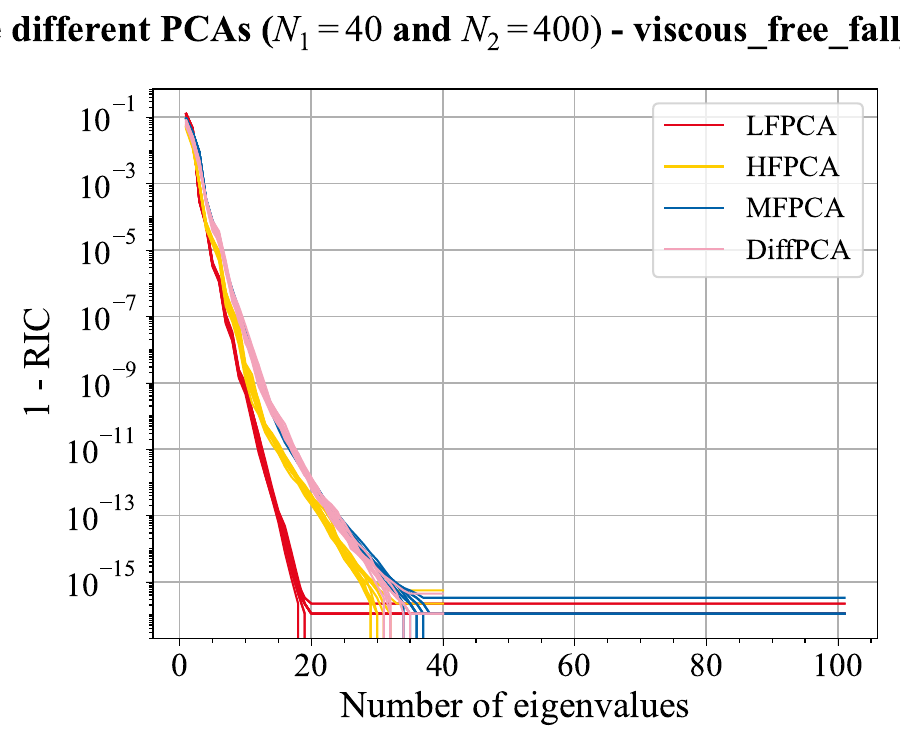}
    \end{subfigure}
    \begin{subfigure}{0.49\textwidth}
        \includegraphics[trim={0 0 0 1cm},clip,width=\textwidth]{figures/decay_per_podtype_viscous_free_fall_ground.pdf}
    \end{subfigure}
    
    \caption{Decay of 1~-~RIC for the different PCAs for the viscous free fall without ground (left) and with a ground (right) test cases with $n_1=40$ and $n_2=400$ for 10 repetitions each.}
    \label{fig:decay-podtype-vff}
\end{figure}

\begin{figure}[H]
    \centering
    \includegraphics[trim={0 0 0 1cm},clip,width=0.49\textwidth]{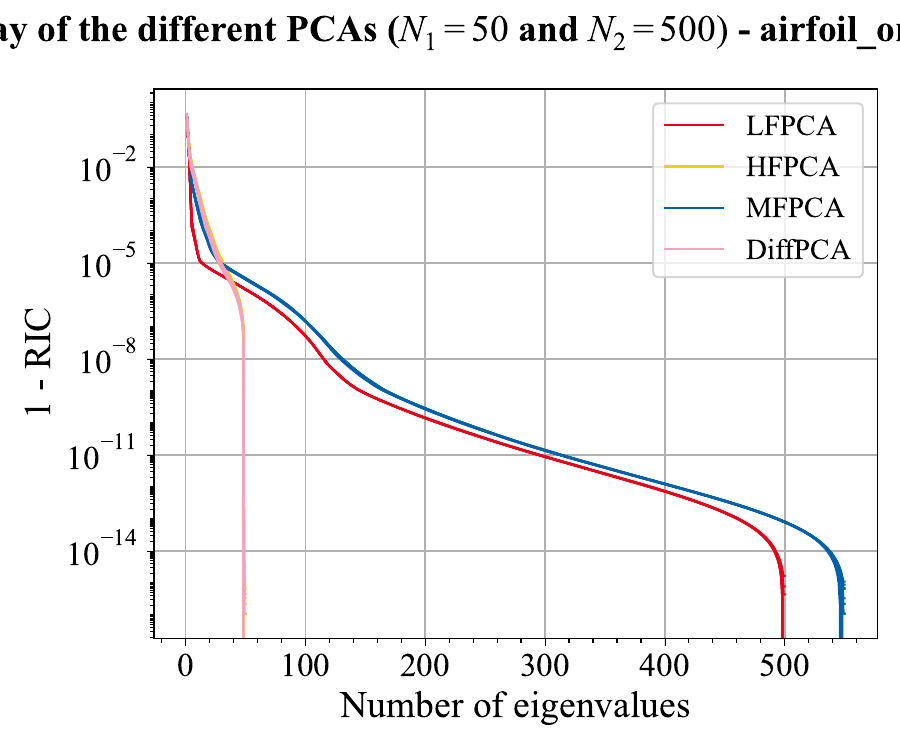}
    \caption{Decay of 1~-~RIC for the different PCAs for the NACA~0015 test case with $n_1=50$ and $n_2=500$ for 10 repetitions each.}
    \label{fig:decay-podtype-ao}
\end{figure}

\begin{figure}[H]
    \centering
    \begin{subfigure}{0.49\textwidth}
        \includegraphics[trim={0 0 0 1cm},clip,width=\textwidth]{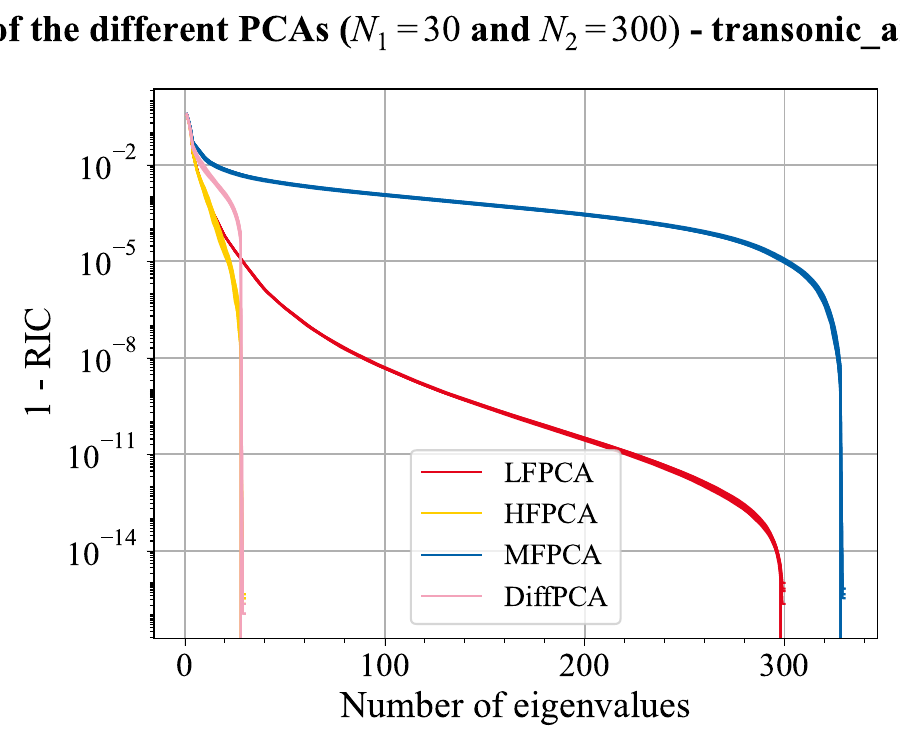}
    \end{subfigure}
    \begin{subfigure}{0.49\textwidth}
        \includegraphics[trim={0 0 0 1cm},clip,width=\textwidth]{figures/decay_per_podtype_transonic_airfoil.pdf}
    \end{subfigure}

    \caption{Decay of 1~-~RIC for the different PCAs for the RAE~2822 airfoil test case RANS/RANS configuration (left) and RANS/Euler configuration (right) with $n_1=30$ and $n_2=300$ for 10 repetitions each.}
    \label{fig:decay-podtype-ta}
\end{figure}

\Cref{tab:latent-dimension} shows the dimensionality of the latent space for the different PCAs over the different study cases.
The RIC is set to 99.9\%.
Only the configuration with the most snapshots is considered, \textit{i.e.}, $n_1=10\times\operatorname{dim}(U)$ and $n_2=10\times n_1$.
As there are ten repetitions, the results are given as intervals or singletons (when the latent space dimensionality is the same for every repetition).
To give some context, the dimensionality of the original spaces of the fields is reminded in the second column.

\begin{table}[H]
    \centering
    \scriptsize
    \caption{Intervals of the dimensionality of the latent space for the different types of PCA over the different study cases, with $n_1=10\times\operatorname{dim}(U)$ and $n_2=10\times n_1$ and a RIC of 99.9\%. In the table the results are intervals or singletons when the latent space dimensionality is the same for all repetitions.}
    \begin{tabular}{l c c c c c}\toprule
         Test case & $d_{\mathbf{y}_1}/d_{\mathbf{y}_2}$ & HFPCA & LFPCA & MFPCA & DiffPCA \\
         \midrule
         Viscous free fall without ground & 101/101 & [2,3] & \{2\} & \{3\} & \{3\} \\
         Viscous free fall with ground & 101/101 & [2,3] & [3,4] & \{4\} & [3,4] \\
         NACA 0015 & 1{,}001/1{,}001 & [10,12] & \{4\} & [7,8] & [10,12] \\
         RAE 2822 RANS/RANS & 41{,}796/10{,}530 & [9,11] & \{10\} & [104,114] & [20,21] \\
         RAE 2822 RANS/Euler & 41{,}796/8{,}910 & [9,11] & \{11\} & [73,80] & [19,21] \\
         \bottomrule
    \end{tabular}
    \label{tab:latent-dimension}
\end{table}

\section{Illustration of the influence of the relative information content threshold on the results of the benchmark}\label{app:ric}

\begin{figure}[H]
        \begin{subfigure}{0.45\textwidth}
            \centering
            \includegraphics[width=0.9\textwidth]{figures/viscous_free_fall_no_groundboxplot_Nhf_40_Nlf_400.pdf}
            \subcaption{\centering Viscous free fall without ground, RIC=99.9\%, $n_1=40$ and $n_2=400$}
        \end{subfigure}
        \hfill
        \begin{subfigure}{0.45\textwidth}
            \centering
            \includegraphics[width=0.9\textwidth]{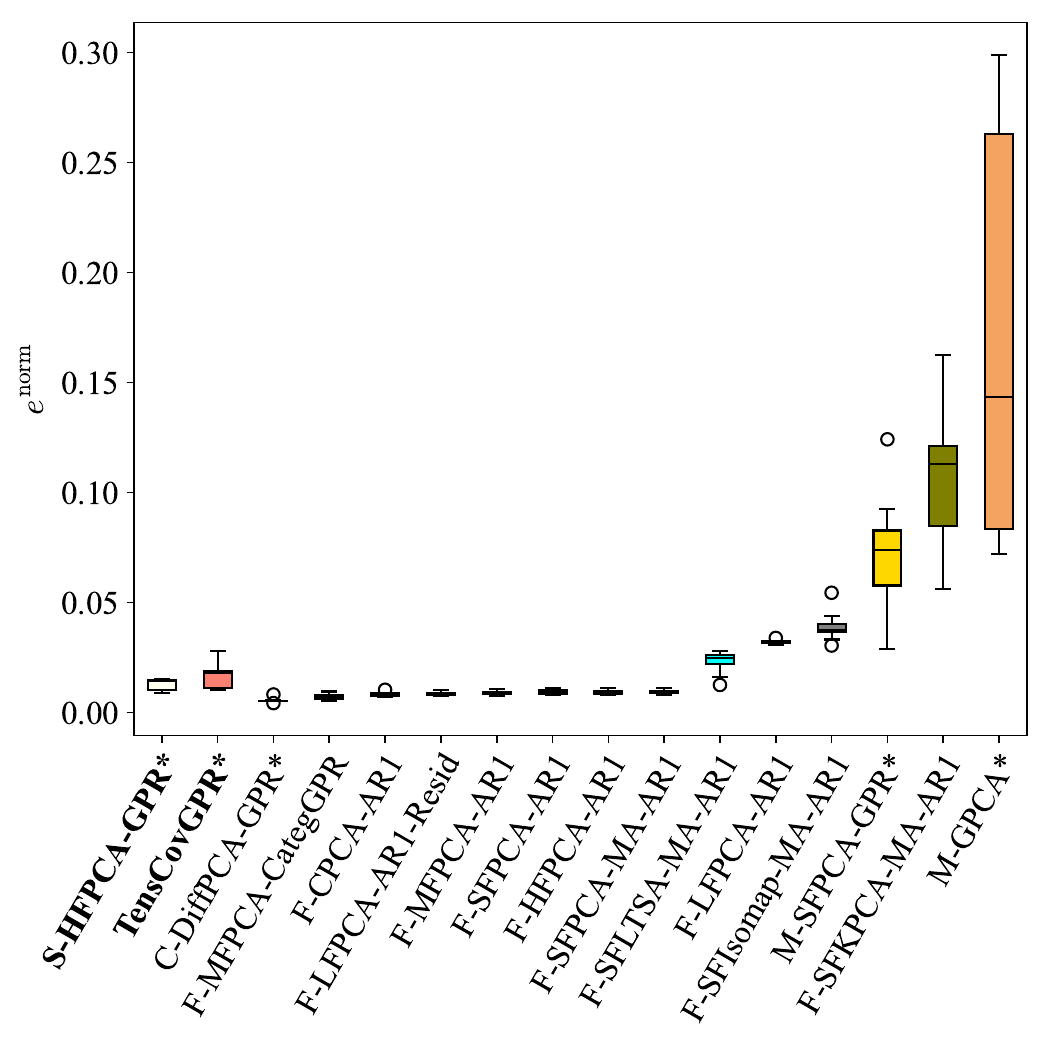}
            \subcaption{\centering Viscous free fall without ground, RIC=99.9999\%, $n_1=40$ and $n_2=400$}
        \end{subfigure}
        \\
        \\
        \begin{subfigure}{0.45\textwidth}
            \centering
            \includegraphics[width=0.9\textwidth]{figures/transonic_airfoil_2boxplot_Nhf_30_Nlf_300.pdf}
            \subcaption{\centering RAE~2822 RANS/RANS, RIC=99.9\%, $n_1=30$ and $n_2=300$}
        \end{subfigure}
        \hfill
        \begin{subfigure}{0.45\textwidth}
            \centering
            \includegraphics[width=0.9\textwidth]{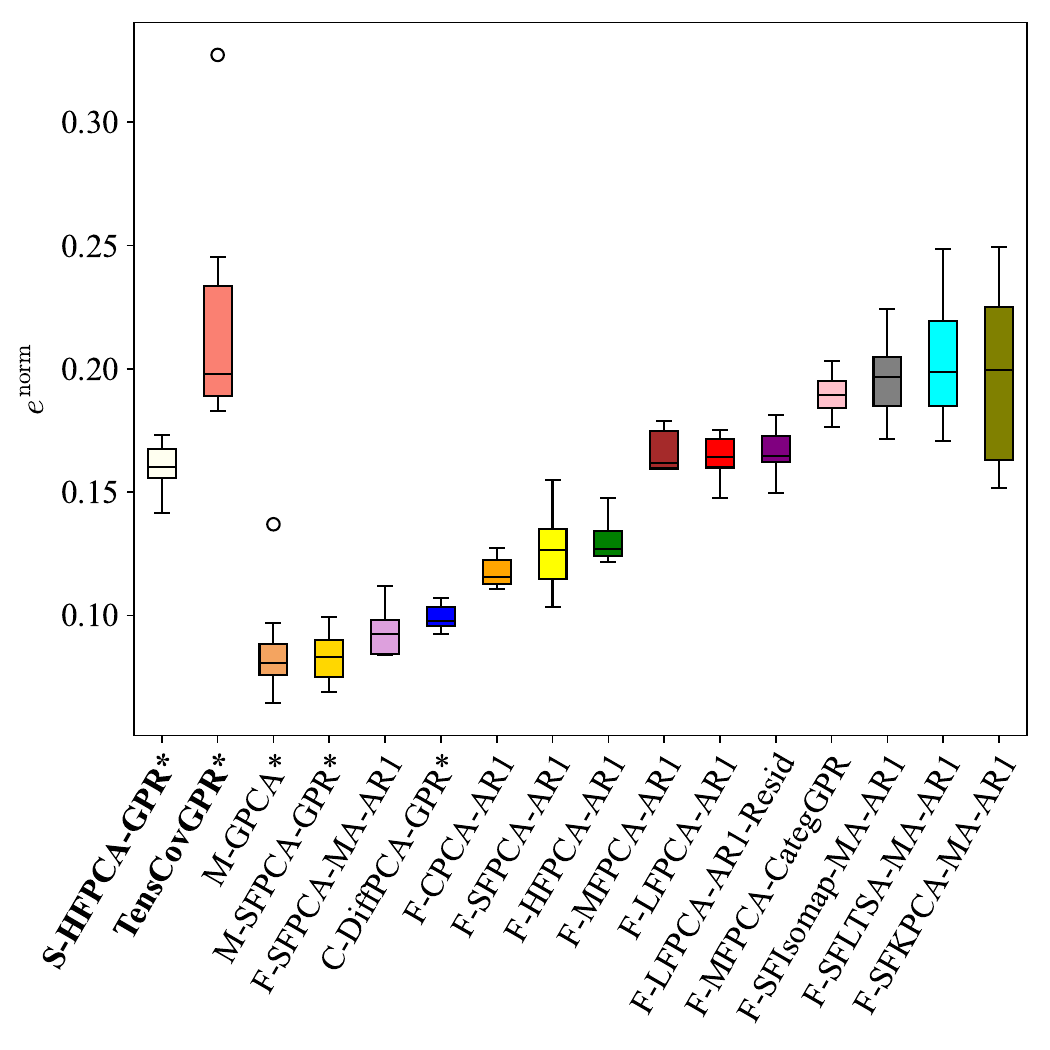}
            \subcaption{\centering RAE~2822 RANS/RANS, RIC=99.9999\%, $n_1=30$ and $n_2=300$}
        \end{subfigure}
        \caption{Boxplots of the normed RMSE for two test cases with RIC=99.9\% (left column) and RIC=99.9999\% (right column) for the viscous free fall case with ground (top row) and RAE 2822 RANS/RANS case (bottom row). Single-fidelity surrogates (bold label) are stacked to the left and multi-fidelity surrogates are sorted by median normed RMSE.}
        \label{fig:ric}
    \end{figure}

    Given that many of the benchmarked surrogates use PCA with a truncation on the RIC value, this is perhaps the most notable numerical setting.
    \Cref{fig:ric} shows the boxplots of the normed RMSE for the viscous free fall without ground (top row) and the RAE 2822 RANS/RANS (bottom row) case for a RIC of 99.9\% (left column) and 99.9999\% (right column).
    It shows that the first case is significantly influenced by the change of RIC, whereas the second is much less affected.
    This highlights the dependence of the conclusions of this paper on the choice of the numerical setting provided in \ref{app:numerical-settings}.
    
\section{Decomposition of the normed RMSE into normed DR and intermediate surrogate modeling RMSE}\label{app:error-decomposition}

    \Cref{tab:error-decomposition-vffng,tab:error-decomposition-vffg,tab:error-decomposition-ao} show that for the viscous free fall and NACA~0015 test cases, the main contribution to the prediction normed RMSE is the error due to the intermediate surrogate.
    For the two RAE~2822 test cases (\Cref{tab:error-decomposition-ta2,tab:error-decomposition-ta}), the breakdown of the prediction RMSE is more even, although the intermediate surrogate remains the main contributor.

    \begin{table}[H]
        \centering
        \caption{Decomposition of the normed RMSE $e^{\text{norm}}$ into normed DR RMSE $e_\text{dr}^{\text{norm}}$ and normed intermediate surrogate modeling RMSE $e_\text{ism}^{\text{norm}}$ for the viscous free fall without ground test case, for every configuration of $(n_1,n_2)$ combined.}
        \begin{tabular}{l c c c}\toprule
             Surrogate & $e^{\text{norm}}$ (\%) & $e_\text{dr}^{\text{norm}}$ (\%) & $e_\text{ism}^{\text{norm}}$ (\%) \\
             \midrule
             F-HFPCA-AR1 & 13.4 & 8.7 & 8.0 \\
             F-LFPCA-AR1 & 15.8 & 11.4 & 8.2 \\
             F-MFPCA-AR1 & 12.3 & 7.2 & 8.1 \\
             F-MFPCA-CategGPR & 16.7 & 7.2 & 13.0 \\
             F-SFPCA-AR1 & 13.4 & 8.7 & 8.1 \\
             F-SFPCA-MA-AR1 & 13.3 & 8.7 & 8.0 \\
             S-HFPCA-GPR & 19.2 & 8.7 & 14.4 \\
             \bottomrule
        \end{tabular}
        \label{tab:error-decomposition-vffng}
    \end{table}

    \begin{table}[H]
        \centering
        \caption{Decomposition of the normed RMSE $e^{\text{norm}}$ into normed DR RMSE $e_\text{dr}^{\text{norm}}$ and normed intermediate surrogate modeling RMSE $e_\text{ism}^{\text{norm}}$ for the viscous free fall with ground test case, for every configuration of $(n_1,n_2)$ combined.}
        \begin{tabular}{l c c c}\toprule
             Surrogate & $e^{\text{norm}}$ (\%) & $e_\text{dr}^{\text{norm}}$ (\%) & $e_\text{ism}^{\text{norm}}$ (\%) \\
             \midrule
            F-HFPCA-AR1 & 13.7 & 8.7 & 8.7 \\
            F-LFPCA-AR1 & 11.6 & 5.3 & 9.4 \\
            F-MFPCA-AR1 & 10.3 & 3.3 & 9.1 \\
            F-MFPCA-CategGPR & 15.8 & 3.3 & 15.2 \\
            F-SFPCA-AR1 & 9.7 & 1.7 & 9.4 \\
            F-SFPCA-MA-AR1 & 9.5 & 1.7 & 9.2 \\
            S-HFPCA-GPR & 19.2 & 8.7 & 14.4 \\
            \bottomrule
        \end{tabular}
        \label{tab:error-decomposition-vffg}
    \end{table}

    \begin{table}[H]
        \centering
        \caption{Decomposition of the normed RMSE $e^{\text{norm}}$ into normed DR RMSE $e_\text{dr}^{\text{norm}}$ and normed intermediate surrogate modeling RMSE $e_\text{ism}^{\text{norm}}$ for the NACA~0015 test case, for every configuration of $(n_1,n_2)$ combined.}
        \begin{tabular}{l c c c}\toprule
             Surrogate & $e^{\text{norm}}$ (\%) & $e_\text{dr}^{\text{norm}}$ (\%) & $e_\text{ism}^{\text{norm}}$ (\%) \\
             \midrule
             F-HFPCA-AR1 & 19.7 & 7.7 & 18.5 \\
             F-LFPCA-AR1 & 21.1 & 15.4 & 14.1 \\
             F-MFPCA-AR1 & 19.2 & 10.0 & 16.4 \\
             F-MFPCA-CategGPR & 25.3 & 10.0 & 23.4 \\
             F-SFPCA-AR1 & 20.1 & 7.7 & 19.0 \\
             F-SFPCA-MA-AR1 & 19.8 & 7.7 & 18.7 \\
             S-HFPCA-GPR & 20.4 & 7.7 & 19.2 \\
             \bottomrule
        \end{tabular}
        \label{tab:error-decomposition-ao}
    \end{table}

    \begin{table}[H]
        \centering
        \caption{Decomposition of the normed RMSE $e^{\text{norm}}$ into normed DR RMSE $e_\text{dr}^{\text{norm}}$ and normed intermediate surrogate modeling RMSE $e_\text{ism}^{\text{norm}}$ for the RANS/RANS configuration of the RAE~2822 airfoil test case, for every configuration of $(n_1,n_2)$ combined.}
        \begin{tabular}{l c c c}\toprule
             Surrogate & $e^{\text{norm}}$ (\%) & $e_\text{dr}^{\text{norm}}$ (\%) & $e_\text{ism}^{\text{norm}}$ (\%) \\
             \midrule
             F-HFPCA-AR1 & 22.8 & 19.3 & 13.7 \\
             F-LFPCA-AR1  & 22.3 & 13.3 & 18.5 \\
             F-MFPCA-AR1 & 21.4 & 10.1 & 18.7 \\
             F-MFPCA-CategGPR & 22.0 & 10.1 & 19.5 \\
             F-SFPCA-AR1 & 23.8 & 18.3 & 17.3 \\
             F-SFPCA-MA-AR1 & 22.3 & 18.3 & 13.8 \\
             S-HFPCA-GPR & 25.7 & 19.3 & 21.0 \\
             \bottomrule
        \end{tabular}
        \label{tab:error-decomposition-ta2}
    \end{table}

    \begin{table}[H]
        \centering
        \caption{Decomposition of the normed RMSE $e^{\text{norm}}$ into normed DR RMSE $e_\text{dr}^{\text{norm}}$ and normed intermediate surrogate modeling RMSE $e_\text{ism}^{\text{norm}}$ for the RANS/Euler configuration of the RAE~2822 airfoil test case, for every configuration of $(n_1,n_2)$ combined.}
        \begin{tabular}{l c c c}\toprule
             Surrogate & $e^{\text{norm}}$ (\%) & $e_\text{dr}^{\text{norm}}$ (\%) & $e_\text{ism}^{\text{norm}}$ (\%) \\
             \midrule
             F-HFPCA-AR1 & 24.6 & 19.3 & 17.6 \\
             F-LFPCA-AR1 & 30.9 & 21.7 & 22.2 \\
             F-MFPCA-AR1 & 27.4 & 11.1 & 24.7 \\
             F-MFPCA-CategGPR & 28.3 & 11.1 & 26.1 \\
             F-SFPCA-AR1 & 25.4 & 18.1 & 19.9 \\
             F-SFPCA-MA-AR1 & 24.7 & 18.1 & 18.9 \\
             S-HFPCA-GPR & 25.7 & 19.3 & 21.0 \\
             \bottomrule
        \end{tabular}
        \label{tab:error-decomposition-ta}
    \end{table}

\bibliographystyle{elsarticle-num}
\bibliography{main.bib}

\end{document}